\documentclass[iop,apj,tighten]{emulateapj}
\usepackage{apjfonts} 
\usepackage{rotating}
\usepackage{array}
\usepackage{afterpage}
\usepackage{float}
\usepackage{morefloats}
\usepackage{graphicx}
\usepackage{amsmath}
\usepackage{color}
\usepackage[breaklinks,colorlinks,citecolor=blue,linkcolor=magenta]{hyperref}

\newcommand{\OIII}{\mbox{O\,\textsc{iii}}}

\newcommand{\OI}{\mbox{O\,\textsc{i}}}
\newcommand{\NII}{\mbox{N\,\textsc{ii}}}

\newcommand{\SII}{\mbox{S\,\textsc{ii}}}

\newcommand{\kms}{km s$^{-1}$}
\newcommand{\ergs}{erg s$^{-1}$}
\newcommand{\Ha}{H$\alpha$}    
\newcommand{\degree}{$^\circ$}

\slugcomment{Accepted for publication in ApJ (2016/02/06)}

\shorttitle{The limited impact of outflow}
\shortauthors{Bae et al.}

\begin{document}

\title{The limited impact of outflows: Integral-field spectroscopy of 20 local AGN\lowercase{s}}

\author{Hyun-Jin Bae$^{1,2}$}
\author{Jong-Hak Woo$^{2,}$\altaffilmark{3}}
\author{Marios Karouzos$^{2}$}
\author{Elena Gallo$^{4}$}
\author{Helene Flohic$^{5}$}
\author{Yue Shen$^{6}$}
\author{Suk-Jin Yoon$^{1}$}

\affil{$^{1}$Department of Astronomy and Center for Galaxy Evolution Research, Yonsei University, Seoul 03722, Republic of Korea; hjbae@galaxy.yonsei.ac.kr} 
\affil{$^{2}$Astronomy Program, Department of Physics and Astronomy, Seoul National University, Seoul 08826, Republic of Korea; woo@astro.snu.ac.kr}
\altaffiltext{3}{Author to whom any correspondence should be addressed}
\affil{$^{4}$Department of Astronomy, University of Michigan, Ann Arbor, MI 48109-1042, USA}
\affil{$^{5}$Department of Physics, University of the Pacific, 3601 Pacific Avenue, Stockton, CA 95211, USA}
\affil{$^{6}$Department of Astronomy, University of Illinois at Urbana-Champaign, Urbana, IL 61801, USA}

\begin{abstract}
To investigate AGN outflows as a tracer of AGN feedback on star-formation, we perform integral-field spectroscopy of 20 type 2 AGNs at z$<$0.1,
which are luminous AGNs with the [\OIII] luminosity 
$>$10$^{41.5}$ \ergs, and exhibit strong outflow signatures in the [\OIII] kinematics. By decomposing the emission-line profile, we obtain the maps of the narrow and broad components of [\OIII] and \Ha\ lines, respectively. The broad components in both [\OIII] and \Ha\ represent the non-gravitational kinematics, i.e., gas outflows, while the narrow components, especially in \Ha, represent the gravitational kinematics, i.e., rotational disk.
By using the integrated spectra within the flux-weighted size of the narrow-line region, 
we estimate the energetics of the gas outflows. The ionized gas mass is 1.0--38.5$\times 10^5 M_{\odot}$, and the mean mass outflow rate is 4.6$\pm$4.3 $M_\odot$ yr$^{-1}$, which is a factor of $\sim$260 higher than the mean mass accretion rate 0.02$\pm$0.01 $M_\odot$ yr$^{-1}$. The mean energy injection rate
of the sample is 0.8$\pm$0.6\% of the AGN bolometric luminosity, while the momentum flux is (5.4$\pm$3.6)$\times$$L_{\text{bol}}/c$ on average, except for two most kinematically energetic AGNs with low $L_{\text{bol}}$, which are possibly due to the dynamical timescale of the outflows.
The estimated outflow energetics are consistent with the theoretical expectations for energy-conserving outflows from AGNs, yet we find no supporting evidence of instantaneous quenching of star formation due to the outflows.  
\end{abstract}

\keywords{galaxies: active --- galaxies: kinematics and dynamics --- techniques: imaging spectroscopy}

\section{Introduction}
\label{intro}
The relatively tight scaling relationships between the mass of supermassive black holes (SMBHs) and host galaxy properties suggest SMBH-galaxy co-evolution \citep[for a review, see][]{2013ARA&A..51..511K}. Hydro-dynamic simulations show that powerful mechanical and/or radiative feedback from mass accreting SMBHs, or active galactic nuclei (AGN), sweeps a large fraction of interstellar medium (ISM), and therefore regulates the growth of both SMBHs and their host galaxies \citep{2006MNRAS.365...11C, 2012ApJ...745L..34Z, 2012MNRAS.425..605F, 2014MNRAS.440.2625Z,2015MNRAS.449.4105C}. The AGN feedback is now generally considered as one of the key ingredients of galaxy evolution scenarios, yet it is still unclear how AGN feedback affects the ISM at galactic scales. 

AGN-driven gas outflows can be a good tracer of the AGN feedback in action \citep[e.g.,][]{2005ApJ...632..751R,2012ApJ...746...86G}, as energetic outflows may significantly influence the surrounding ISM and also star formation in host galaxies. Observational studies have shown that gas outflows are prevalent among AGNs 
based on statistical investigations of the gas kinematics in the narrow-line region (NLR) \citep[e.g.,][]{2005AJ....130..381B,2011ApJ...737...71Z,2011ApJ...741...50W,2014ApJ...795...30B,2016ApJ...817..108W}. Thus, it is important to investigate AGN outflows to understand AGN feedback as well as the SMBH-galaxy co-evolution. 

Spectroscopic observations with integral-field units (IFUs) open a new window for studying AGN outflows by providing spatially resolved information on the gas and stellar kinematics. Extensive optical and near-infrared spectroscopic studies have been performed for low-z and high-z AGNs with IFU over the past decades \citep[e.g.,][]{2011ApJ...739...69M,2012ApJ...746...86G,2013A&A...549A..43H,2013MNRAS.436.2576L,2014A&A...562A..21C}. 

The main purpose of the IFU studies of AGN outflows is to obtain detailed kinematic information of AGNs and their surrounding gas. Extensive studies of nearby Seyfert galaxies have revealed the kinematics of gas outflows in the narrow-line region (NLR) \citep[e.g.,][]{2009MNRAS.396....2B, 2014MNRAS.438.3322S, 2014ApJ...780L..24R,2015ApJ...799..234F}, while other studies focused on the most luminous quasars (or QSOs) at higher z in order to catch the energetic AGN feedback in action \citep{2006ApJ...650..693N,2011ApJ...729L..27R,2012ApJ...746...86G,2013MNRAS.436.2576L,2014MNRAS.443..755H}. These studies have revealed the structure of gas outflows in nearby Seyfert galaxies and probed the energetics of the luminous AGNs, yet there is a lack of systematic investigation on the strong outflows based on a statistical sample, particularly in the local universe at z$<\sim$0.1.  

Recently, \citet[][see also Bae \& Woo 2014, Woo et al. 2017]{2016ApJ...817..108W} performed a census of ionized gas outflows using a large sample of $\sim$39,000 type 2 AGNs out to z$\sim$0.3, by investigating the luminosity-weighted velocity and velocity dispersion of the [\OIII] line at 5007\AA. Using the spatially-integrated SDSS spectra, they find that AGN outflows are prevalent among local AGNs. The sample of AGNs with detected gas outflows, particularly with extreme velocities, is very useful for follow-up spatially-resolved studies in investigating the nature of gas outflows and understanding the role of outflows as one of the potential AGN feedback mechanisms. As a pilot study, \citet{2016ApJ...819..148K,2016ApJ...833..171K} performed an IFU spectroscopy of six luminous type 2 AGNs using Gemini/GMOS-IFU, successfully demonstrating the power of IFU spectroscopy for studying both gas outflows and star-formation in local AGNs. 

In this paper, we present the IFU results of a luminosity-limited sample of 20 type 2 AGNs using the Magellan/IMACS-IFU and the VLT/VIMOS-IFU. 
The sample is selected as the best candidates with extreme gas kinematics from our previous statistical study  \citet{2016ApJ...817..108W}. 
By performing emission-line decomposition into narrow and broad components, we  try to understand the complex nature of the NLR and its kinematics, and estimate the energetics of the outflows (i.e., energy injection rate, momentum flux, mass outflow rate, and so on) based on simple physical assumptions.
In Section \ref{obs}, we describe the sample selection, observation, and reduction process. In Section \ref{analysis}, we describe the analysis method 
and 2-D map construction. In Section \ref{nlr_prop}, we present the NLR properties based on  [\OIII] and \Ha, respectively, and the size-luminosity relation for the NLR of AGNs. In Section \ref{energetics}, we present the energetics of the gas outflows, and we discuss the results in Section \ref{discussion}. Finally, summary and conclusions follow in Section \ref{summary}. We adopt a standard $\Lambda$CDM cosmology, i.e., Ho = 70 \kms\ Mpc$^{-1}$, $\Omega_{\Lambda}$ = 0.73, and $\Omega_{m}$ = 0.27, unless noted otherwise. 

\begin{figure}
\centering
\includegraphics[width=0.45\textwidth]{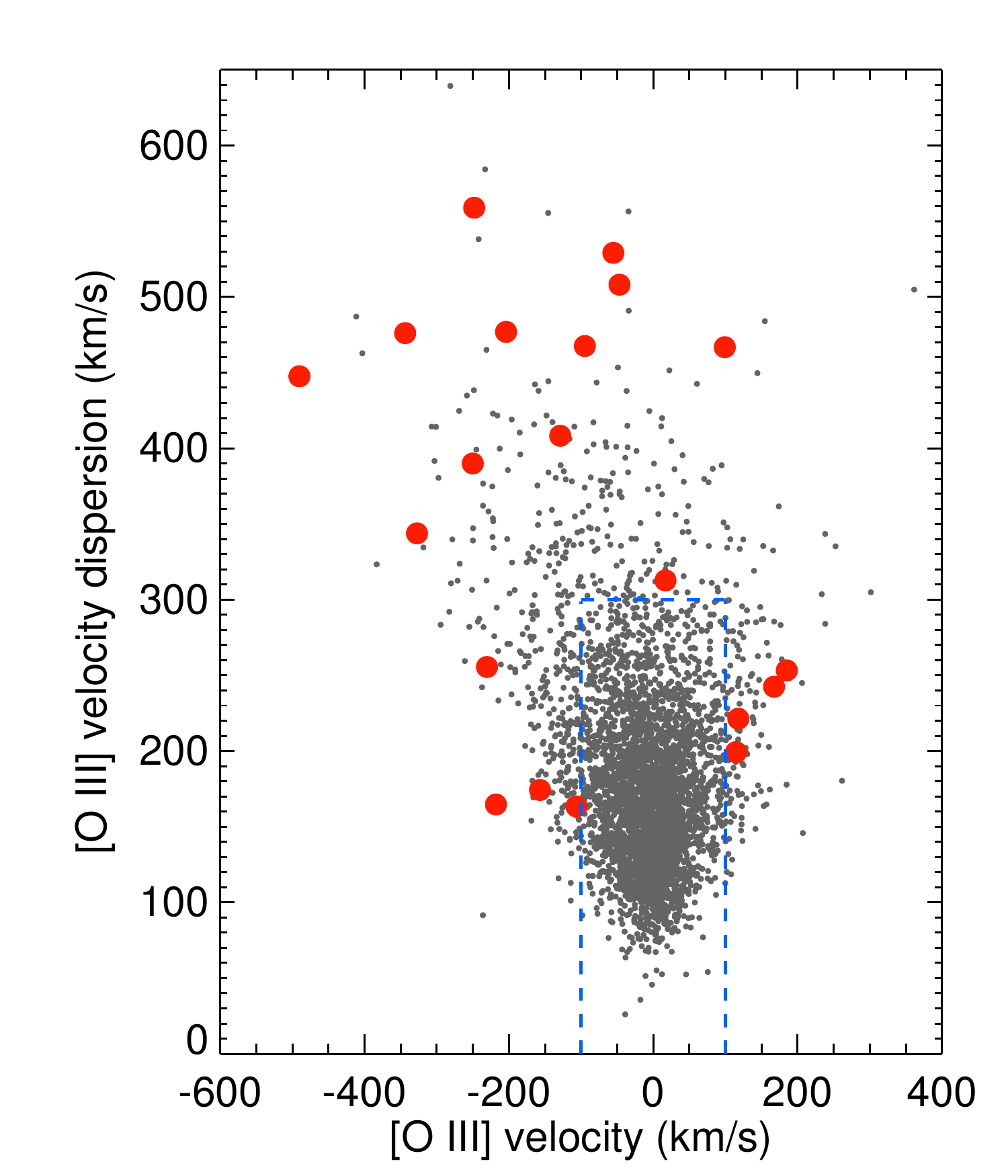}
\caption{The velocity-velocity dispersion diagram for 3396 luminous ($L_{[\text{O III}], cor} > 10^{41.5}$ \ergs) type 2 AGNs at z $<$ 0.1 (gray dots), and the 20 AGNs observed with IFU (red dots). The blue dashed lines indicate our selection criteria for the AGNs with signs of gas outflows, i.e., [\OIII] velocity offset $>$ 100 \kms\ or [\OIII] velocity dispersion $>$ 300 \kms. }
\label{fig:target}
\end{figure}

\begin{table*}
\center
\caption{Targets and observations\label{tbl-1}}
\begin{tabular}{cccccclccc}
\tableline
\tableline
SDSS name & $\alpha_{2000}$ & $\delta_{2000}$ & Redshift & log $L_{\text{[O III],cor}}$ & Class &  Obs. date  & Telescope/Instrument & seeing & $t_{exp}$    \\
           & hh mm ss.s & dd mm ss & &  & &  &  &   &\\
(1) & (2) & (3) & (4) & (5) & (6)  & (7) & (8) & (9)  & (10)\\
\tableline
J0130+1312  & 01 30 37.8 & +13 12 52 & 0.07272 & 42.01 & S & 2014 Dec 26 & Magellan/IMACS-IFU & 0.9 & 2400 \\
J0341$-$0719  &03 41 34.9 & $-$07 19 25 & 0.06603 & 41.90 & S & 2014 Dec 26 & Magellan/IMACS-IFU &0.7 & 7200 \\
J0806+1906  & 08 06 01.5 & +19 06 15 & 0.09799 & 42.34 & S &  2014 Dec 26 & Magellan/IMACS-IFU &0.8 & 3600 \\
J0855+0047 & 08 55 47.7 & +00 47 39 & 0.04185 &  41.63 & S & 2013 Apr  & VLT/VIMOS-IFU & 1.0  & 10050 \\
J0857+0633  & 08 57 59.0 & +06 33 08 & 0.07638 & 41.74 & S & 2014 Apr 06 & Magellan/IMACS-IFU &1.0 & 7200  \\
J0911+1333  & 09 11 24.2 & +13 33 20 & 0.07966 & 42.17 & S & 2014 Dec 27 & Magellan/IMACS-IFU &0.8 & 5400  \\
J0952+1937  & 09 52 59.0 & +19 37 55 & 0.02445 & 41.52 & S & 2014 Apr 05 & Magellan/IMACS-IFU &1.0 & 3600  \\
J1001+0954  & 10 01 40.5 & +09 54 32 & 0.05638 & 42.11 & S & 2014 Apr 06 & Magellan/IMACS-IFU &1.0 & 5400  \\
J1013$-$0054  & 10 13 46.8 & $-$00 54 51 & 0.04258 & 42.18 & S & 2014 Apr 04 &Magellan/IMACS-IFU & 1.0 & 3600  \\
J1054+1646  & 10 54 23.8 & +16 46 53 & 0.09498 &42.39 & L & 2014 Apr 27 & Magellan/IMACS-IFU &0.7 & 7200  \\
J1100+1321 &  11 00 03.9 & +13 21 50 & 0.06420 & 42.10 & S & 2014 Dec 26 & Magellan/IMACS-IFU &0.8 & 7200  \\
J1100+1124 & 11 00 37.2 & +11 24 55 & 0.02700  & 40.88 & S & 2013 Apr    & VLT/VIMOS-IFU & 0.8 & 8040 \\
J1106+0633 & 11 06 30.7 & +06 33 34 & 0.04044 & 41.90  & S &  2013 Apr  & VLT/VIMOS-IFU & 0.8 & 8040 \\
J1147+0752 &  11 47 20.0 & +07 52 43 & 0.08271 & 42.33 & L & 2014 Dec 28 & Magellan/IMACS-IFU &0.9 & 7200  \\
J1214$-$0329 &  12 14 51.2 & $-$03 29 22 & 0.03382 & 42.41 & S & 2014 Apr 04 & Magellan/IMACS-IFU & 0.7 & 3600  \\
J1310+0837 &  13 10 57.3 & +08 37 39 & 0.05211 & 41.76 & S & 2014 Apr 04 & Magellan/IMACS-IFU &1.0 & 3600  \\
J1440+0556 &  14 40 18.0 & +05 56 34 & 0.06105 & 42.45 & S & 2014 Apr 06 & Magellan/IMACS-IFU &1.1 & 5400  \\
J1448+1055 &  14 48 38.5 & +10 55 36 & 0.08930 & 42.83 & S & 2014 Apr 06 & Magellan/IMACS-IFU &1.3 & 5400  \\
J1520+0757 &  15 20 33.7 & +07 57 12 & 0.04343 & 41.11 & S & 2014 Apr 06 & Magellan/IMACS-IFU &1.3 & 5400  \\
J2039+0033 &  20 39 07.0 & +00 33 16 & 0.04835 & 42.15 & S & 2014 May 21 & Magellan/IMACS-IFU &1.0 & 10800  \\
\tableline
\end{tabular}
\tablecomments{(1) SDSS name of AGN; (2) Right ascension (J2000); (3) Declination (J2000); (4) Redshift derived from the brightest spaxel of observed spectra; (5) Extinction-corrected [\OIII] luminosity in logarithm (\ergs); (6) Class from the emission-line diagnostics, i.e., Seyfert (S), \& LINER (L); (7) Date of observation (local); (8) Telescope and instrument; (9) Seeing size (\arcsec); (11) Exposure time (second). }
\end{table*}

\begin{figure}
\centering
\includegraphics[width=0.48\textwidth]{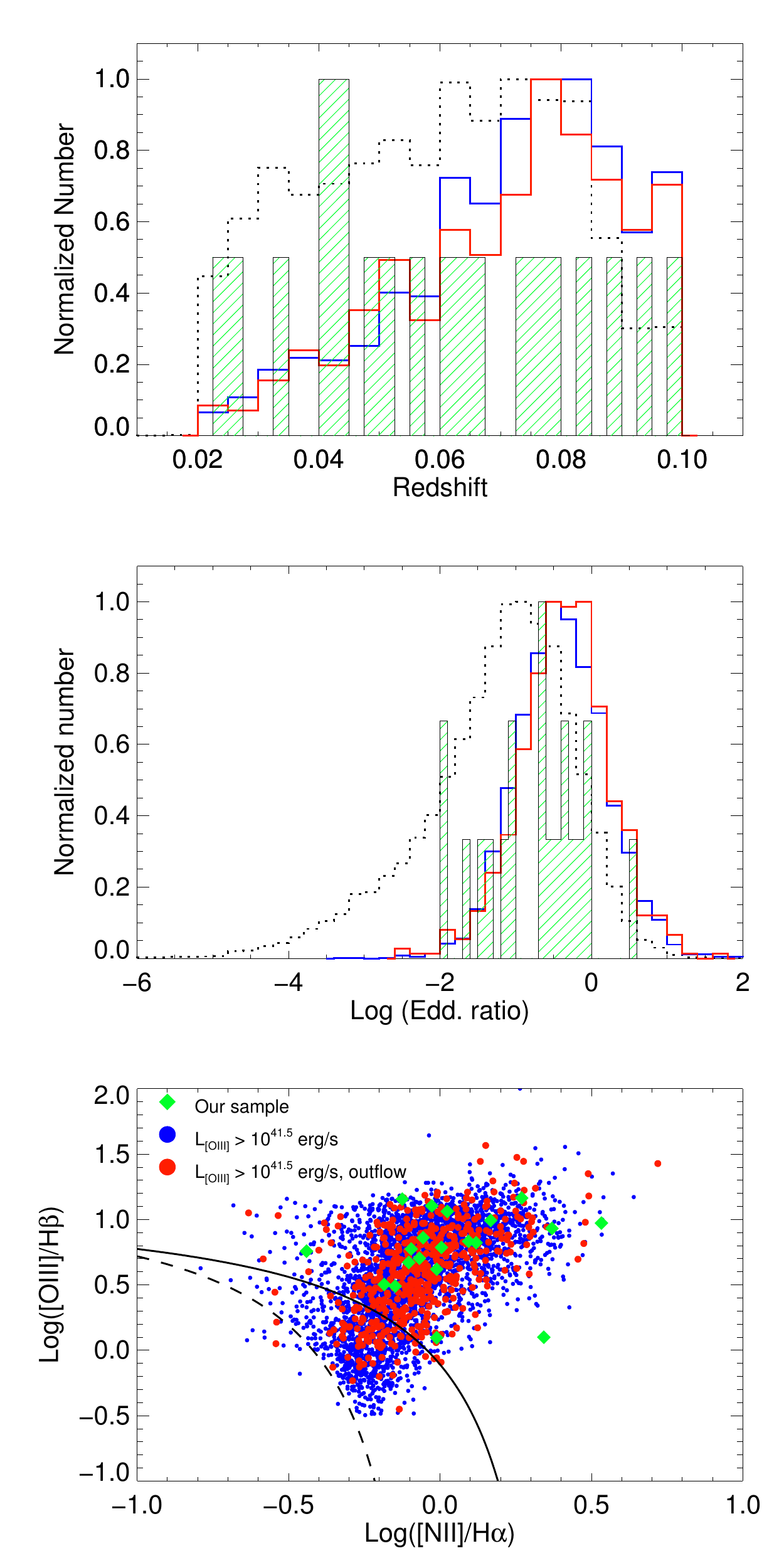}
\caption{The distributions of redshifts (top) and Eddington ratios (middle) for type 2 AGNs at z $<$ 0.1 (dotted line), for the luminosity-limited ($L_{\text{[O III]}} > 10^{41.5}$ \ergs) sample (blue lines), and for the strong outflow AGNs from the luminosity-limited sample (red lines, $v_{\text{[O III]}} > 100$ \kms\ and/or $\sigma_{\text{[O III]}} > 300$ \kms). Green histograms represent the distributions of 20 IFU-observed AGNs. The emission-line diagnostic diagram for the samples is at the bottom. The 20 IFU-observed AGNs are denoted with green diamonds.}
\label{fig:prop1}
\end{figure}

\section{Observations and Reduction}
\label{obs}

\begin{figure}
\centering
\includegraphics[width=0.48\textwidth]{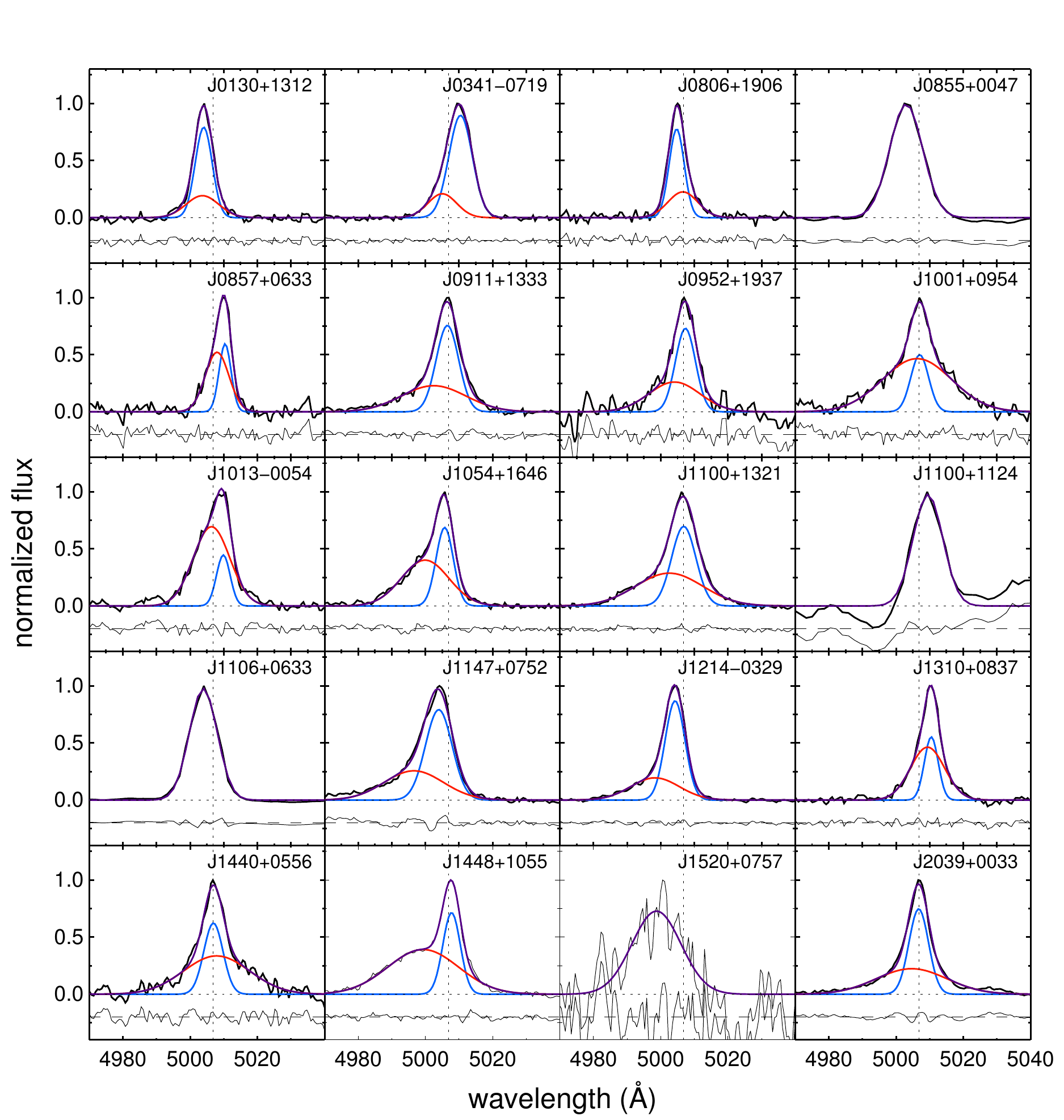}
\includegraphics[width=0.48\textwidth]{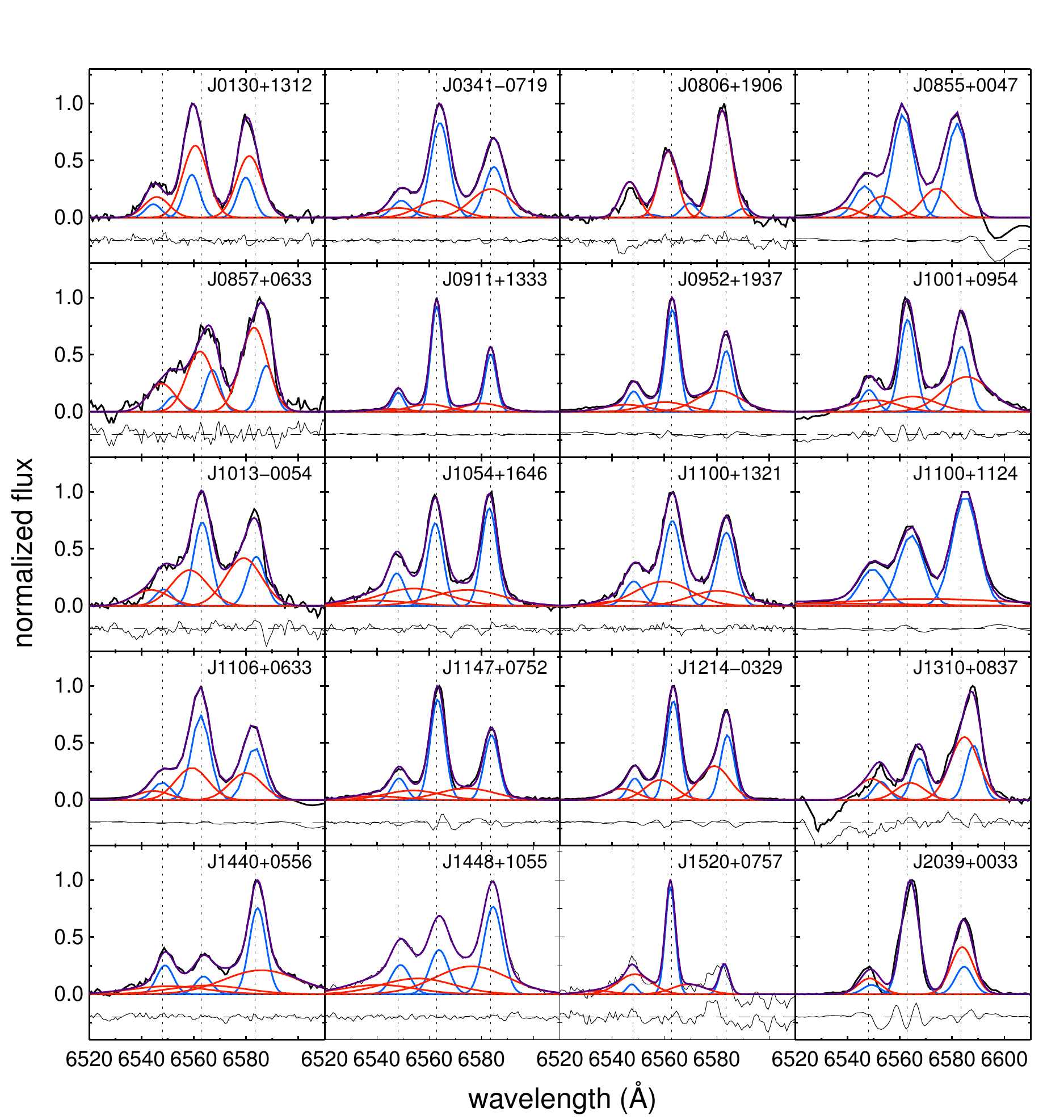}
\caption{The kinematic decompositions for the [\OIII] (top panel) and \Ha\ lines (bottom panel) based on the integrated spectra of 20 AGNs within the NLR size ($R_{\text{NLR}}$, see Section \ref{size}). The black thick lines are the residual spectra after stellar-component subtraction from the raw spectra. The red (blue) lines show the broad (narrow) component of the line profile, while the purple lines show the total (narrow+broad) profile. The black thin lines below the spectrum are the difference between the residual spectra and the best-fit models. The AGNs are listed in ascending R.A from top-left to bottom-right. We note that J1520+0757 has a noisy spectrum due to bad weather conditions during the observation of the target. }
\label{fig:fitting}
\end{figure}

\subsection{Sample Selection}
\label{sample}
We used the results of the ionized gas kinematics of $\sim$39,000 type 2 AGNs at z $<$ 0.3 from \citet{2016ApJ...817..108W}. The study has several advantages as an outflow census of type 2 AGNs. First, \citet{2016ApJ...817..108W} uniformly selected a large number of type 2 AGNs from the SDSS DR7 \citep{2009ApJS..182..543A} using the emission-line diagnostics. Second, they measured the systemic velocity based on the stellar absorption lines, since the systemic velocity provided by the SDSS pipeline is uncertain due to the fact that redshift measurements are partly based on emission-line features. Third, they provided the ionized gas kinematics
by measuring the first moment (velocity) and the second moment (velocity dispersion) of [\OIII]. 

We selected AGNs with strong outflow signatures from the parent sample of \citet{2016ApJ...817..108W} (Figure \ref{fig:target}). First, we limited the redshift range to z $<$ 0.1 in order to have enough spatial resolution and extent of outflows for given the IFU field-of-view (FoV). Then, we selected 3396 AGNs ($\sim$9\%) with a relatively high dust-corrected [\OIII] luminosity (log $L_{\text{[O III],cor}} > 41.5$ \ergs) since the outflow fraction increases with the [\OIII] luminosity \citep[e.g.,][]{2014ApJ...795...30B,2016ApJ...817..108W, 2016ApJ...817..108W}. We further selected 491 AGNs ($\sim$14\% of the luminosity-limited sample) with strong outflow signatures in [\OIII] kinematics, i.e., [\OIII] velocity offset $v_{\text{[O III]}}$ $>$ 100 \kms, or [\OIII] velocity dispersion $\sigma_{\text{[O III]}}$$>$ 300 \kms. The selected AGNs are only $\sim$1\% of the parent sample, consisting of the best candidates for outflow studies with local AGNs. Among the selected AGNs, we observed 18 AGNs with IFUs (see Table \ref{tbl-1}). Note that we observed 2 additional targets (J1100+1124 and J1520+0757), which satisfy the selection criteria for gas kinematics but with a slightly lower [\OIII] luminosity ($L_{\text{[O III],cor}} \sim 10^{41}$ \ergs). 
We also note that the AGNs presented in the pilot study by \citet{2016ApJ...819..148K,2016ApJ...833..171K} is a subsample with stronger outflow signatures than ours (e.g., $L_{\text{[O III],cor}} > 10^{42.0}$ \ergs). 

We calculate the AGN bolometric luminosity $L_{\text{bol}}$ as $\text{log} L_{\text{bol}} = 3.8 + 0.25\text{log} L_{\text{[O III],cor}} + 0.75\text{log} L_{\text{[\OI],cor}}$, where $L_{\text{[\OI],cor}}$ is the extinction-corrected [\OI]$\lambda$6300 luminosity \citep{2009MNRAS.399.1907N}. We infer the black hole mass ($M_{\text{BH}}$) by adopting the $M_{\text{BH}}-\sigma_{\star}$ relation \citep{2012ApJS..203....6P}, where $\sigma_{\star}$ is the stellar velocity dispersion obtained from the MPA-JHU catalog of SDSS DR7\footnote{http://www.mpa-garching.mpg.de/SDSS/}. 
We find that AGNs with strong outflows show no significant difference in the distributions of redshift, Eddington ratio, and  the emission-line ratios \citep[e.g.,][]{1981PASP...93....5B}, compared to the AGNs in the luminosity-limited sample (Figure \ref{fig:prop1}). Thus, we assume that the selected AGNs are a random sub-sample of the luminosity-limited sample.

\subsection{Magellan/IMACS-IFU Observations \& Reduction}
We observed 17 out of 20 type 2 AGNs with the IFU of the Inamori Magellan Areal Camera (IMACS-IFU) on the Magellan telescope at Las Campanas Observatory. We used the $f/4$ (long mode) camera with a recently upgraded Mosaic 3 CCD (8k$\times$8k). We used the 300 lines/mm grating with a tilt angle of 6.0$^\circ$, providing a large wavelength range of 3420-9550\AA\ with a spectral resolution $R\approx$1600. The choice of $f/4$ camera provides an FoV of 5\arcsec$\times$4\farcs 5 with 600 fibers, and each spaxel (spatial pixel) has a size of 0\farcs 2 in diameter on the sky. Given the range of redshift of our targets, we explore the central region within a several kpc from the nucleus, which provides detailed information on the gas outflows in the central region of the host galaxies. 

The observations were performed during 7 nights in April, May, and December 2014 with a seeing 0\farcs 7 -- 1\farcs 3 in April and May, and 0\farcs 7 -- 0\farcs 9 in the December run. In each afternoon, we obtained bias images, HeNeAr arcs, dome flats, and sky flats if possible. During the night, we initially took an acquisition image with the IFU to make sure the target was located at the center of the FoV. During our first run (April 6), however, the acquisition was poorly performed due to an unexpected spatial offset of the IFU images on the CCD. Hence the targets of the night failed to be located at the center of the FoV. For the other nights, after locating the target at the center of the FoV, we took science exposures and an arc image at the same position of the telescope. For each observing run, we observed spectrophotometric standard stars for flux calibration. 

For the IMACS-IFU data reduction, we utilized an IDL-based software P3D\footnote{http://p3d.sourceforge.net}, which is a general-purpose data reduction tool for fiber-fed IFUs \citep{2010A&A...515A..35S}. Since P3D does not officially support the IMACS-IFU data reduction yet, we modified the parameter files in the code for the IMACS-IFU instrument setup. First, we constructed a master bias frame then subtracted the bias frame from all exposures. We also constructed a combined dome flat image with the IFU as a reference frame for tracing of the dispersed light. Using the combined dome flat image, we obtained a tracing solution for each target. Since we did not obtain a dome flat image after each science exposure, we manually shifted the dome flat image for each science exposure to properly trace the dispersed light. The shift is, on average, $\sim$5 pixels on the CCD in the cross-dispersion direction (y-axis). After we obtained the individual tracing solutions, we applied the solution to the HeNeAr arc frame obtained after each target in order to get a wavelength solution. We used a fifth order polynomial function to obtain the wavelength solution, resulting in a residual r.m.s. of $\sim$0.02\AA. 

Then we removed cosmic rays on each science frame by using the PyCosmic routine, developed for a robust cosmic-ray removal of fiber-fed IFU data \citep{2012A&A...545A.137H}. After the cosmic-ray removal, we extracted the object and sky spectra by using the tracing solutions. Among several methods for spectrum extraction, we used the modified optimal extraction with a predefined Gaussian function from the tracing solution, which gives less noise in spectra and more robust results than the top-hat extraction. \citep{1986PASP...98..609H, 2010A&A...515A..35S}. IMACS has an internal atmospheric dispersion corrector, hence we did not correct the atmospheric dispersion during the reduction as we do not find any wavelength-dependent spatial offset in the final constructed image. After the spectra extraction, we obtained a mean spectrum of sky background by using dedicated sky fibers, and subtracted the mean sky spectrum from each science spectrum. Finally, we combined all science exposures and calibrated the flux with a sensitivity function obtained from standard star observations. In addition, to increase the signal-to-noise ratio (S/N) in the outskirts of the FoV, we binned 7 adjacent spaxels into a single spectrum. Since the combined spaxels have a size of 0\farcs 6 in diameter on the sky, we still have spaxels smaller than the seeing (0\farcs 7--1\farcs 3) during the observing runs. 

\subsection{VLT/VIMOS-IFU Observations \& Reduction}
For 3 out of 20 AGNs, we obtained the data with the multi-purpose optical instrument VIMOS in IFU mode on the VLT-UT3 (Program ID: 091.B-0343(A), PI: Flohic). We used the MR grating with the GG475 filter, providing a spectral range of 4900 -- 10150\AA\ with a spectral resolution of $\sim$ 720. The VIMOS-IFU has 6400 fibers (80$\times$80) without dedicated sky fibers. With the adjustable scales of 0.67\arcsec per fiber, we have a large FoV of 27\arcsec$\times$27\arcsec, which is a factor of $\sim$30 larger than our IMACS-IFU FoV. Thus, the VIMOS-IFU data allowed us to investigate the impact of gas outflows on galactic scales of $\sim$10 kpc for the redshift range of our targets. The observing run was executed as a service-mode observation in April 2013. The exposure time was $\sim$2.8 hours per target, with 15$\times$670 seconds exposures. We applied a spatial offset of $\pm$2\arcsec\ for science exposures to compensate for dead/bad fibers in the array. 

The reduction for the VLT/VIMOS-IFU data was performed with P3D in the same manner as the IMACS-IFU data. After that, we combined the multiple exposures by considering the spatial offset of $\pm$2\arcsec. For both J1100+1124 and J1106+0633, we discarded three exposures that suffered from a wrong spatial offset during the observations. 

\begin{figure}
\centering
\includegraphics[width=0.48\textwidth]{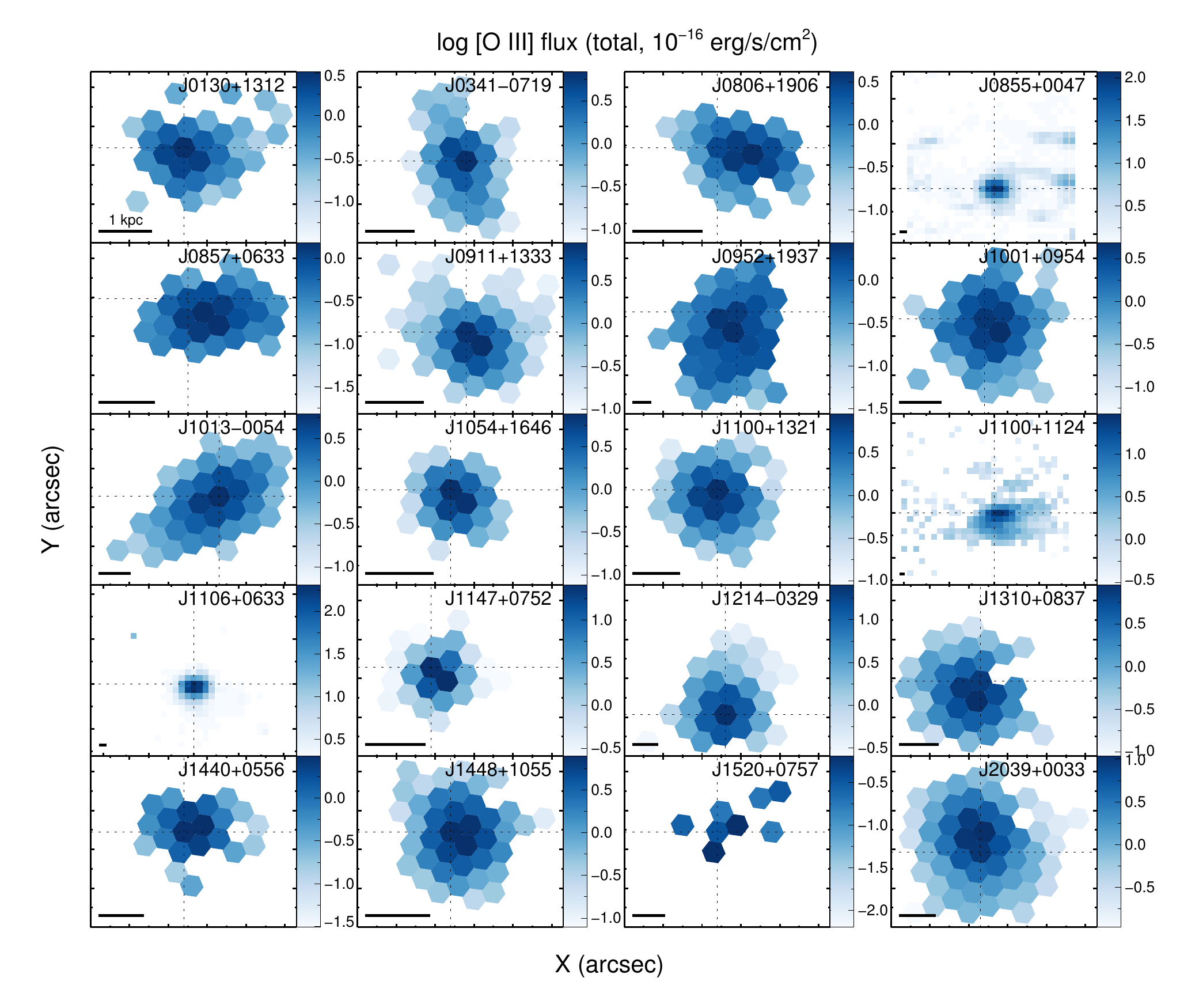}
\includegraphics[width=0.48\textwidth]{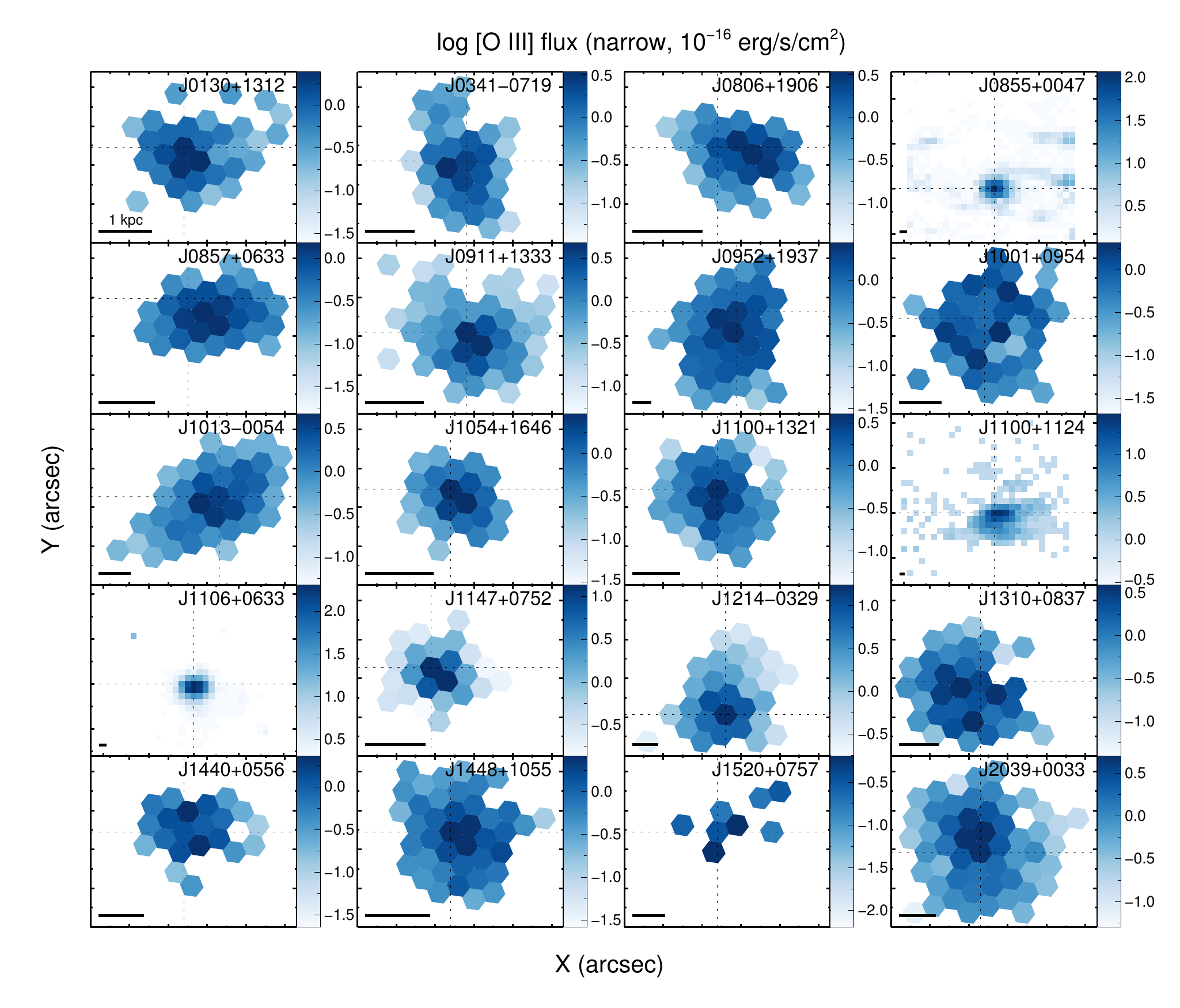}
\includegraphics[width=0.48\textwidth]{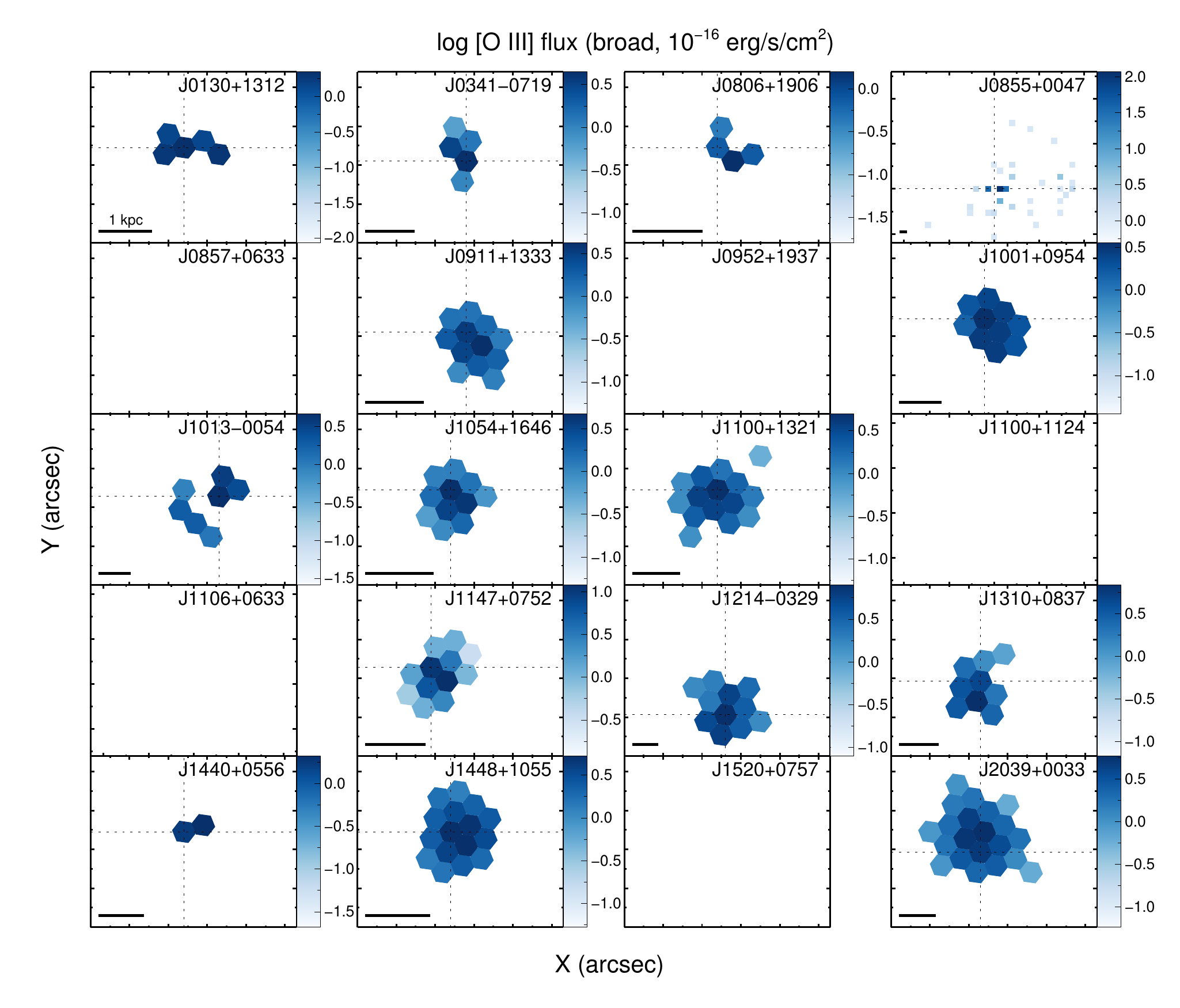}
\caption{The [\OIII] flux maps for the total (top panel), the narrow (middle panel), and the broad component (bottom panel). Blank maps for the broad component indicate no detection of the AGNs. The major ticks in both x- and y-axes represent 1\arcsec\ for the Magellan targets, while the major ticks denote 5\arcsec\ for the VLT targets. The AGNs are listed in order of ascending R.A from top-left to bottom-right.}
\label{fig:oiii_flux}
\end{figure}

\begin{figure}
\centering
\includegraphics[width=0.48\textwidth]{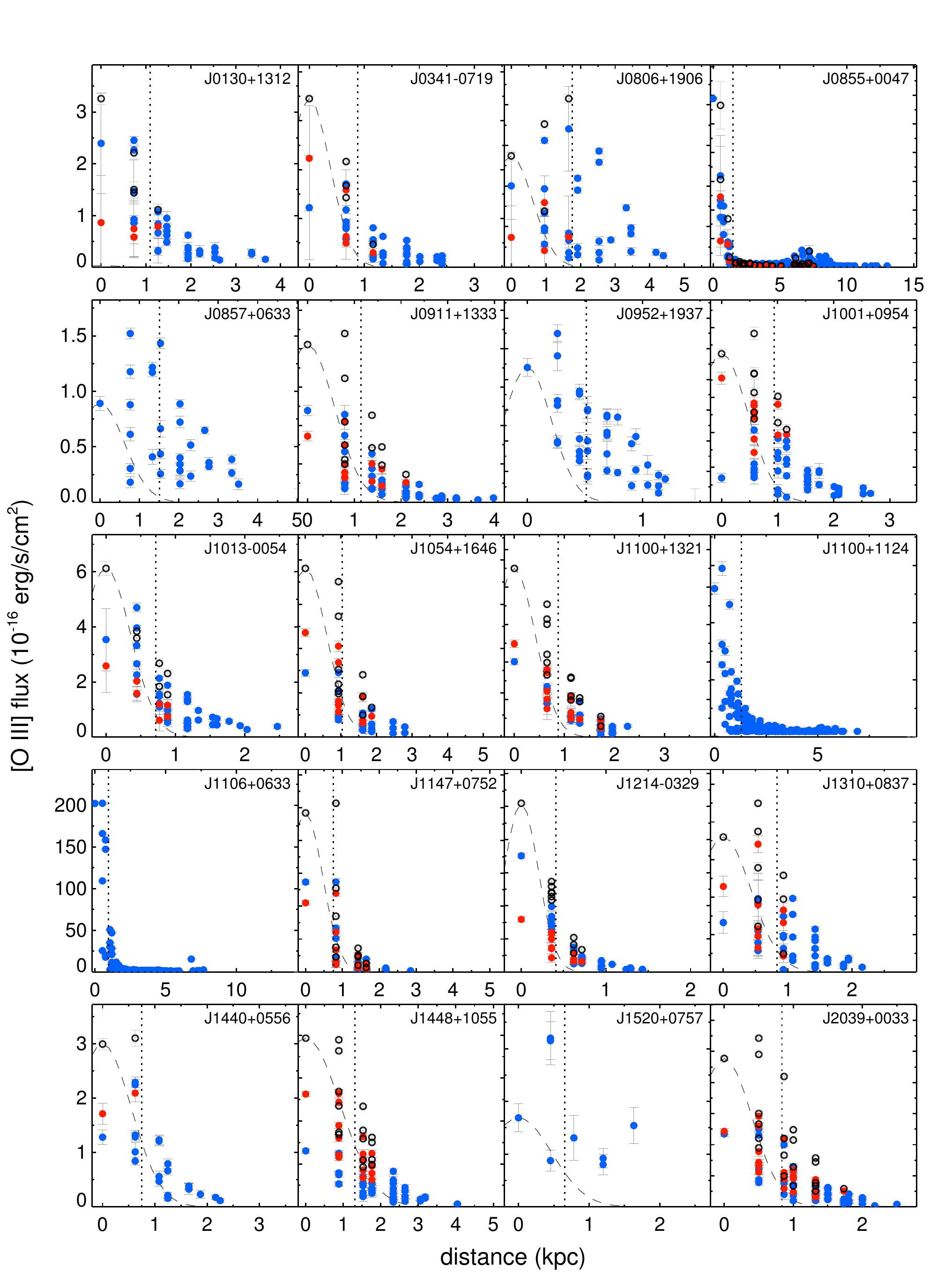}
\caption{The radial flux distributions of [\OIII]. Dotted lines in the vertical direction indicate the effective radius of the NLR based on [\OIII]. The dashed curves show the point-spread-function based on the seeing size. Each panel contains the radial distributions of the narrow component (blue), the broad component (red), if it exists, and the total (narrow+broad) component (black) with 1$\sigma$ uncertainties.}
\label{fig:dist_oiii_flux}
\end{figure}

\section{Analysis}
\label{analysis}
\subsection{Emission-line properties}
\label{decomposition}
For each spaxel, we measured the flux and velocities of emission lines after subtracting stellar continuum, as similarly performed by \citet{2016ApJ...817..108W}. For stellar continuum, we utilized the pPXF code \citep{2004PASP..116..138C} to construct the best-fit model using the MILES simple stellar population models with solar metallicity \citep{2006MNRAS.371..703S}. In this procedure we measured the velocity of the stellar component in each spaxel. 

Then, we fit the narrow emission-lines, e.g., the Balmer lines, [\OIII]$\lambda$5007, [\NII]-doublet, and [\SII]-doublet, by utilizing the MPFIT code \citep{2009ASPC..411..251M}. Since AGNs with gas outflows generally show a broad wing component, especially in the [\OIII] line profile \citep{2005ApJ...627..721G, 2013MNRAS.433..622M,2014Natur.513..210S,2016ApJ...817..108W}, we used two Gaussian functions to represent the broad and the narrow components. For the \Ha+[\NII] region, we assumed that each narrow and broad wing component  of \Ha\ and [\NII] has the same kinematics, i.e., velocity and velocity dispersion. We also fixed the flux ratio of [\NII]$\lambda$6549 and [\NII]$\lambda$6583 as one third for both broad and narrow components (Figure \ref{fig:fitting}). 

To ensure that the broad component is not fitting the noise in the spectrum, we adopted two conditions for accepting the broad wing component. First, the peak amplitude of the broad component should be a factor of three larger than the standard deviation of the residual spectra at 5050 -- 5150\AA. For the \Ha+[\NII] region, we choose a larger amplitude of either \Ha\ or [\NII] broad component as the peak amplitude. Second, the sum of the width ($\sigma$) of the two Gaussian components should be smaller than the distance between the peaks of the two Gaussian components. If the broad component is not detected in the fitting, we alternatively fit the emission lines with a single Gaussian function. 

Based on the model fit of the emission lines, we calculated the first moment $\lambda_0$ and the second moment $\sigma_{\text{line}}$ of each emission line, which respectively represent the luminosity-weighted velocity and velocity dispersion of the emission line, as 
\begin{eqnarray}
\lambda_0 = {\int \lambda f(\lambda) d\lambda \over \int f(\lambda) d\lambda}, 
\end{eqnarray}

\begin{eqnarray}
\sigma_{\text{line}}^2 = {\int \lambda^2 f(\lambda) d\lambda \over \int f(\lambda) d\lambda} - \lambda_0^2, 
\end{eqnarray}
where $f(\lambda)$ is the flux at each wavelength $\lambda$.
We calculated the velocity shift of emission lines by comparing the first moment and the systemic velocity measured from the center of host galaxies. The second moment was corrected for the instrumental resolution measured from the sky emission lines. To estimate the uncertainties for the measurement, we adopted a Monte Carlo realization generating 100 mock spectra by randomizing the flux with noise, and obtained the measurements from each spectrum. Then we adopted 1$\sigma$ dispersion of the distribution of each measurement as the uncertainty.

\section{the narrow-line region properties}
\label{nlr_prop}
In this section, we present the NLR properties based on the spatial distributions of [\OIII] (Section \ref{oiii}) and \Ha\ flux and kinematics (Section \ref{ha}). Then we examine the sizes of the NLR and outflows and the size-luminosity relationship (Section \ref{size}).

\begin{figure}
\centering
\includegraphics[width=0.48\textwidth]{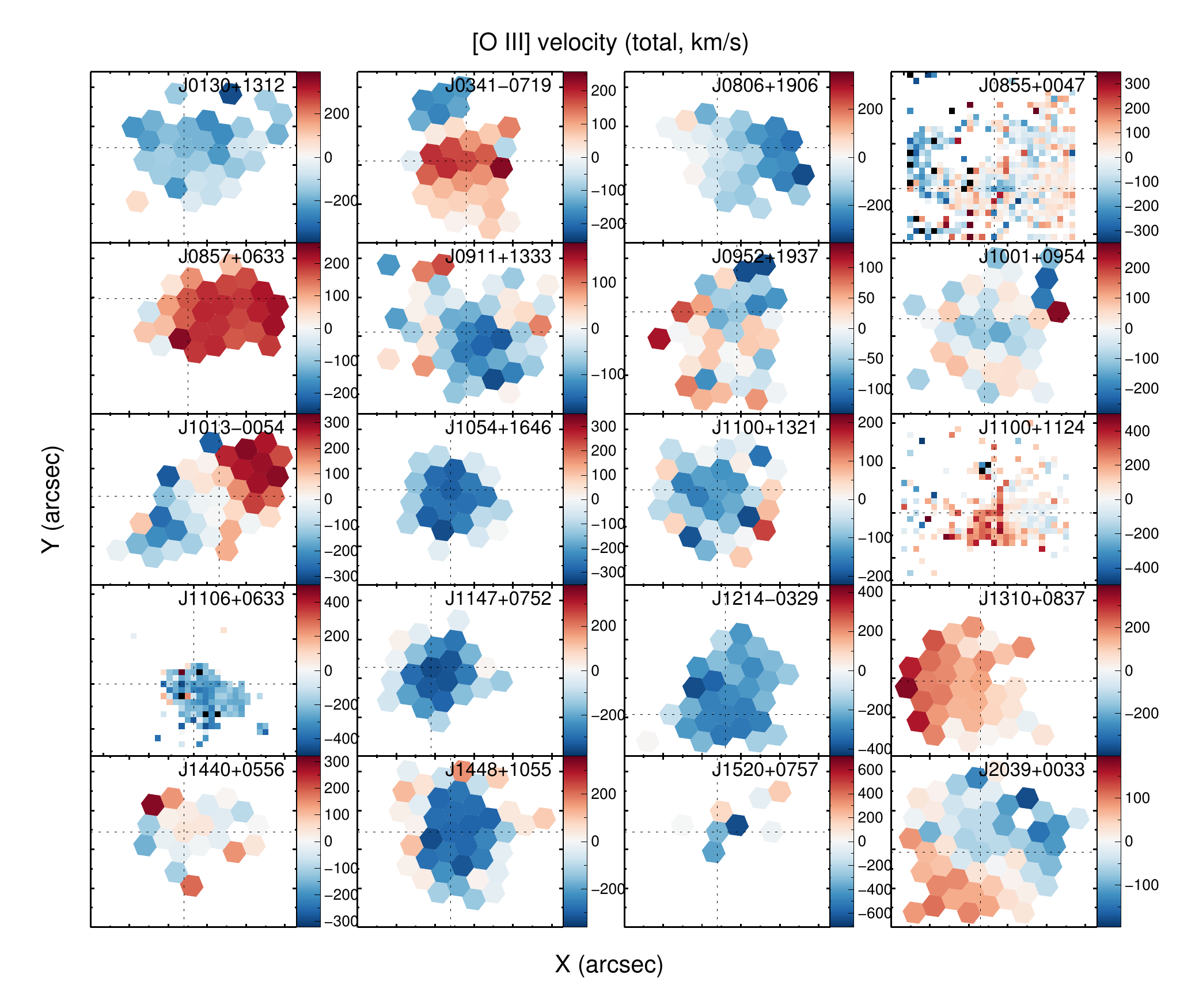}
\includegraphics[width=0.48\textwidth]{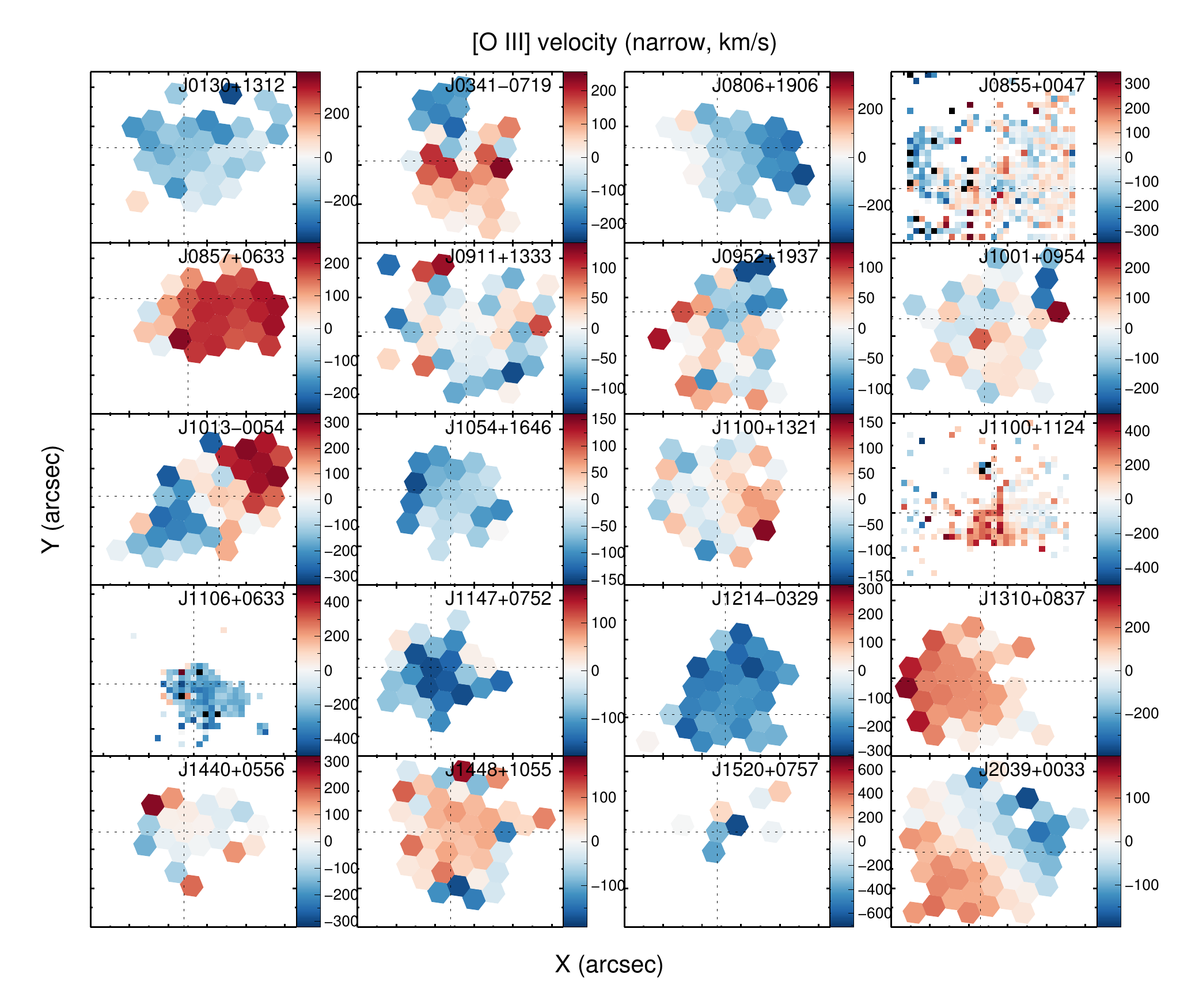}
\includegraphics[width=0.48\textwidth]{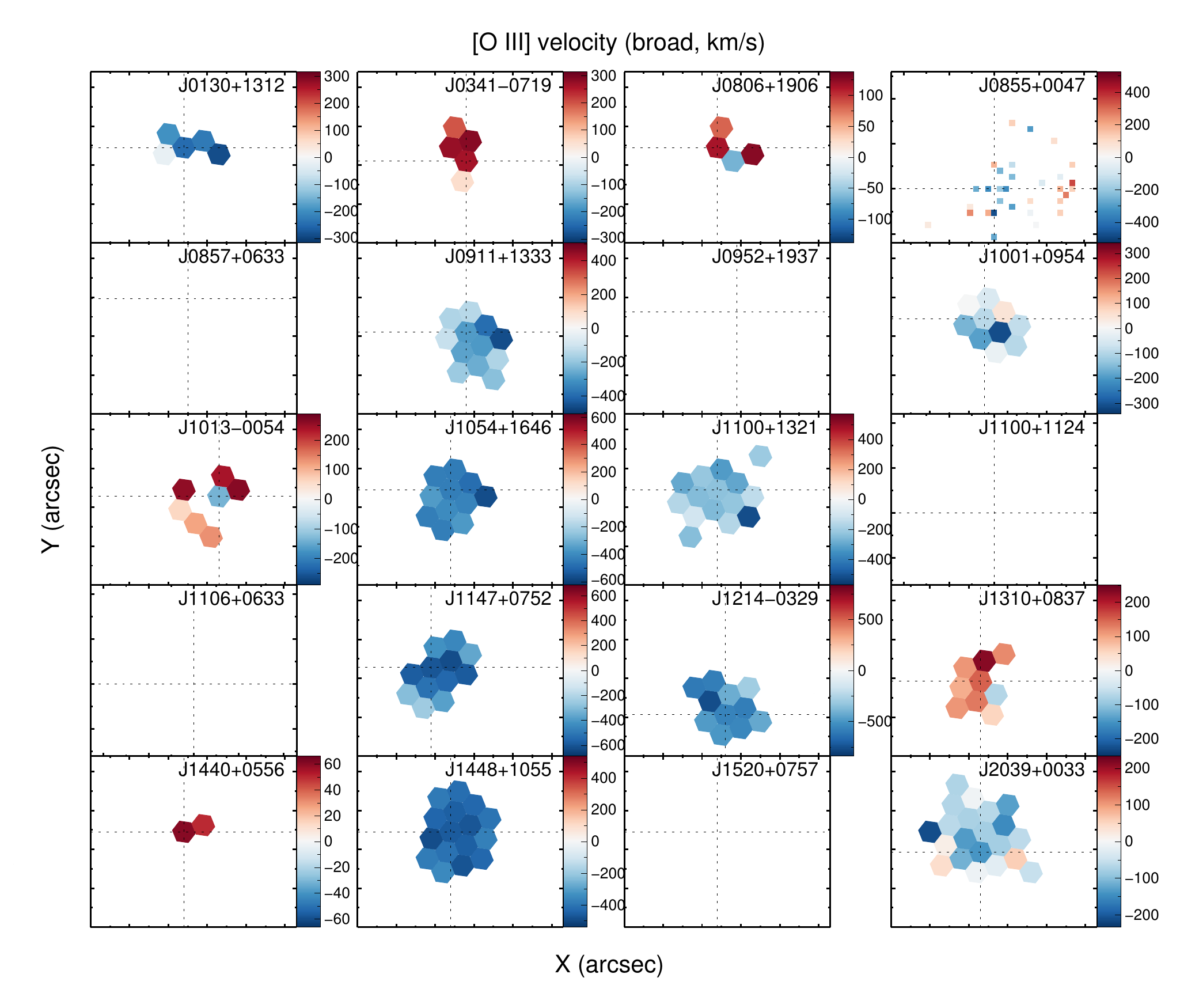}
\caption{The [\OIII] velocity maps for the total (top panel), the narrow (middle panel), and the broad component (bottom panel). Blank maps for the broad component indicate no detection of the AGNs. The major ticks in both x- and y-axes represent 1\arcsec\ for the Magellan targets, while the major ticks denote 5\arcsec\ for the VLT targets. The AGNs are listed in order of accending R.A from top-left to bottom-right.}
\label{fig:oiii_vel}
\end{figure}

\begin{figure}
\centering
\includegraphics[width=0.48\textwidth]{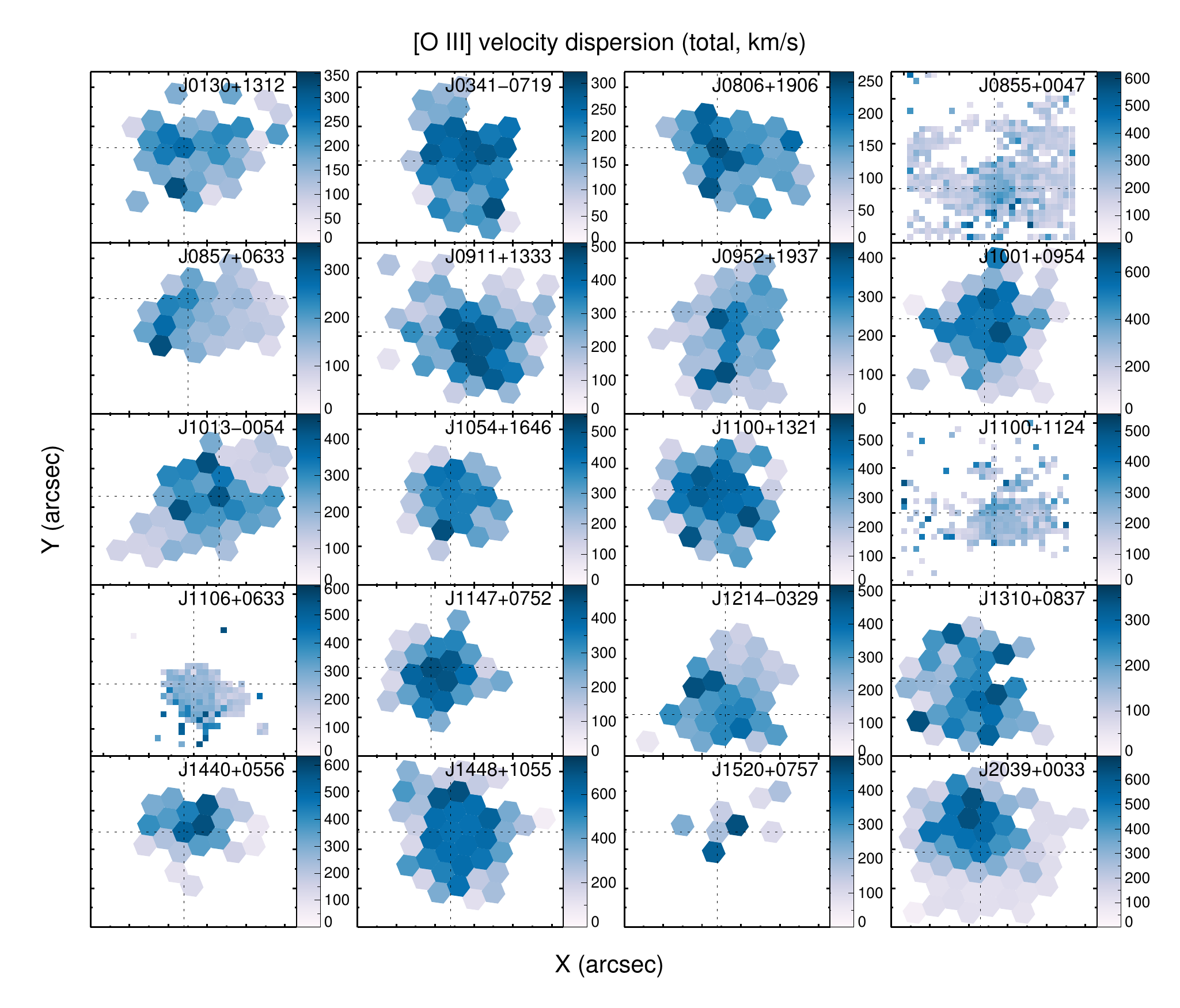}
\includegraphics[width=0.48\textwidth]{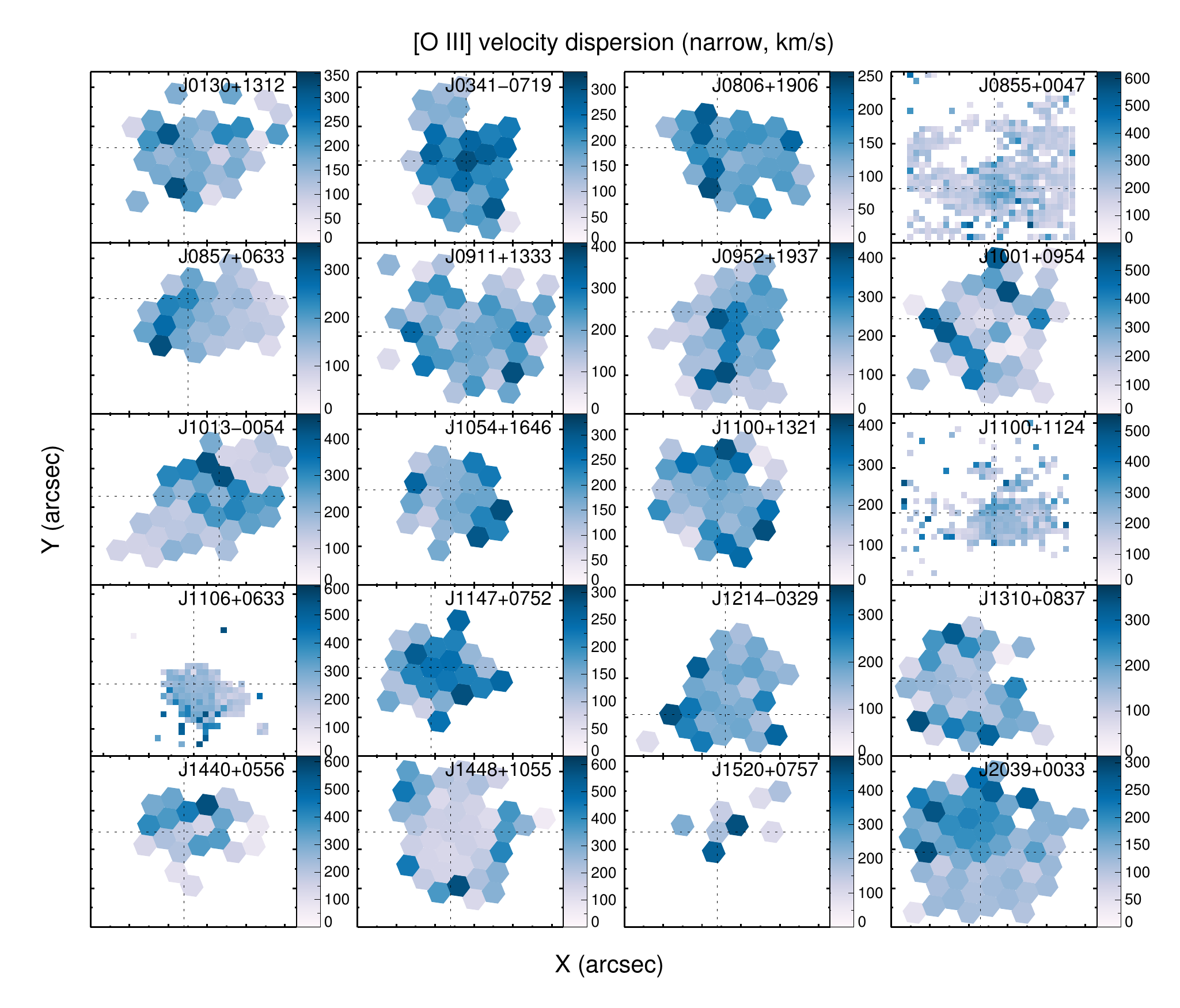}
\includegraphics[width=0.48\textwidth]{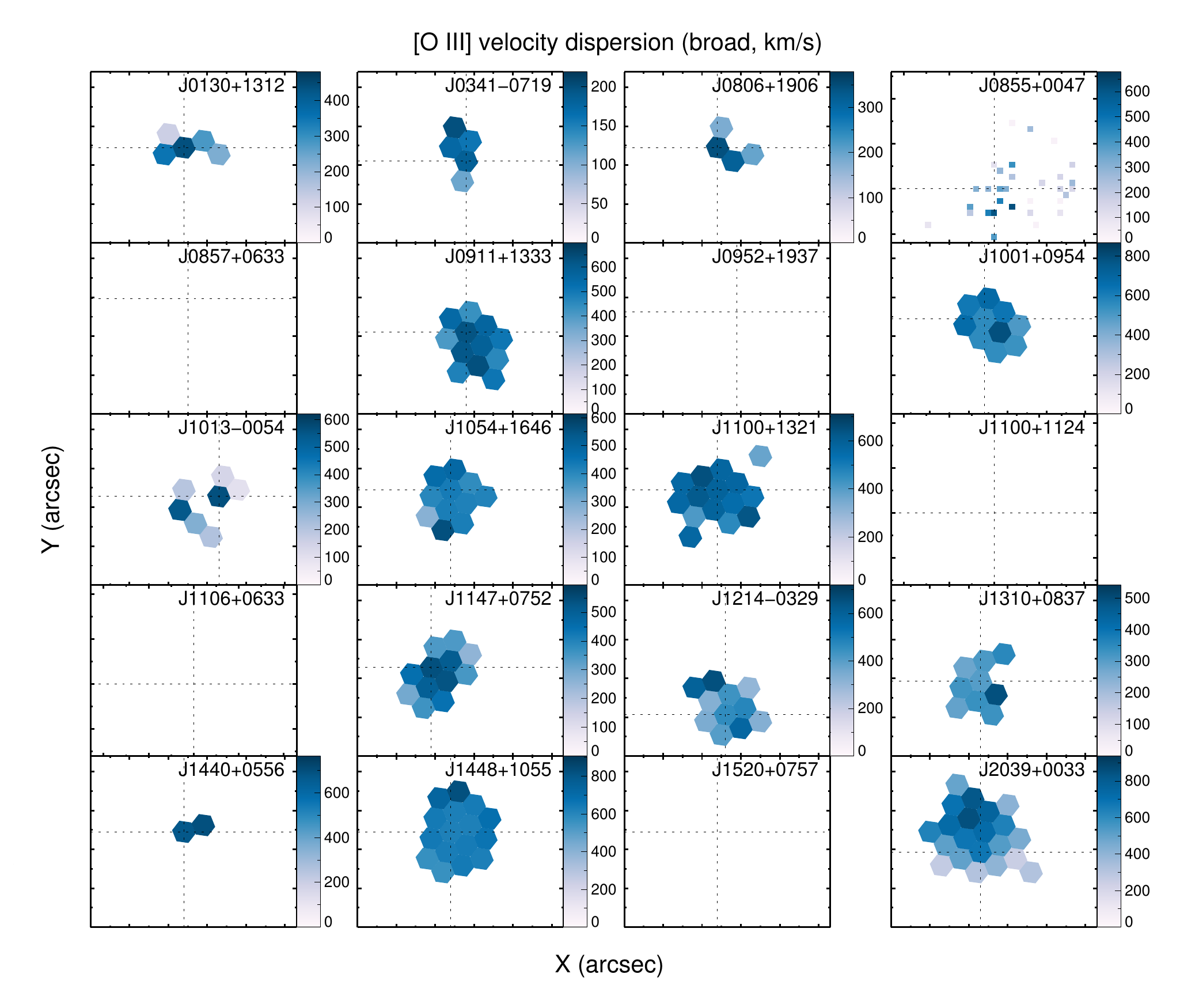}
\caption{The [\OIII] velocity dispersion maps for the total (top panel), the narrow (middle panel), and the broad component (bottom panel). Blank maps for the broad component indicate no detection of the AGNs. The major ticks in both x- and y-axes represent 1\arcsec\ for the Magellan targets, while the major ticks denote 5\arcsec\ for the VLT targets. The AGNs are listed in order of accending R.A from top-left to bottom-right.}
\label{fig:oiii_veldisp}
\end{figure}
\subsection{the [OIII]-emitting region}
\label{oiii}

\begin{figure*}
\centering
\includegraphics[width=0.48\textwidth]{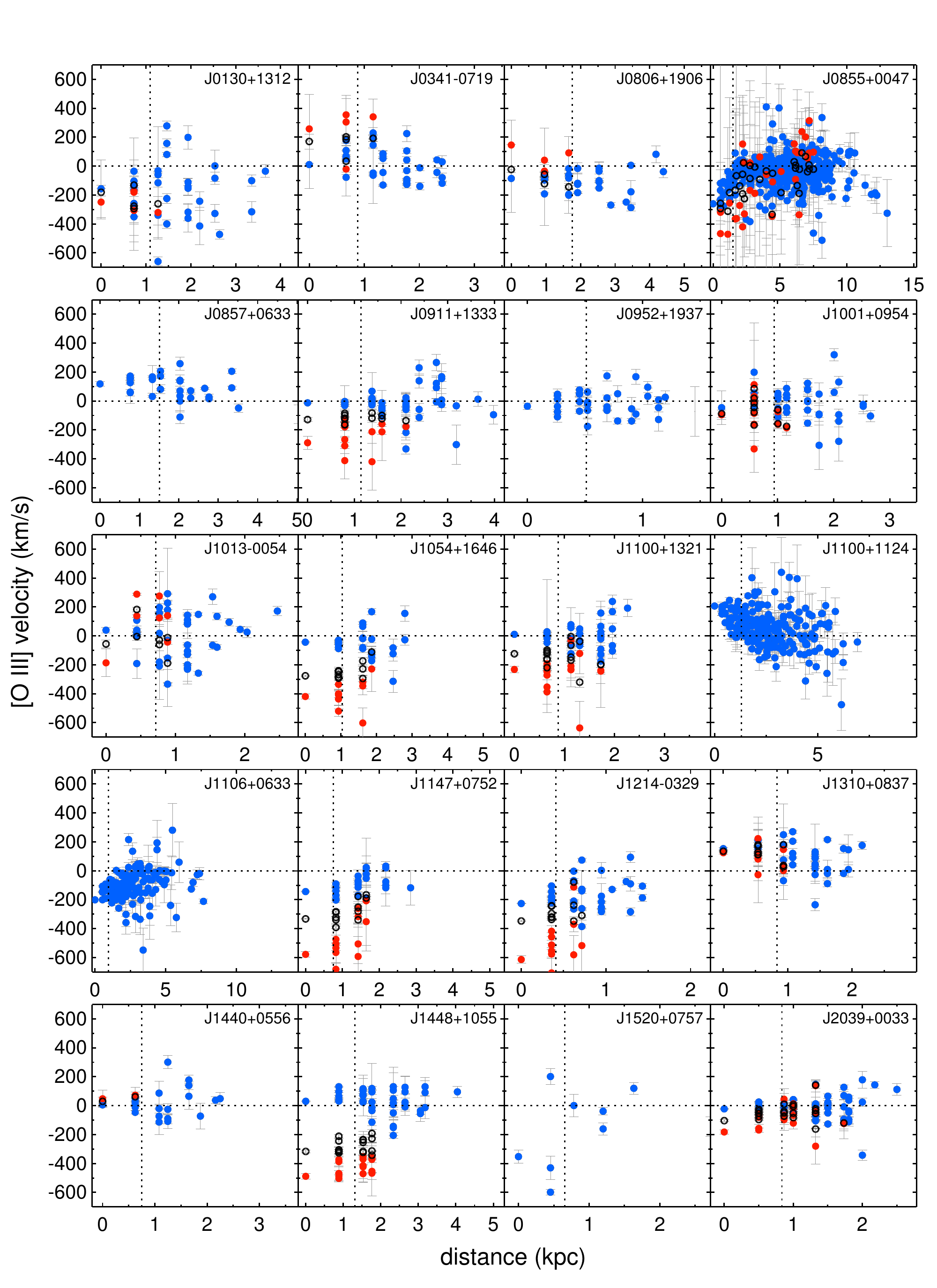}
\includegraphics[width=0.48\textwidth]{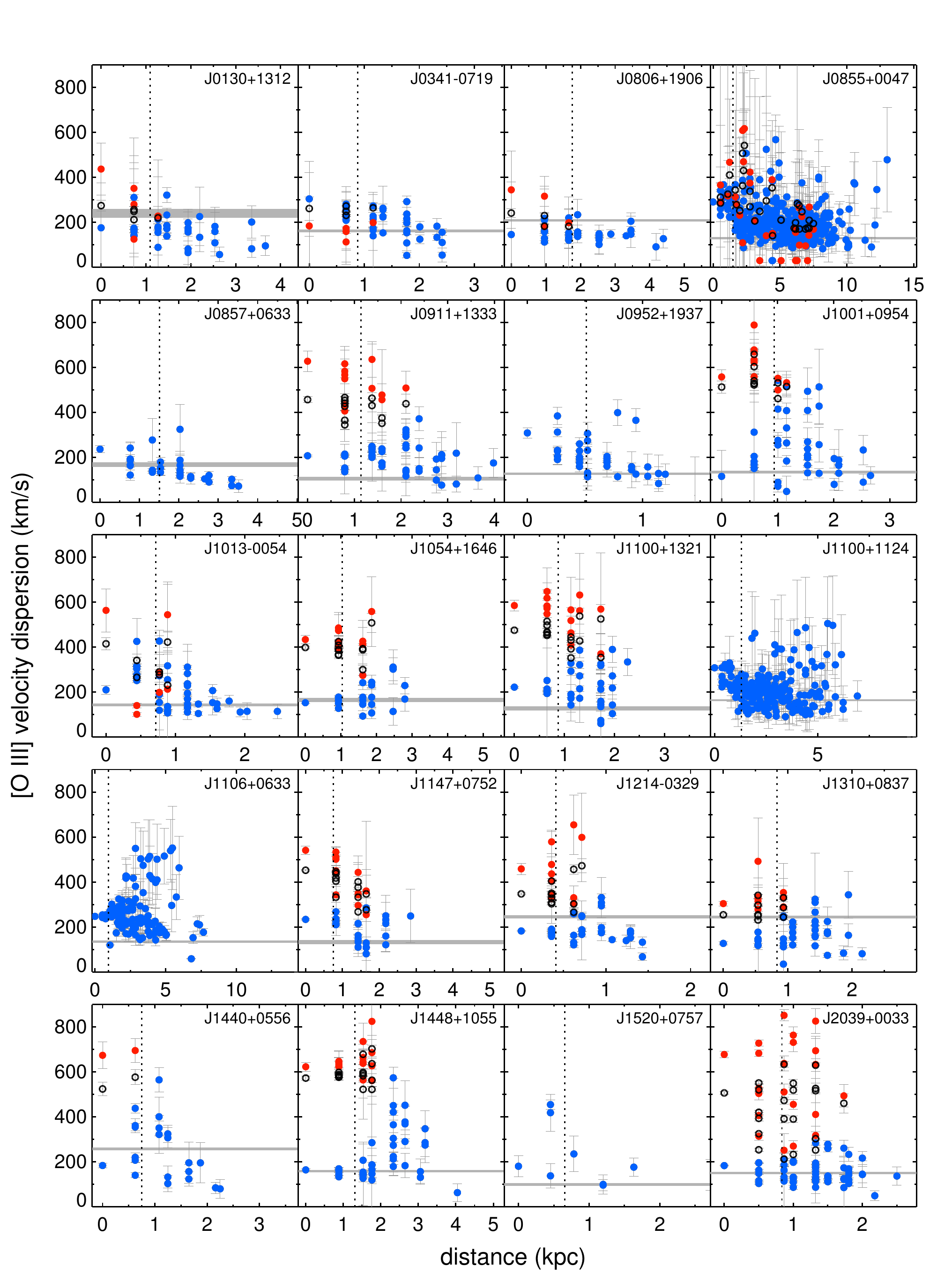}
\caption{The radial distributions of the [\OIII] velocity (left) and velocity dispersion (right). Dotted lines in the vertical direction indicate the effective radius of the NLR based on [\OIII] (see Section 4.1). The dashed curves show the point-spread-function based on the seeing size. Each panel contains the radial distributions of the narrow component (blue), the broad component (red), if it exists, and the total (narrow+broad) component (black) with 1$\sigma$ uncertainties. The gray shaded areas in the velocity dispersion panels indicate the range of stellar velocity dispersion from the SDSS with 1$\sigma$ uncertainty. The AGNs are listed in order of accending R.A from top-left to bottom-right.}
\label{fig:dist_oiii_vvd}
\end{figure*}

\begin{figure}
\centering
\includegraphics[width=0.48\textwidth]{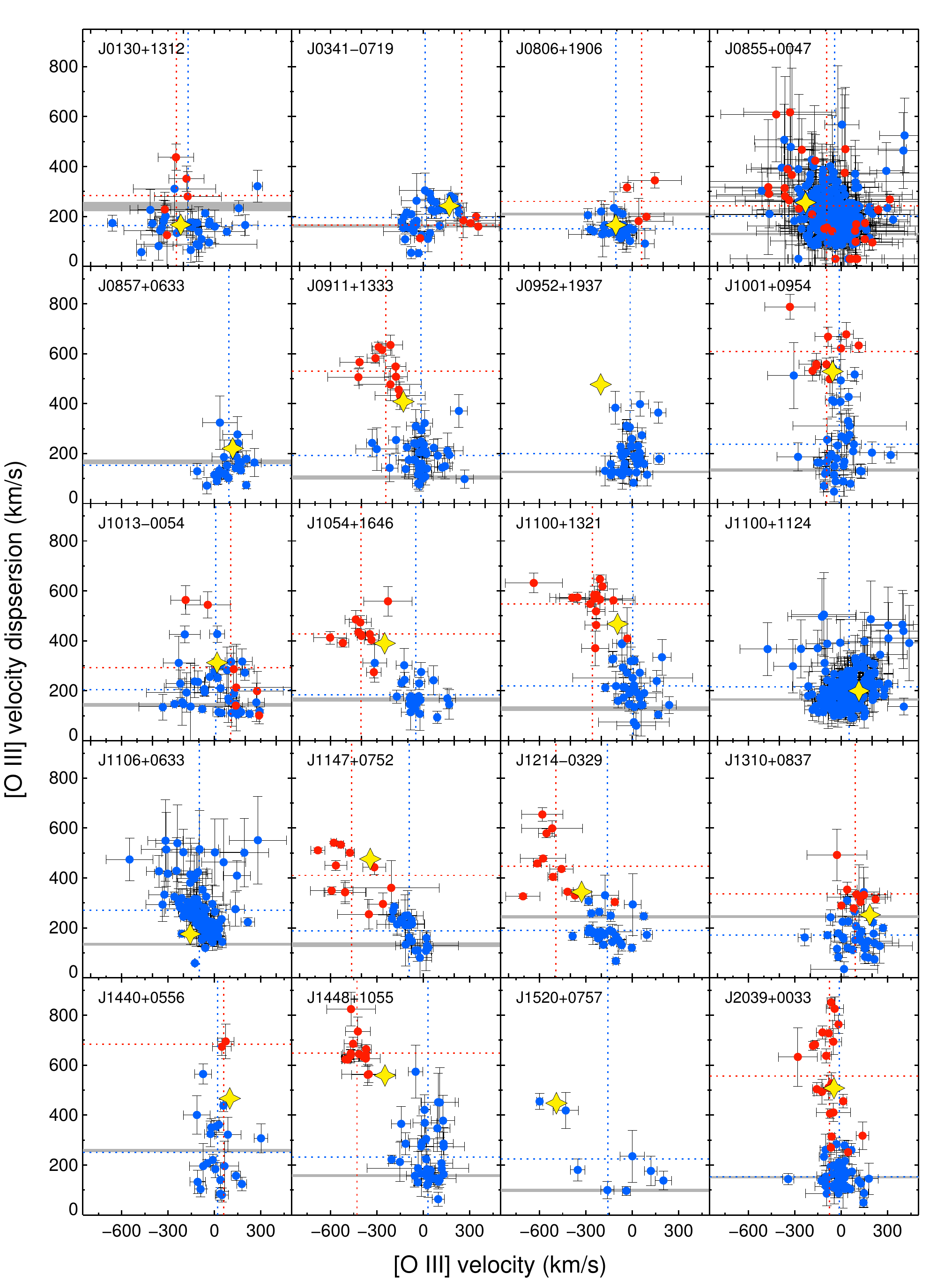}
\caption{The velocity-velocity dispersion diagrams for the narrow (blue dots) and broad components (red dots) of the [\OIII] line. The error bars represent 1$\sigma$ uncertainties in the measurement. Blue (red) dotted lines are the mean values of the velocity and velocity dispersion for the narrow (broad) component. Yellow stars are the values obtained from the SDSS spectra. The horizontal gray shaded areas indicate the range of stellar velocity dispersion from the SDSS with 1$\sigma$ uncertainty. The AGNs are listed in order of accending R.A from top-left to bottom-right.}
\label{fig:vvd_oiii}
\end{figure}

\subsubsection{Morphology}
\label{oiii_morph}
We perform a visual inspection of the spatial distribution of the [\OIII] flux. The morphologies are, in general, in good agreement with the flux distribution of the stellar component, showing no noticeable biconical outflow features (Figure \ref{fig:oiii_flux}). For 12 out of 20 AGNs, however, the [\OIII] flux distribution shows a lopsidedness from the stellar center of the host galaxy (Figure \ref{fig:dist_oiii_flux}), although it is relatively uncertain due to the poor sampling (0\farcs 6 spaxel in diameter, see Section \ref{obs}) and the seeing size (0\farcs 7 -- 1\farcs 3). In general, the spaxel with the maximum [\OIII] flux has an offset of $\sim$1 spaxel from the center of the galaxy. Such a lopsidedness in the [\OIII] flux distribution is possibly due to the dust obscuration of either the approaching or receding component of biconical outflows \citep[e.g.,][]{2010ApJ...708..419C, 2014ApJ...785...25F, 2014ApJ...795...30B,2016ApJ...828...97B}. 

By performing the kinematic decomposition of [\OIII] as described in Section \ref{decomposition}, we find that 15 our of 20 AGNs have a detectable broad component in the line profile. After separating the narrow and broad components from the line profile, we examine the morphologies of the [\OIII] flux distribution of the two components separately (middle and bottom panels in Figure \ref{fig:oiii_flux}, respectively). First, the spaxels with a broad component have a smaller extent than those with a narrow component. This is mainly due to difficulties in detecting a broad component from low S/N spectra at the outskirts of the FoV. Second, the spaxels containing the broad component show diverse morphologies. Among the 15 broad-component-detected AGNs, 8 AGNs show a compact, round-shape morphology, while the other 7 AGNs show an irregular (e.g., patchy or elongated) shape. 

\begin{table*}
\center
\caption{The characteristics of the spatially resolved velocity fields of the NLR \label{tbl-2}}
\begin{tabular}{ccccccccccc}
\tableline
\tableline
& \multicolumn{3}{c}{[\OIII]} & \multicolumn{3}{c}{H$\alpha$} \\
& \multicolumn{3}{c}{} & \multicolumn{3}{c}{} \\
\cline{2-7}
SDSS name & $V_{[\OIII], T} $ & $V_{[\OIII], N} $ & $V_{[\OIII], B}$ & $V_{H\alpha, T}  $ & $V_{H\alpha, N} $ & $V_{H\alpha, B} $ & Disk type \\
(1) & (2) & (3) & (4) & (5) & (6) & (7) & (8)\\
\tableline
J0130+1312    &  blue  & blue  & blue  & rotation & rotation & rotation & SF\\
J0341$-$0719 &  red  & rotation & red  & rotation & rotation & rotation & SF\\
J0806+1906    &  blue  & blue  & red  & rotation & rotation & rotation & AGN\\
J0855+0047   &  blue   & blue  & blue  & blue   & rotation & blue  & SF      \\
J0857+0633    &  red  & red  & -- & rotation & rotation & -- & AGN \\
J0911+1333    &  blue  & blue  & blue  & blue  & rotation & blue  & SF\\
J0952+1937    & blue  & blue  & -- & rotation & rotation & blue  & SF \\
J1001+0954    & blue  & rotation & blue  & red  & rotation & red  & SF \\
J1013$-$0054 &  rotation & rotation & ambiguous & rotation & rotation & rotation & SF \\
J1054+1646    &  blue  & blue  & blue  & blue  & rotation & blue  & AGN \\
J1100+1321    &  blue  & rotation & blue  & blue  & rotation & blue  & SF\\
J1100+1124   &  red  & red  &  --  & red  & red  & --  &  --       \\
J1106+0633   &  blue  & blue  & -- & blue  & blue  & blue   & --        \\
J1147+0752    &  blue  & rotation & blue  & blue  & rotation & blue  & SF \\ 
J1214$-$0329 &   blue  & rotation & blue  & blue  & rotation & blue  & SF  \\
J1310+0837    &   red  & red  & red  & rotation & rotation & -- & AGN \\ 
J1440+0556    &   systemic & systemic & red  & red  & red  & red  & -- \\
J1448+1055    &    blue  & red  & blue  & blue  & rotation & blue  & AGN\\
J1520+0757    &   blue  & blue  & -- & blue  & blue  & blue  & --  \\
J2039+0033    &   blue  & rotation & blue  & rotation & rotation & rotation & SF \\
\tableline
\end{tabular}
\tablecomments{(1) the name of AGN; (2) the characteristics of the velocity field based on the total profile of [\OIII]; (3) the same as (2) but based on the narrow component of [\OIII]; (4) the same as (2) but based on the broad component of [\OIII]; (5) the same as (2) but for the total profile of \Ha; (6) the same as (2) but based on the narrow component of \Ha; (7) the same as (2) but based on the broad component of \Ha; (8) type of disk seen in the narrow component of \Ha. }
\end{table*}

\subsubsection{Kinematics}
We present the kinematic properties of [\OIII] based on its total, narrow, and broad components. First, we examine the 2D maps of the line-of-sight velocity structure of the [\OIII]-emitting region (Figure \ref{fig:oiii_vel}). When we examine the total profile of [\OIII], we find that 18 AGNs show blueshifted (14) or redshifted (4) velocity shifts within the central kpc or a more extended region, compared to the systemic velocity of the host galaxy. One AGN (J1013-0054) shows a rotational kinematics, and another AGN (J1440+0556) shows no or little velocity offset beyond the uncertainty in velocity offset measurement within the central kpc region. 

When we focus on the broad and narrow components separately, the velocity structure looks different from the total component. The majority (10 out of 15) of broad-component-detected AGNs show negative velocity offset, while four AGNs show positive velocity offset and one AGN shows ambiguous velocity structure. The result is consistent with the model predictions of biconical outflows and dust obscuration, which expect a larger number of AGNs with negative velocity offset than those with positive velocity offset \citep{2016ApJ...828...97B}. For the 15 broad-component-detected AGNs, the spaxels with a narrow component show mostly either negative velocity offset (5), positive velocity offset (2), rotational features (7), or no/little velocity offset (1) after removing the broad component (see Table \ref{tbl-2}).

Second, we examine the 2D maps of velocity dispersion of [\OIII] (Figure \ref{fig:oiii_veldisp}). The maps of total [\OIII] profile provide a hint for the mixture of narrow and broad component in the FoV. The range of velocity dispersion in the central region is from $\sim$200 \kms to $\sim$600 \kms, which is much larger than the stellar velocity dispersions of the host galaxy. As we separate the broad and narrow components, the velocity dispersion of the broad component becomes larger up to $\sim$800 \kms, while that of the narrow component is, in general, broadly consistent with the stellar velocity dispersion. In the maps of the narrow component, we also find spaxels with a relatively large velocity dispersion at the boundary of spaxels with a broad component, which is possibly due to an un-removed broad component in the line profile.  

Third, we investigate the radial distribution of the [\OIII] velocity and velocity dispersion for each component (Figure \ref{fig:dist_oiii_vvd}). In most cases, the absolute velocities of the broad components tend to become smaller as a function of distance from the center, while those of the narrow components show a flat distribution as a function of distance. The velocity dispersions of the broad component are at least a few times larger than the stellar velocity dispersion, and also decreasing as a function of distance. In J0911+1333, for example, the velocity dispersions of the broad component are a factor of 4$-$6 larger than the stellar velocity dispersion. Such large velocity dispersions indicate non-gravitational origin, i.e., AGN outflows. In comparison, the velocity dispersions of the narrow component are larger than the stellar velocity dispersion by a factor of $\sim$2 at the center, and become comparable to the stellar velocity dispersion at larger distance. We find that the trends in velocity dispersion of narrow and broad components are qualitatively similar for all 15 broad-component-detected AGNs.  

Last, we compare the [\OIII] velocity and velocity dispersion for each component (Figure \ref{fig:vvd_oiii}). We clearly see that the narrow and broad components are located in a different locus on the velocity-velocity dispersion (VVD) diagram. For example, J0911+1333 shows narrow components (blue dots) with larger velocity offsets compared to the broad components (red dot), while the velocities obtained from the total profile show a velocity in the middle of the velocities for narrow and broad component. The narrow components are located within 300 \kms\ of the velocity offset, while the broad components show a blueshifted velocity offset from $-$500 \kms\ to $-$100 \kms. The mean velocity dispersion of the narrow component is $\sim$200 \kms, while that of the broad component is $\sim$550 \kms. The rotational features and smaller range of velocity dispersion indicate that the narrow-component kinematics are related to the gravitational potential of the host galaxy, while the broad-component kinematics are due to non-gravitational phenomena.

\begin{figure}
\centering
\includegraphics[width=0.48\textwidth]{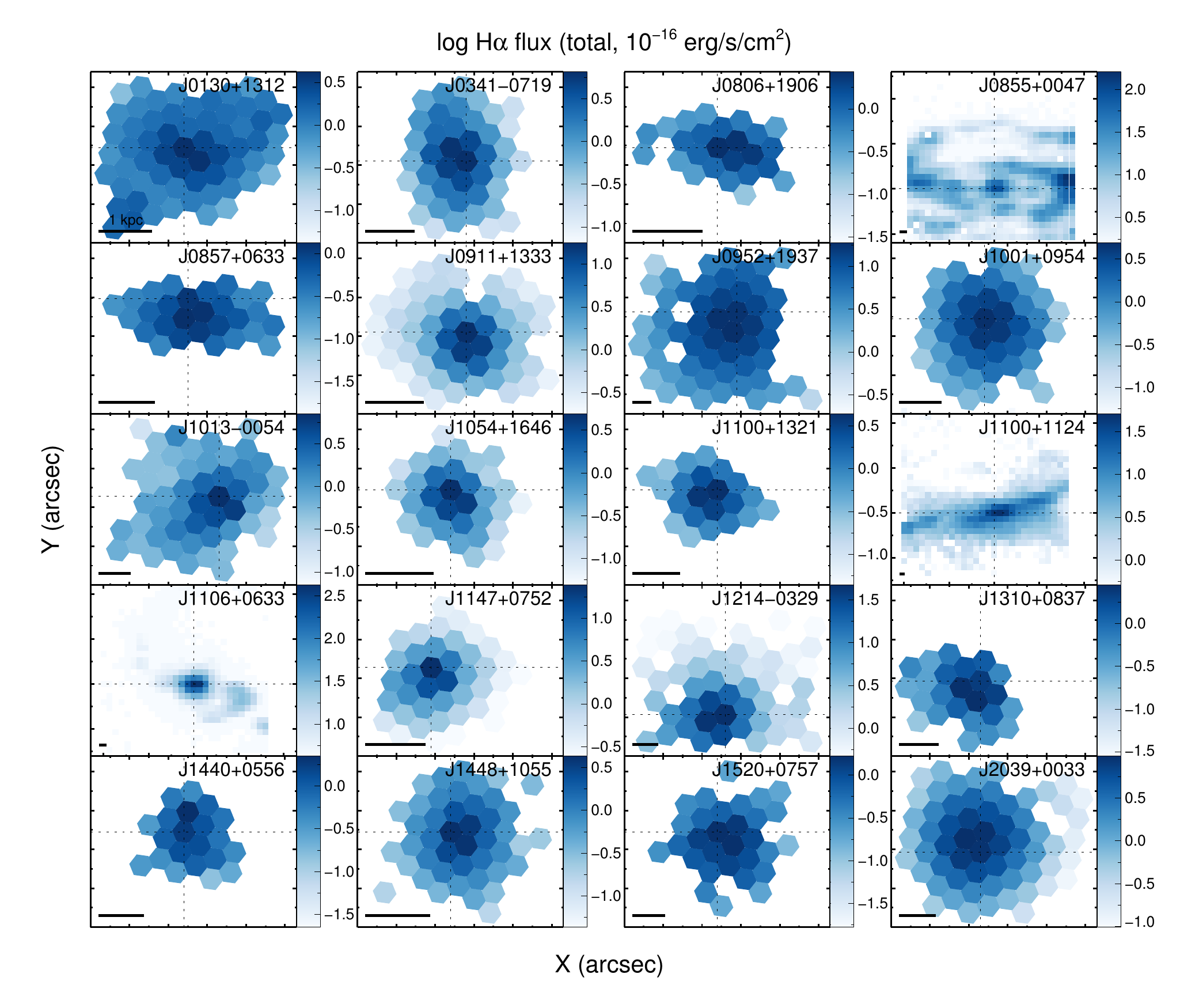}
\includegraphics[width=0.48\textwidth]{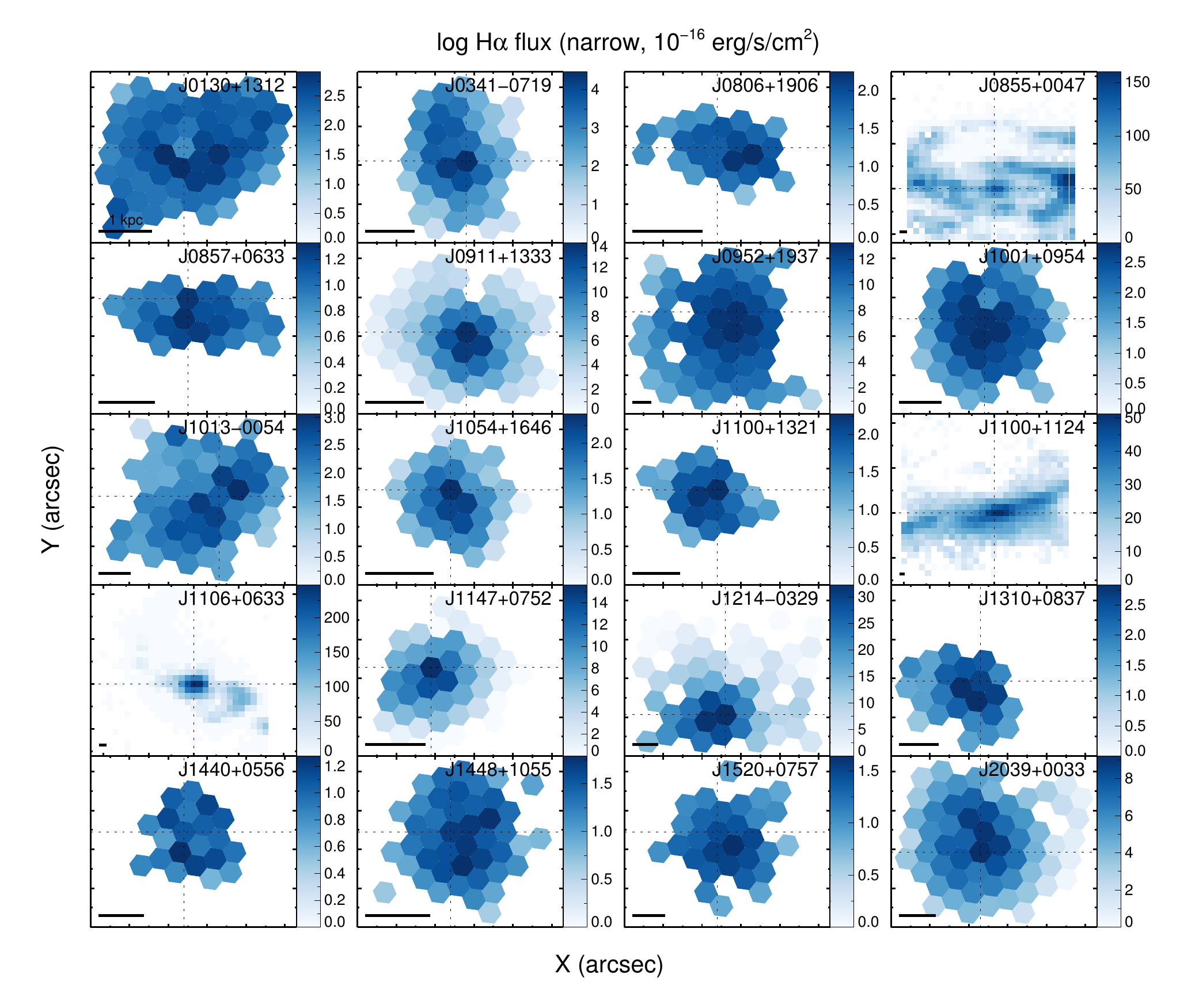}
\includegraphics[width=0.48\textwidth]{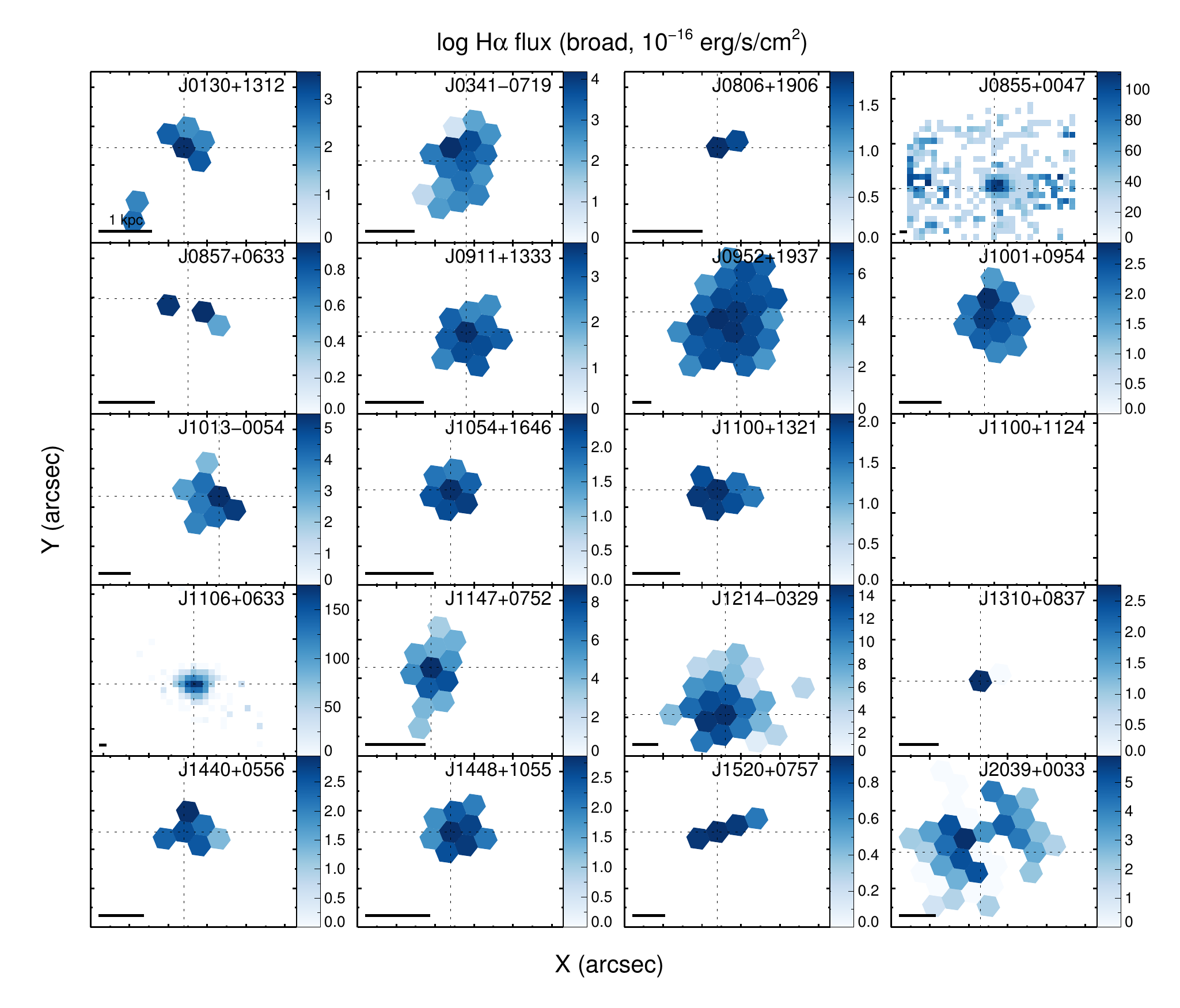}

\caption{The flux maps (same as Figure \ref{fig:oiii_flux}) for \Ha.}
\label{fig:ha_flux}
\end{figure}

\begin{figure}
\centering
\includegraphics[width=0.48\textwidth]{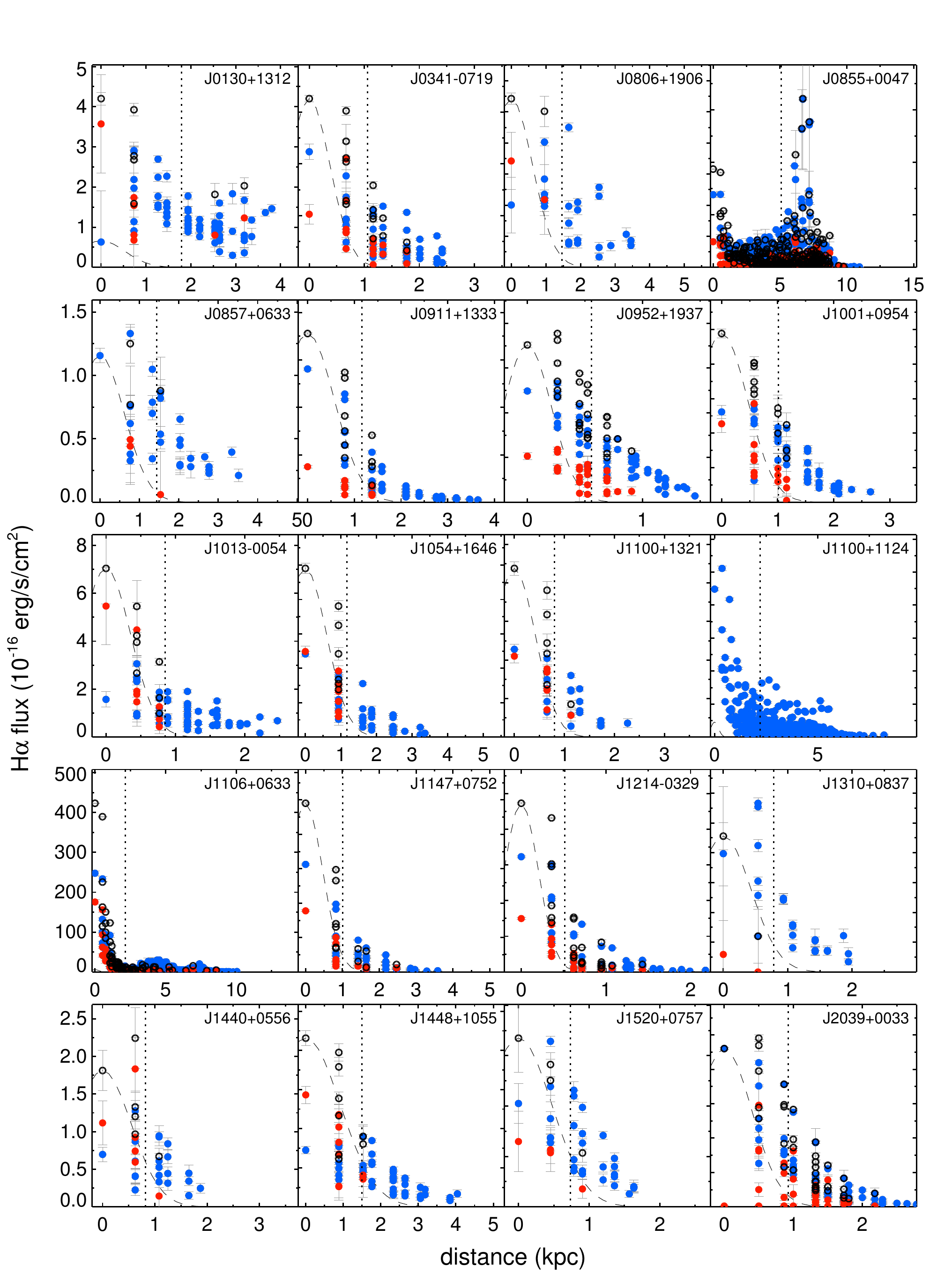}
\caption{The radial flux distributions (same as Figure \ref{fig:dist_oiii_flux}) for \Ha.}
\label{fig:dist_ha_flux}
\end{figure}

\begin{figure}
\centering
\includegraphics[width=0.48\textwidth]{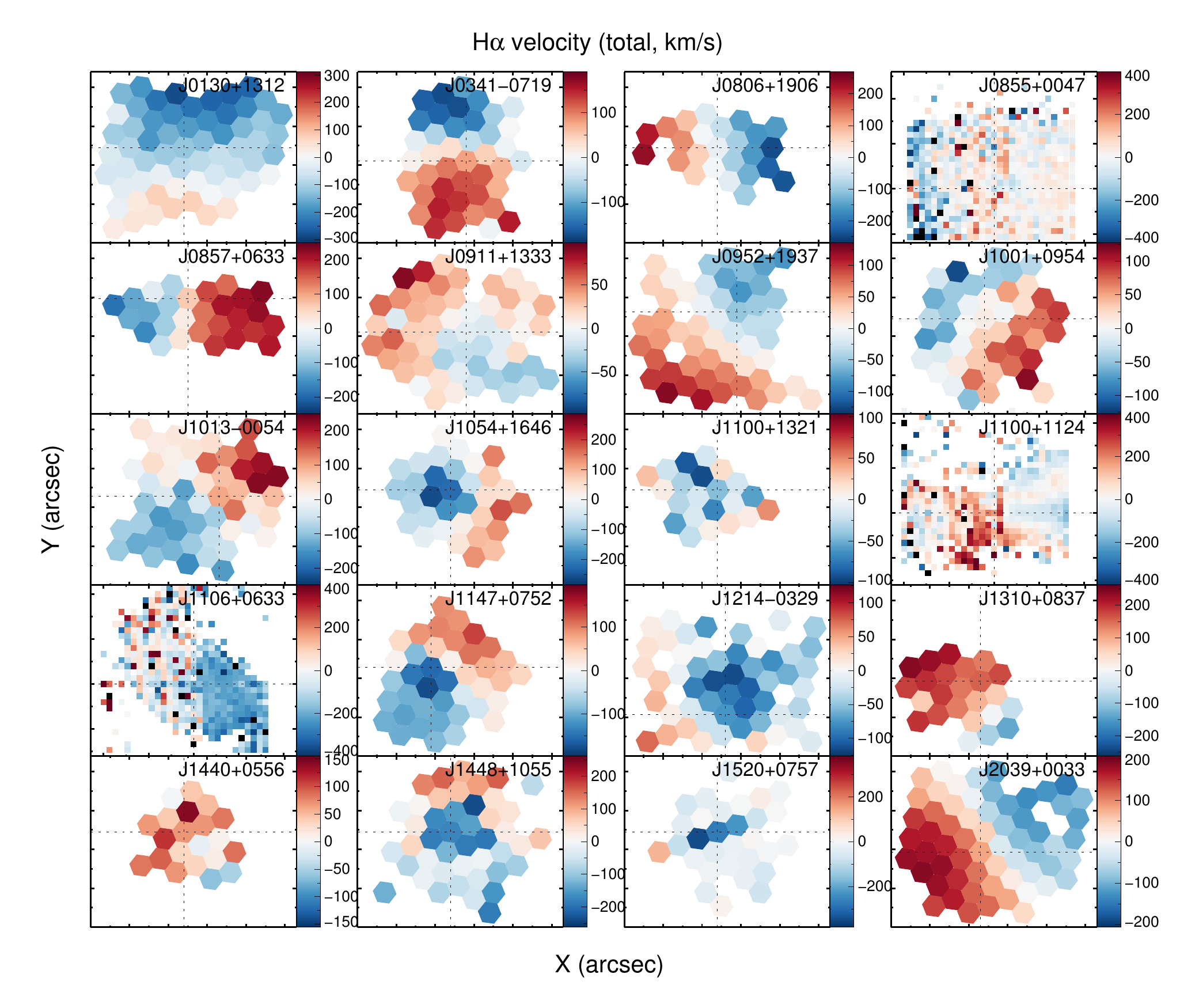}
\includegraphics[width=0.48\textwidth]{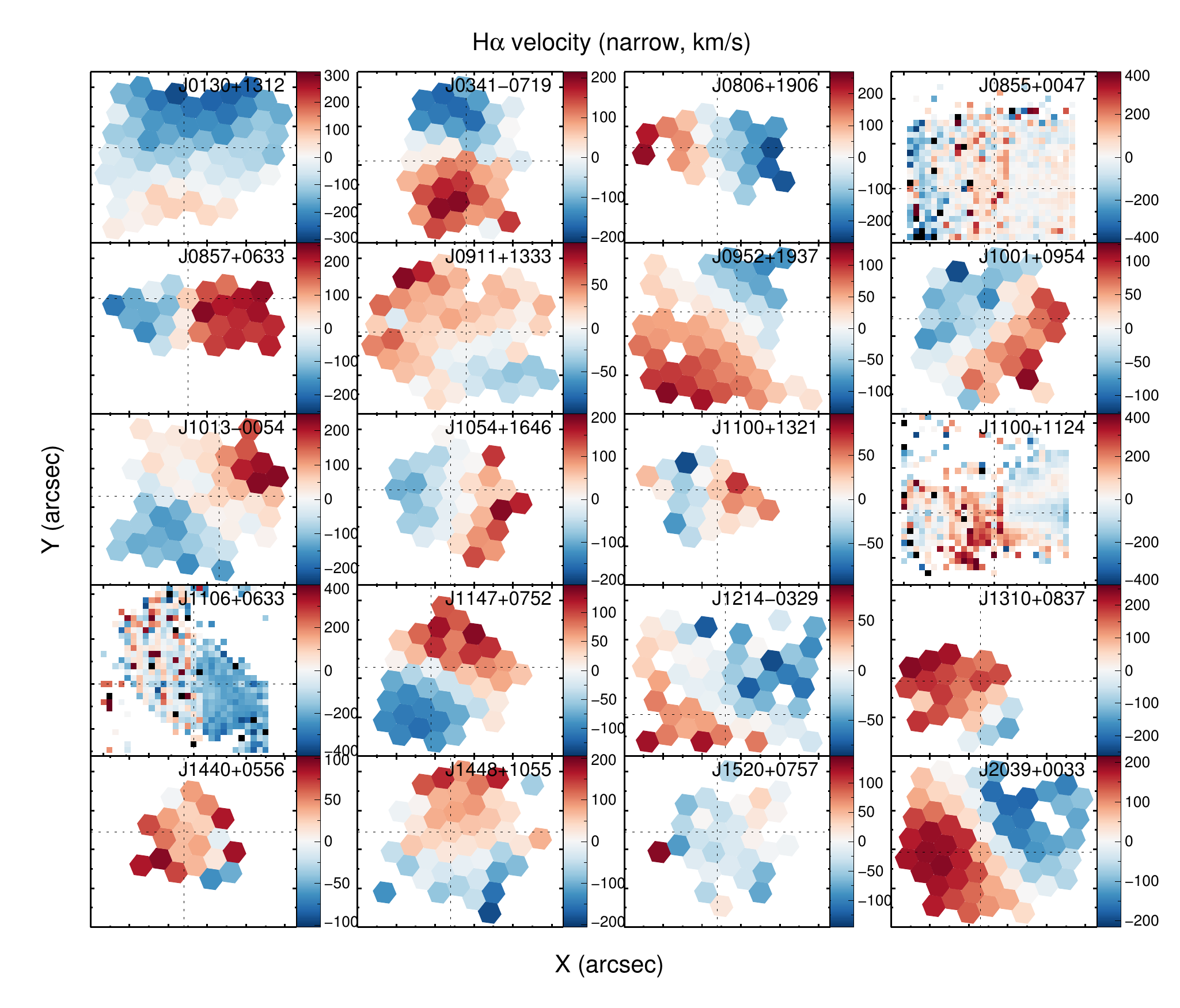}
\includegraphics[width=0.48\textwidth]{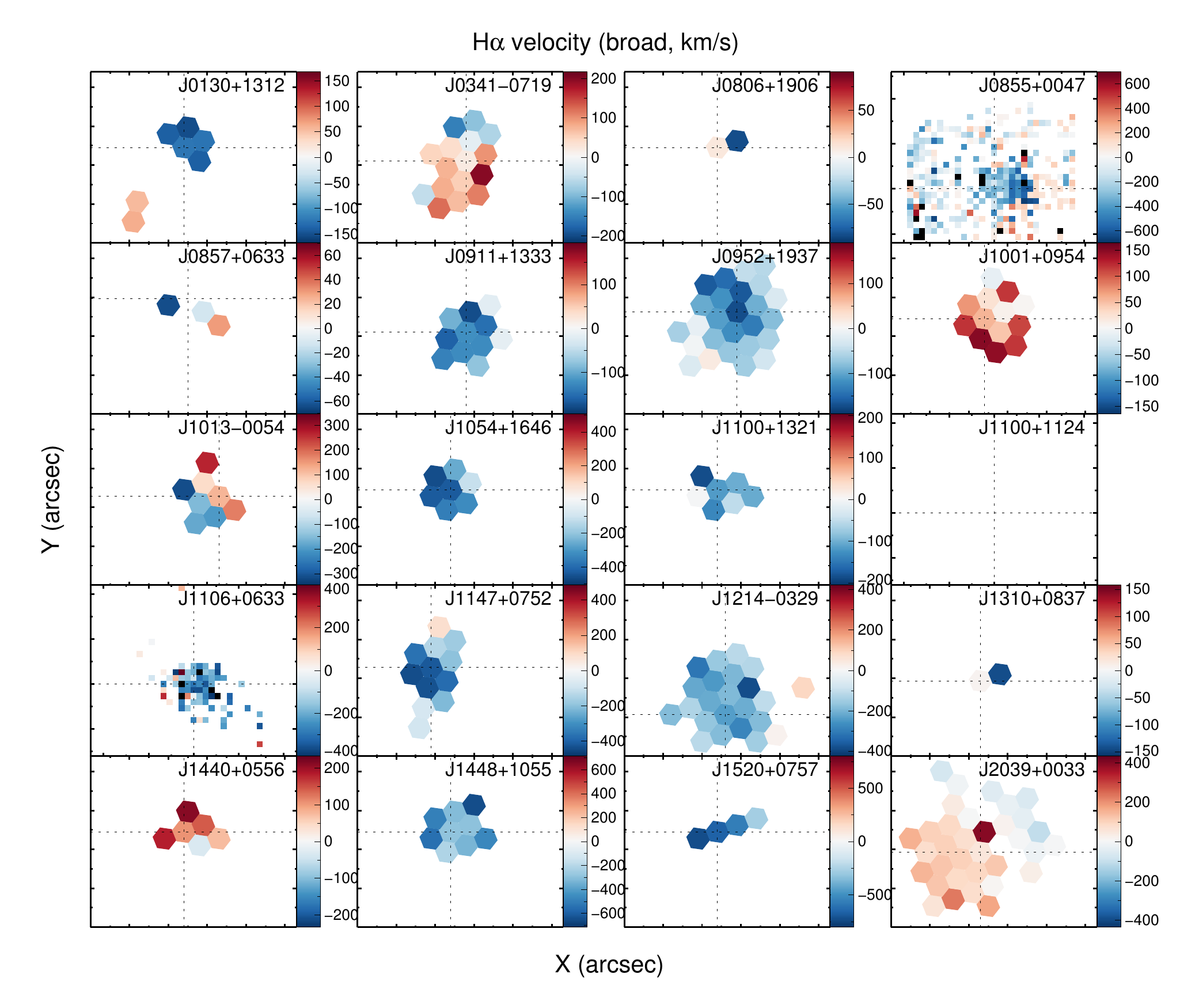}

\caption{The velocity maps (same as Figure \ref{fig:oiii_vel}) for \Ha.}
\label{fig:ha_vel}
\end{figure}

\begin{figure}
\centering
\includegraphics[width=0.48\textwidth]{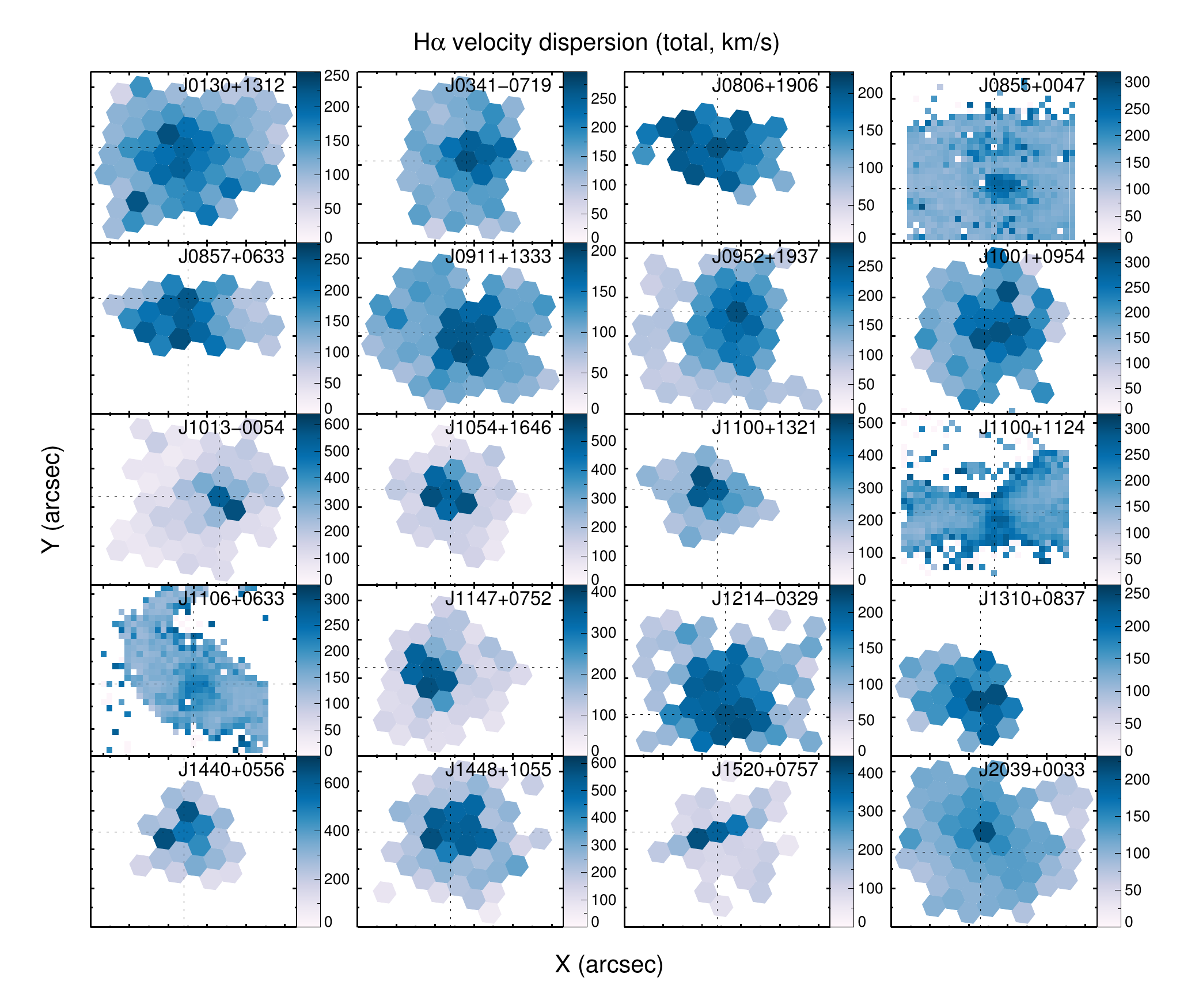}
\includegraphics[width=0.48\textwidth]{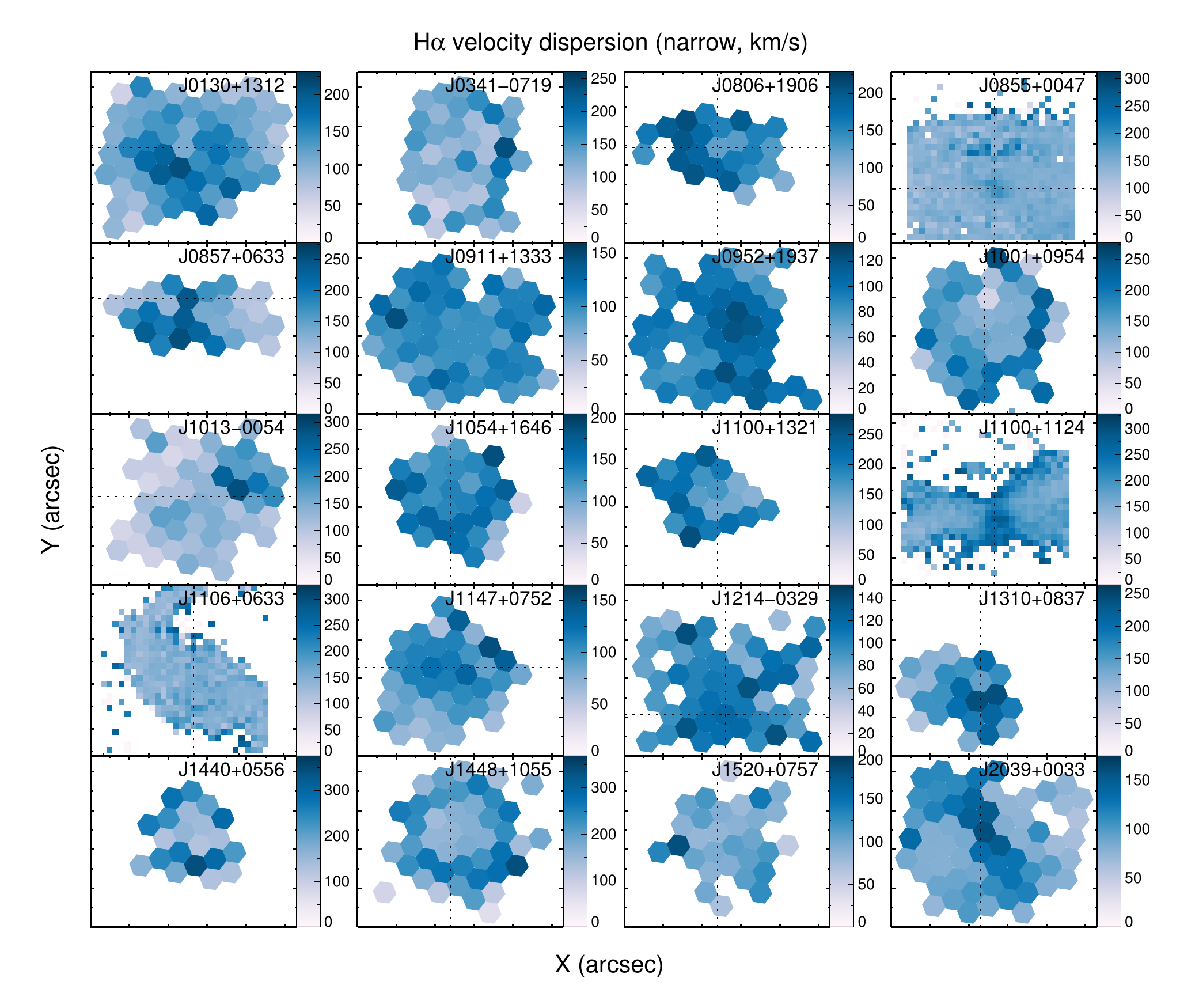}
\includegraphics[width=0.48\textwidth]{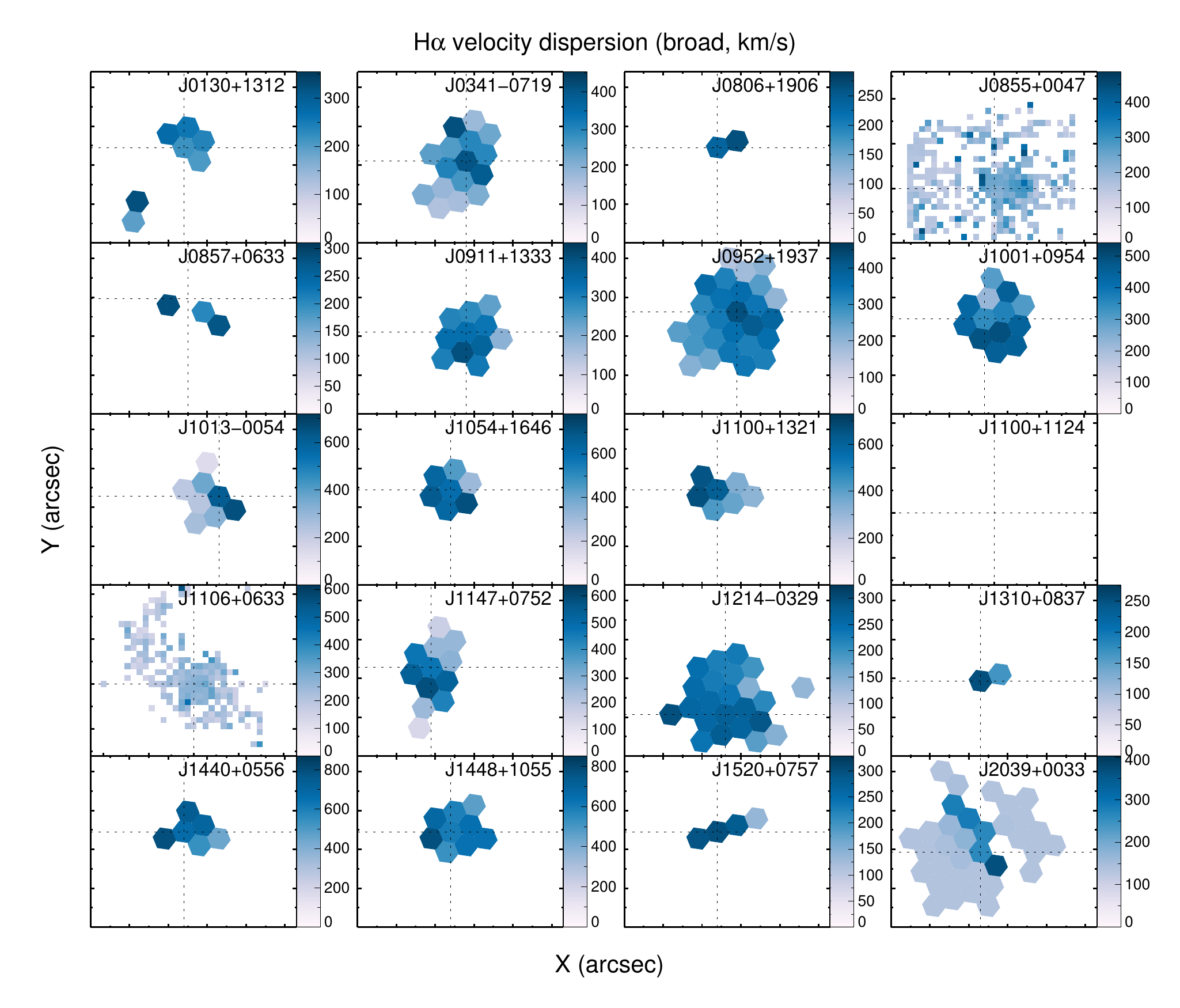}

\caption{The velocity dispersion maps (same as Figure \ref{fig:oiii_veldisp}) for \Ha.}
\label{fig:ha_veldisp}
\end{figure}

\begin{figure*}
\centering
\includegraphics[width=0.48\textwidth]{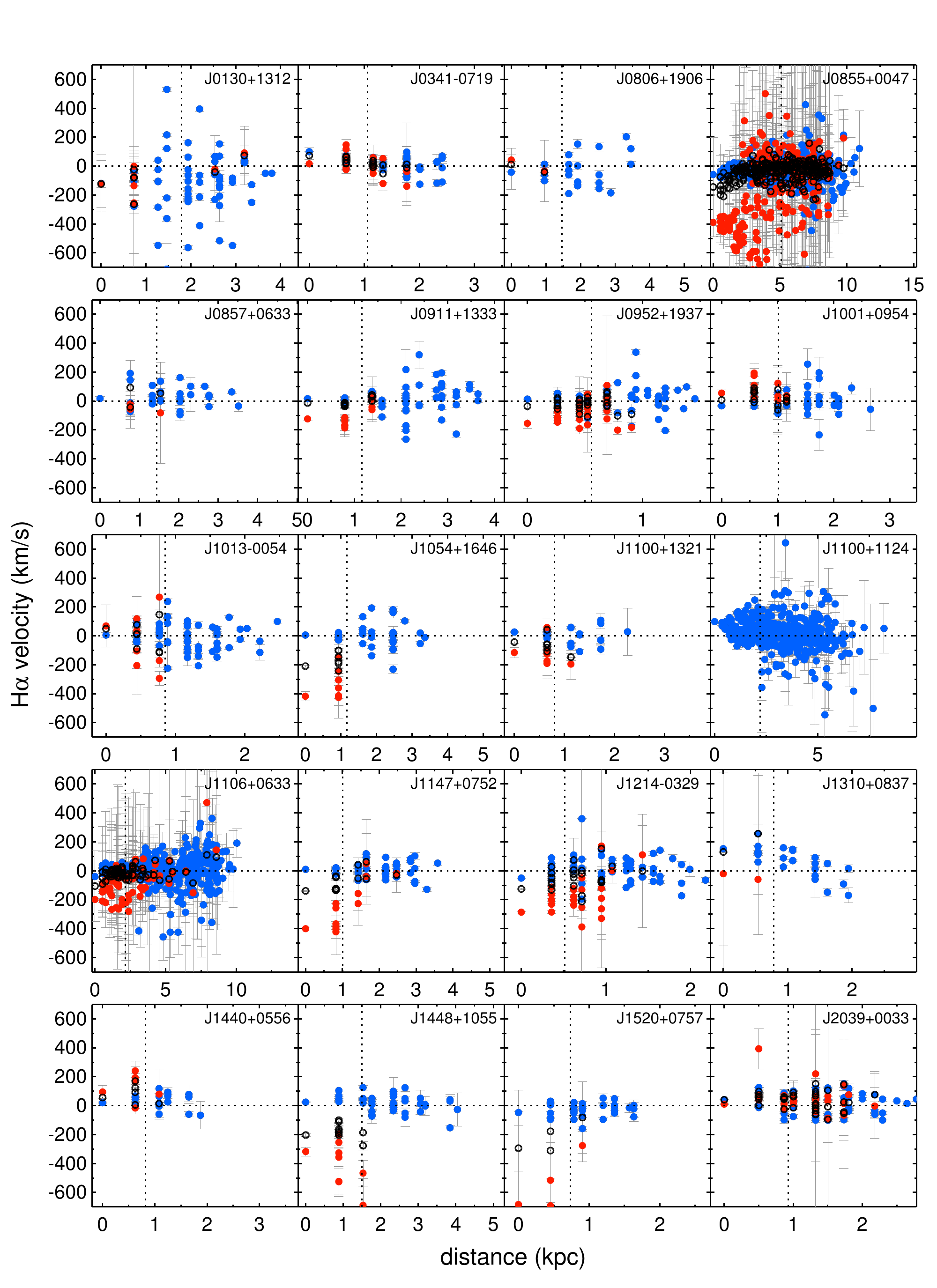}
\includegraphics[width=0.48\textwidth]{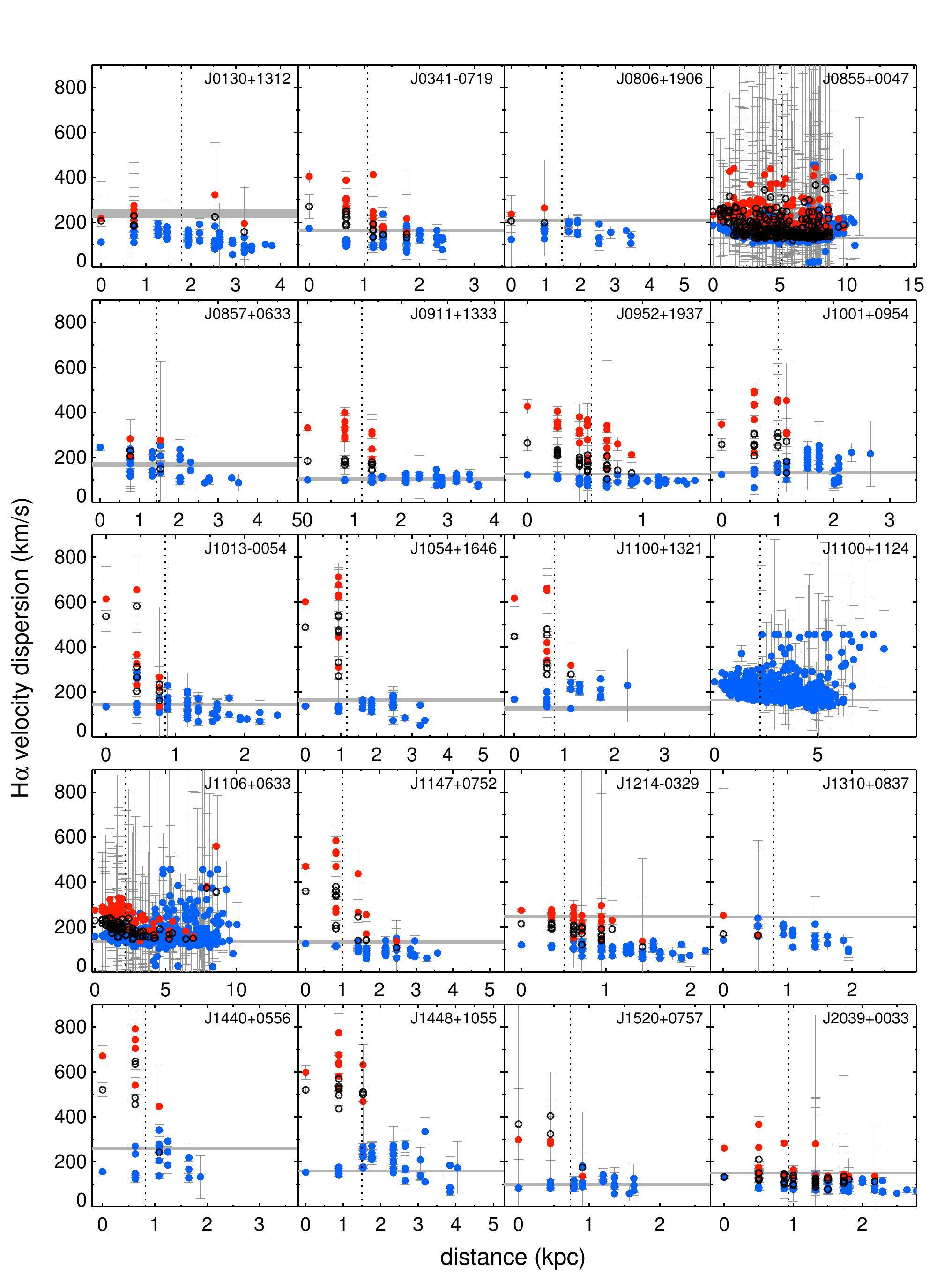}
\caption{The radial distributions of velocity (left) and velocity dispersion (right) (same as Figure \ref{fig:dist_oiii_vvd}) for \Ha.}
\label{fig:dist_ha_vvd}
\end{figure*}

\begin{figure}
\centering
\includegraphics[width=0.48\textwidth]{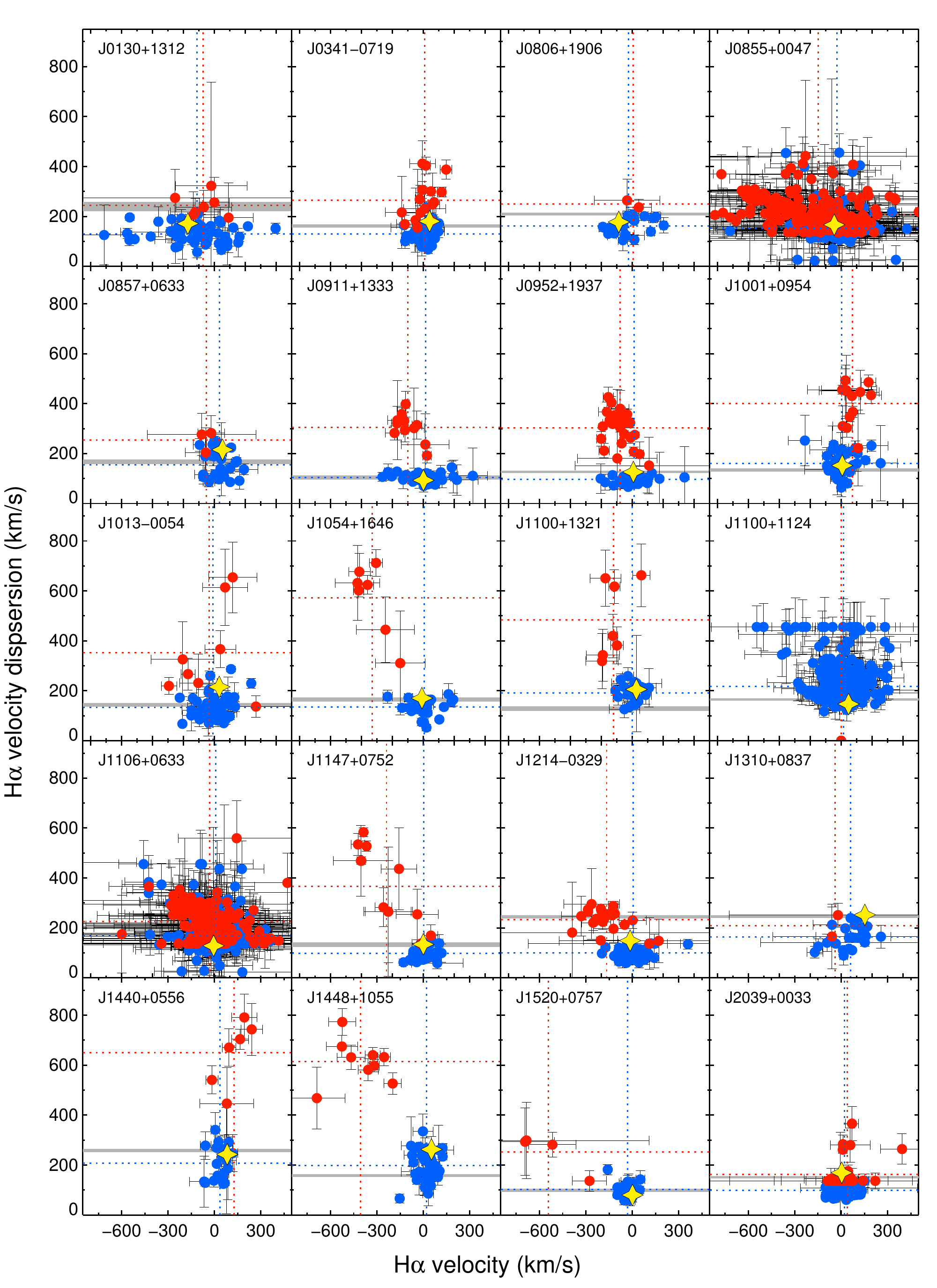}
\caption{The VVD diagrams as the same with the Figure \ref{fig:vvd_oiii}, but for \Ha.}
\label{fig:vvd_ha}
\end{figure}

\subsection{the \Ha-emitting region}
\label{ha}
\subsubsection{Morphology}
\label{ha_morph}
We perform a visual inspection of the morphology of the \Ha\ flux distribution (Figure \ref{fig:ha_flux}). Similarly to [\OIII], the morphology of the \Ha\ flux distribution follows the flux distribution of the stellar component in the host galaxy. We also find a lopsidedness in \Ha\ flux in six out of 20 AGNs (Figure \ref{fig:dist_ha_flux}). Interestingly, such asymmetry is more frequently found in [\OIII] (12 out of 20) than in \Ha. All objects with \Ha\ flux lopsidedness also show [\OIII] flux lopsidedness, implying that what causes the lopsidedness in the flux distribution is commonly affecting both \Ha- and [\OIII]-emitting regions, but is less significant in \Ha.

By performing a multi-Gaussian decomposition for the \Ha+[\NII] lines, we find that 19 out of 20 AGNs have a detectable broad component in the \Ha\ line profile (bottom panel of Figure \ref{fig:ha_flux}). Similarly to [\OIII], the spatial distribution of the broad \Ha\ component is smaller than that of the narrow component. Also, the morphologies of the spaxels with an \Ha\ broad component is mostly irregular, but somewhat comparable to those of the [\OIII] broad component. We find that all AGNs with a [\OIII] broad component (15) also have a broad component in \Ha, implying a common physical origin for both broad components in [\OIII] and \Ha. Also, there are four AGNs having a broad component in \Ha\ but not in [\OIII]. Among them, three AGNs have a lower S/N in [\OIII] than in \Ha\, while the remaining AGN (J1106+0633) has comparable S/N in [\OIII] and \Ha. For this object, it is possible that the [\OIII] broad component was not revealed by our VLT/VIMOS observations, since the configuration has a low spectral resolution ($R\sim$720).

\subsubsection{Kinematics}
\label{ha_kin}
We examine the velocity structure of the \Ha-emitting region (Figure \ref{fig:ha_vel}). When we focus on the total component of \Ha, we find that 13 AGNs show either negative (9) or positive (4) velocity offset, while the other seven AGNs show velocity structures similar to those of the stellar component, i.e., rotational or systemic velocity. If we examine the narrow component of \Ha, we find that the velocity of the narrow \Ha\ is consistent with systemic or rotational velocity in most objects (16/20). Among the remaining four AGNs, two AGNs show negative velocity offset and the other two AGNs show positive velocity offset. After removing the broad component in the total profile, the velocity map of the narrow component of \Ha\ has the signature of Keplerian disk rotation (see Table \ref{tbl-2}). Rotational disk features are found in the narrow component of \Ha\ of 16 AGNs in our sample. Seven AGNs have a rotational disk feature in the narrow component of both \Ha\ and [\OIII], which are consistent with one another. 

Similarly to [\OIII], the maps of the velocity dispersion of the total \Ha\ show clear signs of the mixture of narrow and broad component in the central region (Figure \ref{fig:ha_veldisp}). The large velocity dispersion and spatial concentration of the broad component in \Ha\ also support its non-gravitational origin. After removing the broad component, the narrow components are, in general, consistent with the stellar velocity dispersion of the host galaxy.  

Also, the distributions of \Ha\ velocity and velocity dispersion as a function of distance are qualitatively similar to those of [\OIII] (Figure \ref{fig:dist_ha_vvd}), while the scales of the \Ha\ velocitiy and velocity dispersion are generally smaller than those in [\OIII]. Similarly, the narrow and broad components are clearly separated in the VVD diagram, as we found in the [\OIII] (Figure \ref{fig:vvd_ha})

\begin{figure}
\centering
\includegraphics[width=0.49\textwidth]{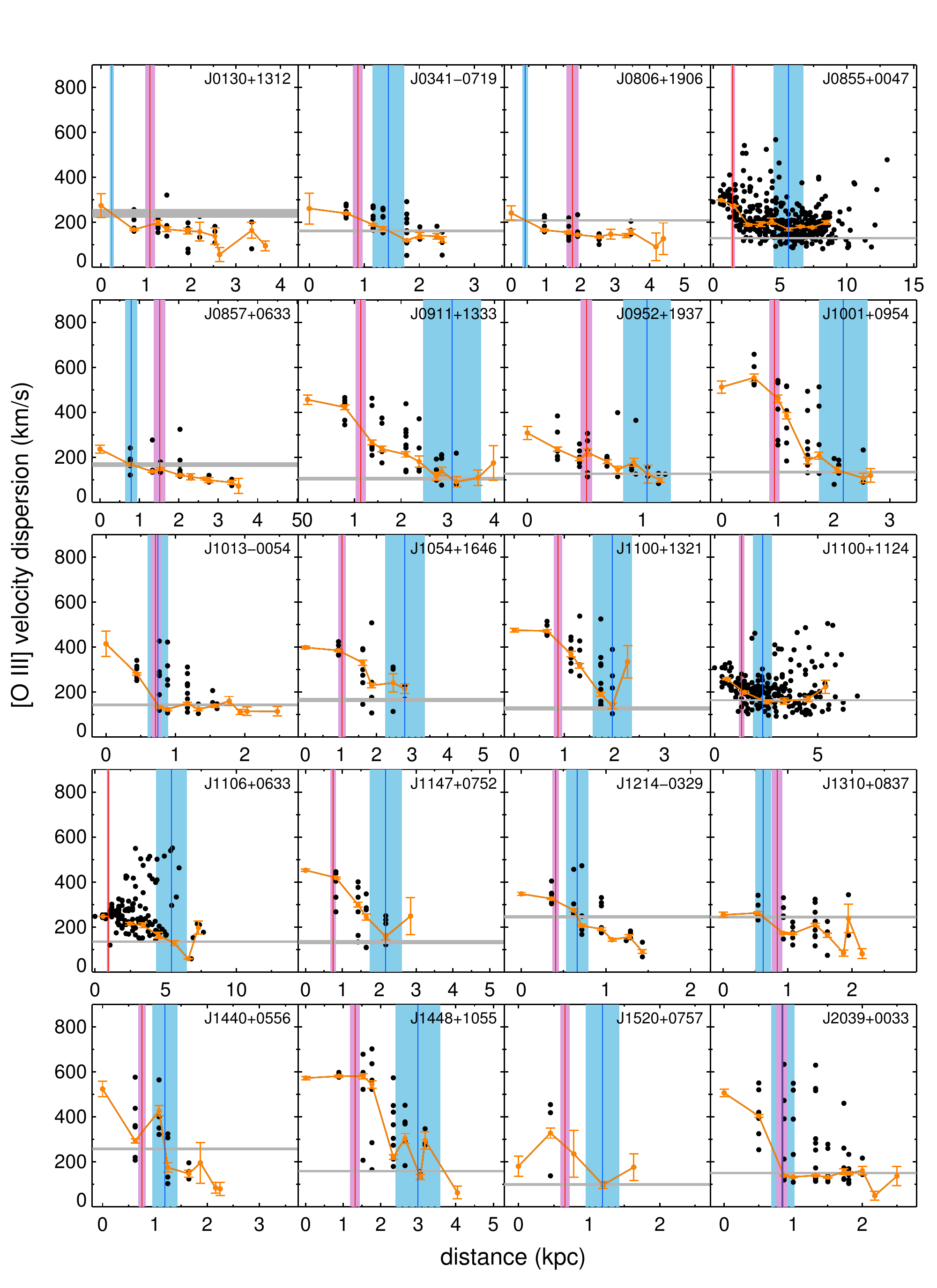}
\caption{The radial distributions of the [\OIII] velocity dispersion (black dots) with the sizes of $R_{\text{NLR}}$ and $R_{\text{out}}$ (see Section \ref{size}). The velocity dispersions were measured from the total profile of [\OIII]. The calculated sizes of $R_{\text{NLR}}$ and $R_{\text{out}}$ are denoted with red and blue vertical lines, respectively. The shaded regions indicate the uncertainties of the sizes.}
\label{fig:sizes}
\end{figure}

\begin{figure*}
\centering
\includegraphics[width=0.95\textwidth]{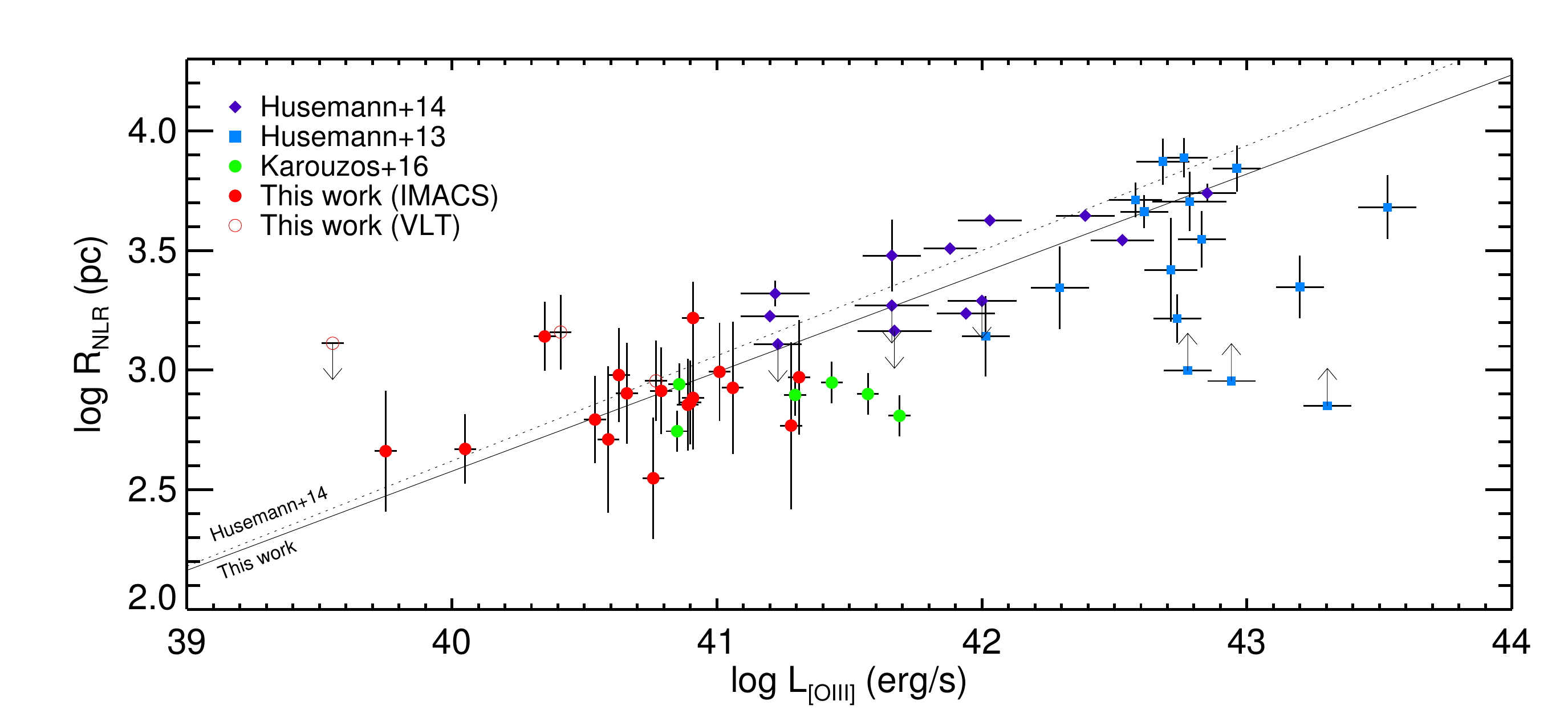}
\caption{The size--luminosity relationship for $R_{\text{NLR}}$ of 55 AGNs. We denote the 17 IMACS-observed AGNs with filled red dots and the 3 VIMOS-observed targets with unfilled red dots. To obtain the relationship for a larger dynamic range of [\OIII] luminosity, we compile the results from 29 luminous type 1 quasars at z $<$ 0.3 \citep[][blue squares and purple diamonds, respectively]{2013A&A...549A..43H,2014MNRAS.443..755H} and six type 2 AGNs \citep[][green dots]{2016ApJ...819..148K}, in which $R_{\text{NLR}}$ is measured in the same manner. The [\OIII] luminosity presented here is the extinction-uncorrected value. The solid line shows the slope of 0.41 resulting from our linear regression, while the other line indicates the slope from the literature as a comparison \citep{2014MNRAS.443..755H}. }
\label{fig:size_lum}
\end{figure*}

\begin{figure*}
\centering
\includegraphics[width=0.95\textwidth]{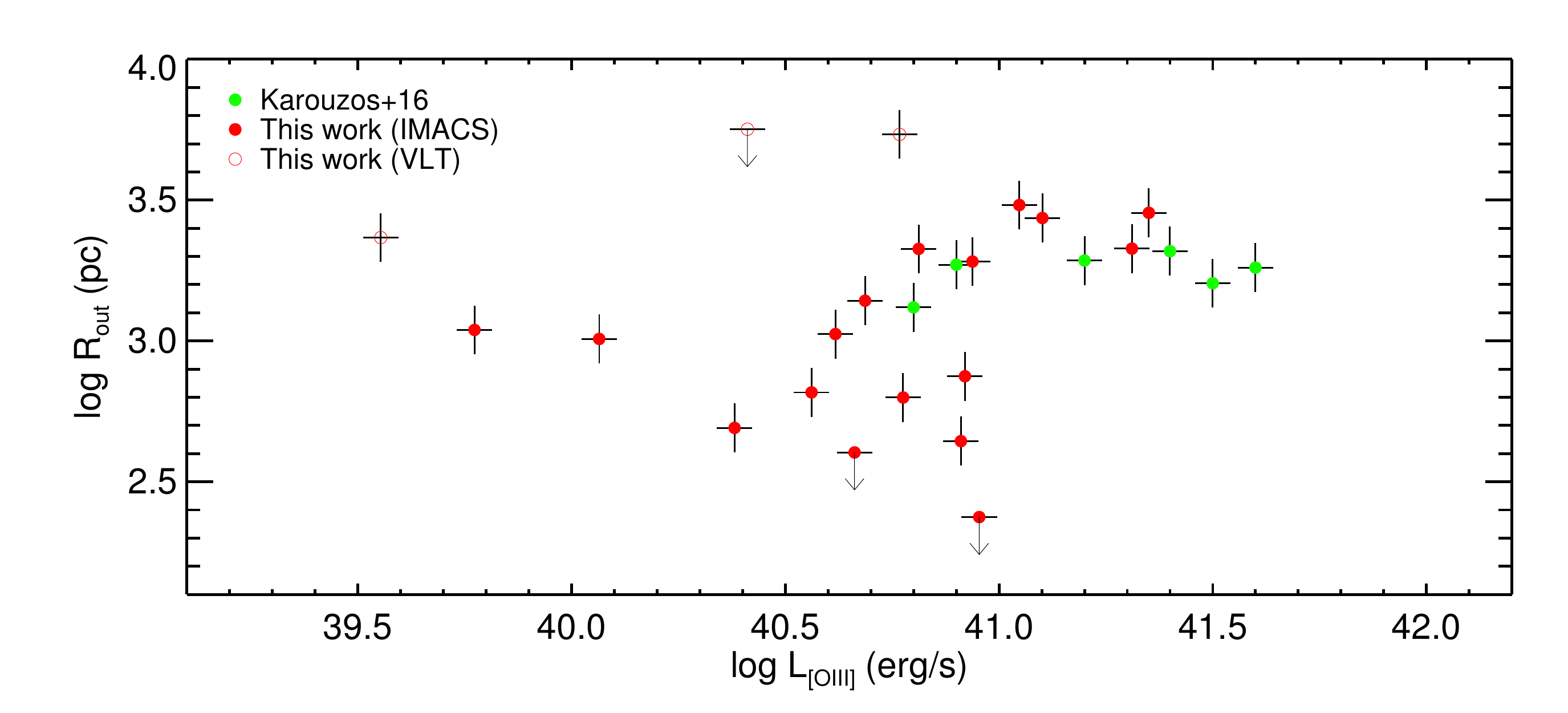}
\caption{The size--luminosity relationship for $R_{\text{out}}$ of 26 type 2 AGNs. We denote the 17 IMACS-observed AGNs with filled red dots, while the 3 VIMOS-observed targets with unfilled red dots. To obtain the relationship for a larger dynamic range of [\OIII] luminosity, we compile the results from six type 2 AGNs \citep[][green dots]{2016ApJ...819..148K}, in which $R_{\text{out}}$ is measured in the same manner. The [\OIII] luminosity presented here is the extinction-uncorrected value.}
\label{fig:size_lum_kin}
\end{figure*}

\begin{table}
\center
\caption{$L_{\text{[O III]}}$ and the sizes of the NLR and outflows of the type 2 AGNs \label{tbl-3}}
\begin{tabular}{ccccccccccc}
\tableline
\tableline
SDSS name &log $L_{\text{[O III]}}$& log $R_{\text{NLR}}$ & log $R_{\text{out}}$     \\
(1) & (2) & (3) & (4)  \\

\tableline
J0130+1312  & 40.63& 2.98$\pm$0.20 & 2.37$^a$  \\
J0341$-$0719& 40.66& 2.90$\pm$0.21 & 3.14$\pm$0.08\\
J0806+1906  & 40.91& 3.22$\pm$0.15 & 2.60$^a$ \\
J0855+0047  & 40.41& 3.16$\pm$0.16 & 3.75$\pm$0.08\\
J0857+0633  & 40.35& 3.14$\pm$0.14 & 2.69$\pm$0.08\\
J0911+1333  & 41.01& 2.99$\pm$0.21 & 3.48$\pm$0.08\\
J0952+1937  & 40.05& 2.67$\pm$0.15 & 3.01$\pm$0.08\\
J1001+0954  & 40.79& 2.91$\pm$0.18 & 3.33$\pm$0.08\\
J1013$-$0054& 40.54& 2.79$\pm$0.18 & 2.82$\pm$0.08\\
J1054+1646  & 41.06& 2.93$\pm$0.28 & 3.44$\pm$0.08\\
J1100+1321  & 40.91& 2.88$\pm$0.22 & 3.28$\pm$0.08\\
J1100+1124  & 39.55& 3.11$^a$          & 3.37$\pm$0.08\\
J1106+0633  & 40.77& 2.96$\pm$0.17 & 3.73$^a$\\
J1147+0752  & 41.28& 2.77$\pm$0.35 & 3.33$\pm$0.08\\
J1214$-$0329& 40.76& 2.55$\pm$0.25 & 2.80$\pm$0.08\\
J1310+0837  & 40.89& 2.86$\pm$0.19 & 2.64$\pm$0.08\\
J1440+0556  & 40.59& 2.71$\pm$0.31 & 3.02$\pm$0.08\\
J1448+1055  & 41.31& 2.97$\pm$0.24 & 3.46$\pm$0.08\\
J1520+0757  & 39.75& 2.66$\pm$0.25 & 3.04$\pm$0.08\\
J2039+0033  & 40.90& 2.86$\pm$0.18 & 2.87$\pm$0.08\\

\tableline
\end{tabular}
\tablecomments{(1) Name of the AGN; (2) extinction-uncorrected [\OIII] luminosity in logarithm with uncertainty (\ergs); (3) the flux-weighted size of the NLR in logarithm with uncertainty (pc); (4) the kinematic size of outflows in logarithm with uncertainty (pc). Note that we adopt a different cosmology for the sizes and luminosity calculation as $H_o$ = 70 km s$^{-1}$ Mpc$^{-1}$, $\Omega_{\Lambda}$ = 0.70, and $\Omega_{m}$ = 0.30, in order to have a consistency with the results in the literature.\\
$^a$Upper-limit of the sizes \\
}
\end{table}

\subsection{Sizes of the NLR and gas outflows}
\label{size}
The size of the NLR $(R_{\text{NLR}})$ is typically measured based on the flux distribution of the emission lines, without taking into account of the outflow properties \citep[e.g.,][]{2002ApJ...574L.105B,2003ApJ...597..768S,2013MNRAS.430.2327L}. To quantify the size of the outflows, we defined outflow size $R_{\text{out}}$, which is the radius where the [\OIII] velocity dispersion is equal to the stellar velocity dispersion of its host galaxy as \citet{2016ApJ...819..148K}.
Here we use the two sizes, $(R_{\text{NLR}})$ and $(R_{\text{out}})$ to investigate the properties of the NLR and outflow kinematics, respectively.  

First, we measure the flux-weighted mean size of the NLR \citep{2014MNRAS.443..755H} for $R_{\text{NLR}}$ as
\begin{eqnarray}
R_{\text{NLR}} = {\int R f(R) dR \over \int f(R) dR}, 
\end{eqnarray}
where $R$ is the distance from the center and $f(R)$ is the flux at a given distance. We only use the spaxels classified as composite or AGN from the emission-line diagnostics \citep[e.g.,][]{1981PASP...93....5B,2009MNRAS.397..135K}. Then, we correct for seeing effects by subtracting the seeing size in quadrature, resulting in a $\sim$14\% decrease in size. Note that $R_{\text{NLR}}$ for all targets is resolved compared to the seeing size. The mean $R_{\text{NLR}}$ is $\sim$880$\pm$360 pc 
for our sample with the mean $L_{\text{[O III]}}$ of $10^{40.7}$ \ergs, where $L_{\text{[O III]}}$ is extinction-uncorrected [\OIII] luminosity (see Table \ref{tbl-3}). 

To estimate the uncertainty in $R_{\text{NLR}}$, we construct 100 mock spatial distributions of the [\OIII] region by randomizing the flux for each spaxel including noise, and we measure $R_{\text{NLR}}$ in the same manner. Then we adopt 1$\sigma$ of the distribution as the uncertainty of $R_{\text{NLR}}$. In addition to this uncertainty, we add 10\% of $R_{\text{NLR}}$ to account for the uncertainty in the seeing size, and also add an half spaxel size ($\sim$0\farcs 3) to take into account 
the uncertainties from the spatial sampling. Note that $R_{\text{NLR}}$ of J1100+1124 is regarded as the upper-limit of $R_{\text{NLR}}$ since the target is observed using VLT/VIMOS, which has a large FoV, and the flux-weighted size includes some spaxels of spiral arms classified as composite region. These spaxels might be contaminated with shock-induced line emissions in the emission-line diagnostics.

Second, we measure the outflow size $R_{\text{out}}$ based on the 1D distributions of the [\OIII] velocity dispersion as a function of distance from center (Figure \ref{fig:sizes}). We obtain the mean value of the [\OIII] velocity dispersion from the spaxels as a function of distance, then determine the radius where the [\OIII] velocity dispersion is equal to the stellar velocity dispersion of the host galaxy, after linearly interpolating the mean values of the [\OIII] velocity dispersion (orange lines). We also correct the measured size by subtracting the seeing size in quadrature. We assume 20\% of uncertainty in $R_{\text{out}}$. We obtain that the mean $R_{\text{out}}$ is $\sim$1800 pc, which is about a factor of two larger than the mean $R_{\text{NLR}}\sim$880 pc (see Table \ref{tbl-3}). The result is in good agreement with the result from \citep{2016ApJ...819..148K}, which reported that $R_{\text{out}}$ is a few times larger than $R_{\text{NLR}}$. Four AGNs (i.e., J0130+1312, J0806+1906, J0857+0633, and J1310+0837), however, show smaller $R_{\text{out}}$ than $R_{\text{NLR}}$ due to their low [\OIII] velocity dispersion compared to the stellar velocity dispersion of the host galaxy, indicating that stellar velocity dispersion may not be the best parameter to separate the non-gravitational outflow signatures from 
the gravitation kinematics. 

Based on our measurements of both $R_{\text{NLR}}$ and [\OIII] luminosity, here we present the size--luminosity relationship (Figure \ref{fig:size_lum}). Since the relationships for type 1 and type 2 AGNs are in good agreement with one another \citep{2003ApJ...597..768S}, we also include 29 high-luminosity type 1 AGNs from the literature with mean $L_{\text{[O III]}}=10^{42.4}$ \ergs\ \citep{2013A&A...549A..43H,2014MNRAS.443..755H} in which the $R_{NLR}$ was measured in a consistent way with our study. We also include six type 2 AGNs obtained from the Gemini/GMOS-IFU \citep{2016ApJ...819..148K,2016ApJ...833..171K}. For this comparison, we use the extinction-uncorrected [\OIII] luminosity and apply the cosmological parameters used in the work of \citet{2014MNRAS.443..755H}, i.e., $\Omega_{\Lambda}$ = 0.70, and $\Omega_{m}$ = 0.30. We assume 10\% uncertainty in $L_{\text{[O III]}}$. 
To fit the size--luminosity relation, we apply a forward regression method using the FITEXY code in the IDL library \citep[e.g.,][]{2012ApJS..203....6P}, obtaining:\begin{equation}
\log R_{\text{NLR}} = (0.41\pm 0.02)\times \log L_{\text{[O III]}} - (14.00\pm 0.77).
\end{equation}
The slope$\sim$0.41$\pm$0.02 is consistent with the slope reported by \citet{2014MNRAS.443..755H} (0.44$\pm$0.06). 

The size--luminosity relation has been reported with different slopes based on different samples and methods, resulting in various physical interpretations for the relation. For example, \citet{2003ApJ...597..768S} found a slope of 0.33$\pm$0.04 for local type 1 and 2 Seyfert galaxies by measuring the size and estimating the [\OIII] luminosity from the HST narrow-band imaging data. In addition, \citet{2013MNRAS.430.2327L} found a slope of 0.25$\pm$0.02 for type 2 quasars and Seyfert galaxies based on heterogeneous data from IFU and long-slit observations. To explain the physical conditions in the NLR of the sources, they argued that the pressure inside of NLR clouds ($P$) as well as the gas density ($n$) drops as radius $r$ increases, i.e., $P(r) \propto r^{-2}$ and $n(r) \propto r^{-2}$, resulting in an ionization parameter ($U$) independent of radius. In contrast, other studies reported a slope of $\sim$0.5 \citep{2002ApJ...574L.105B,2014MNRAS.443..755H,2013ApJ...774..145H}. For example, \citet{2002ApJ...574L.105B} found a slope of 0.52$\pm$0.06 for luminous Seyferts and quasars based on HST narrow-band imaging data. Similarly, \citet{2013ApJ...774..145H} reported a slope of 0.4--0.5 for type 2 quasars and Seyfert galaxies, but they used the luminosity at 8$\mu$m as a proxy of AGNs luminosity rather than [\OIII] luminosity, arguing that the AGN luminosity is more directly traced by the luminosity of 8$\mu$m than of [\OIII]. To explain the slope of $\sim$0.5, these studies adopted a simple model that assumes a constant ionization parameter and the density for the clouds, which is not the case for our sample (see Section \ref{sr_energetics}). 
 
Similar to the photoionization size ($R_{\text{NLR}}$)--luminosity relation, we examine the relationship between $R_{\text{out}}$ and [\OIII] luminosity (Figure \ref{fig:size_lum_kin}). Although the dynamic range of [\OIII] luminosity is rather small in our sample, we find no clear relationship between $R_{\text{out}}$ and [\OIII] luminosity, presumably due to the dynamical timescale of the outflows. If we consider the relatively low outflow velocity$\sim$1000 \kms\ and $R_{\text{out}}\sim$1--2 kpc, it will take (1--2)$\times 10^6$ years to reach $R_{\text{out}}$ for the outflow, which is much larger than the photoionization timescale for the NLR. As a result, $R_{\text{out}}$ may not show a clear relationship with $L_{\text{[O III]}}$ while $R_{\text{NLR}}$ does. To further constrain whether there is a positive relationship between $R_{\text{out}}$ and [\OIII] luminosity, we may need to obtain $R_{\text{out}}$ from higher luminosity type 2 AGNs.

\subsection{Properties of \Ha\ disk}
\label{disk}
In previous sections, we find both gravitational and non-gravitational kinematics in the NLR, and the non-gravitational kinematics are closely related
to rotation as we noticed from the velocity maps of the narrow component of \Ha. The properties of rotational disks in the sample are worth investigating, since they might provide useful hints on AGN feedback and co-evolution. 

By using the integrated spectra within $R_{\text{NLR}}$, we classify the disk represented by the narrow \Ha\ into two groups: 1) SF-type (log [\NII]/H$\alpha \geq$ 0); 2) AGN-type (log [\NII]/H$\alpha$ < 0). Eleven AGNs are classified as SF-type, five AGNs are classified as AGN-type, while four AGNs have no/ambiguous rotation in \Ha\ (summarized in Table \ref{tbl-2}). We use the line ratio of the narrow component of \Ha\ and [\NII] for disk classification, since we clearly see the rotational feature in the narrow component in \Ha\ but not in [\OIII], which strongly represents non-gravitational kinematics.

For the two groups, we compare the $D_{n}(4000)$ and H$\delta_{A}$ indices of the central region of AGN (3\arcsec\ in diameter). We obtained the indices from the MPA-JHU catalog of SDSS DR7 galaxies. The two groups show clearly different ranges of indices. The SF-type group has $D_{n}(4000) = 1.37\pm$0.13 and H$\delta_{A} = 2.83\pm$1.32, while the AGN-type group has $D_{n}(4000) = 1.71\pm 0.15$ and H$\delta_{A} = -0.08\pm$0.92, showing that the SF-type group has smaller $D_{n}(4000)$ and stronger H$\delta_{A}$ than the AGN-type group, as expected. 

\begin{figure}
\centering
\includegraphics[width=0.5\textwidth]{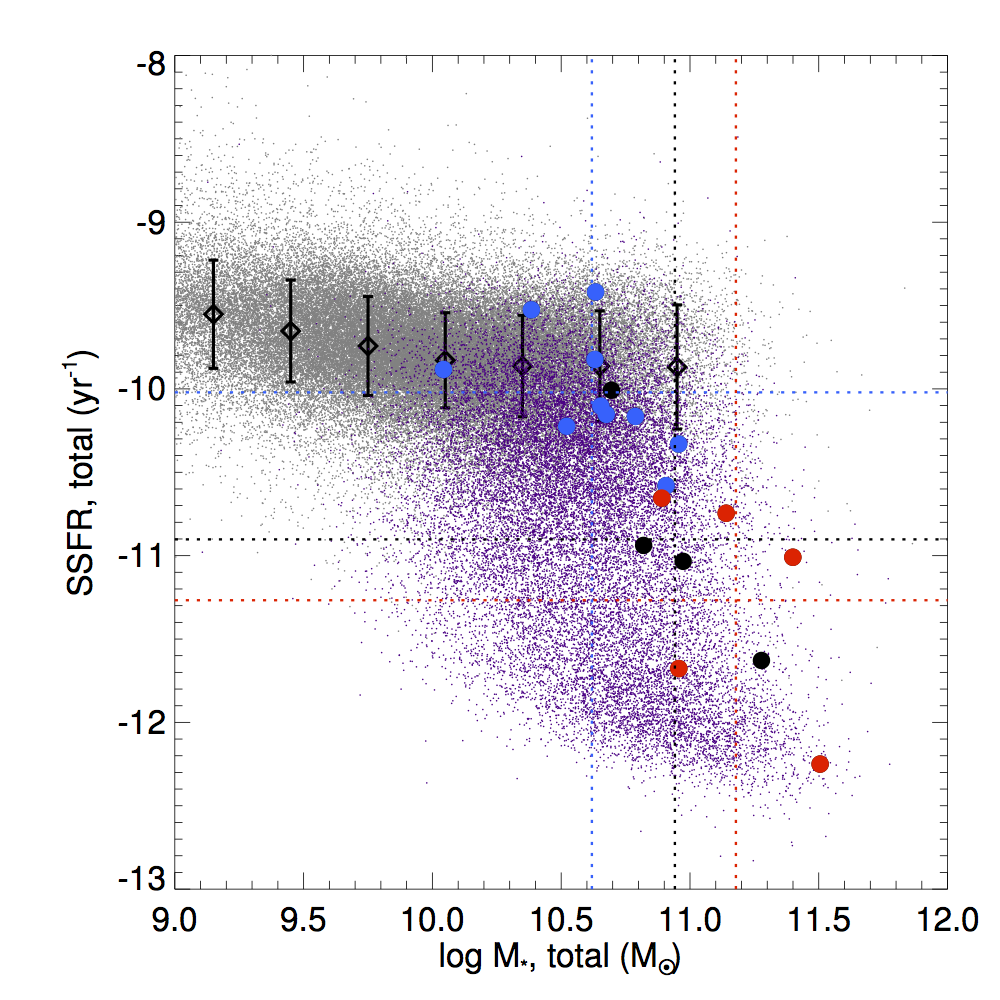}
\caption{The distribution of specific star-formation rate (SSFR) and stellar mass for star-forming galaxies (gray) and AGN-host galaxies (purple). Blue, red, and black dots represent the different types of AGNs with SF-type disk, AGN-type disk, and no/ambiguous rotation, respectively. Black diamonds and error bars represent the mean values SSFR for star-forming galaxies at each bin of stellar mass and their 1$\sigma$ distributions, respectively. Dotted lines represent the mean values of the stellar mass and specific star-formation rate for each group. (see Section \ref{disk}). }
\label{fig:ssfr}
\end{figure}

We also compare the distribution of specific star-formation rate (SSFR) as a function of the stellar mass ($M_{*}$) (Figure \ref{fig:ssfr}). We also adopted the SSFR and $M_{*}$ from the MPA-JHU catalog of SDSS DR7 galaxies. The SSFRs for whole galaxy were estimated by combining the synthetic models on the integrated spectra and photometric information \citep{Brinchmann:2004hyc}, which provides SFRs sensitive over the past 10$^8$--10$^9$ years \citep{1998ARA&A..36..189K}. The AGNs with SF-type disks (blue dots) have higher SSFR (log SSFR = $-$10.0$\pm$0.4 yr$^{-1}$) and smaller stellar mass (log $M_{*}$=10.6$\pm$0.3 $M_{\odot}$), while the AGNs with AGN-type disks (red dots) have smaller SSFR (log SSFR = $-$11.3$\pm$0.7 yr$^{-1}$) and larger stellar mass (log $M_{*}$=11.2$\pm$0.3 $M_{\odot}$). The results consistently indicate that the AGNs with SF-type disks have on-going star formation at a similar level to that of star-forming galaxies of similar stellar mass (see Woo et al. 2017). We will compare the energetics of the AGNs with SF- and AGN-type disks and discuss the feedback scenario for the AGNs in Section \ref{feedback}.

\begin{figure*}
\centering
\includegraphics[width=0.48\textwidth]{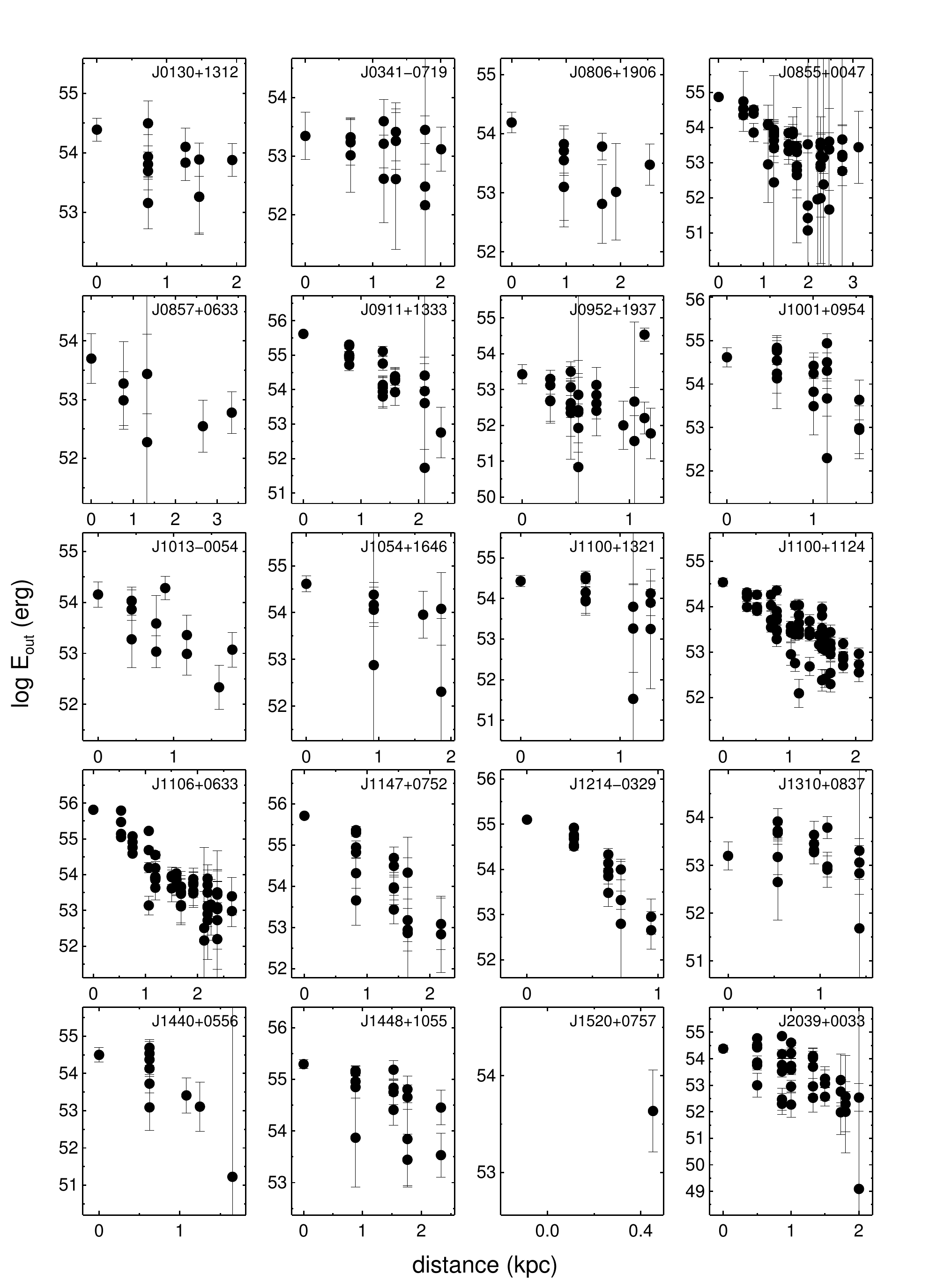}
\includegraphics[width=0.48\textwidth]{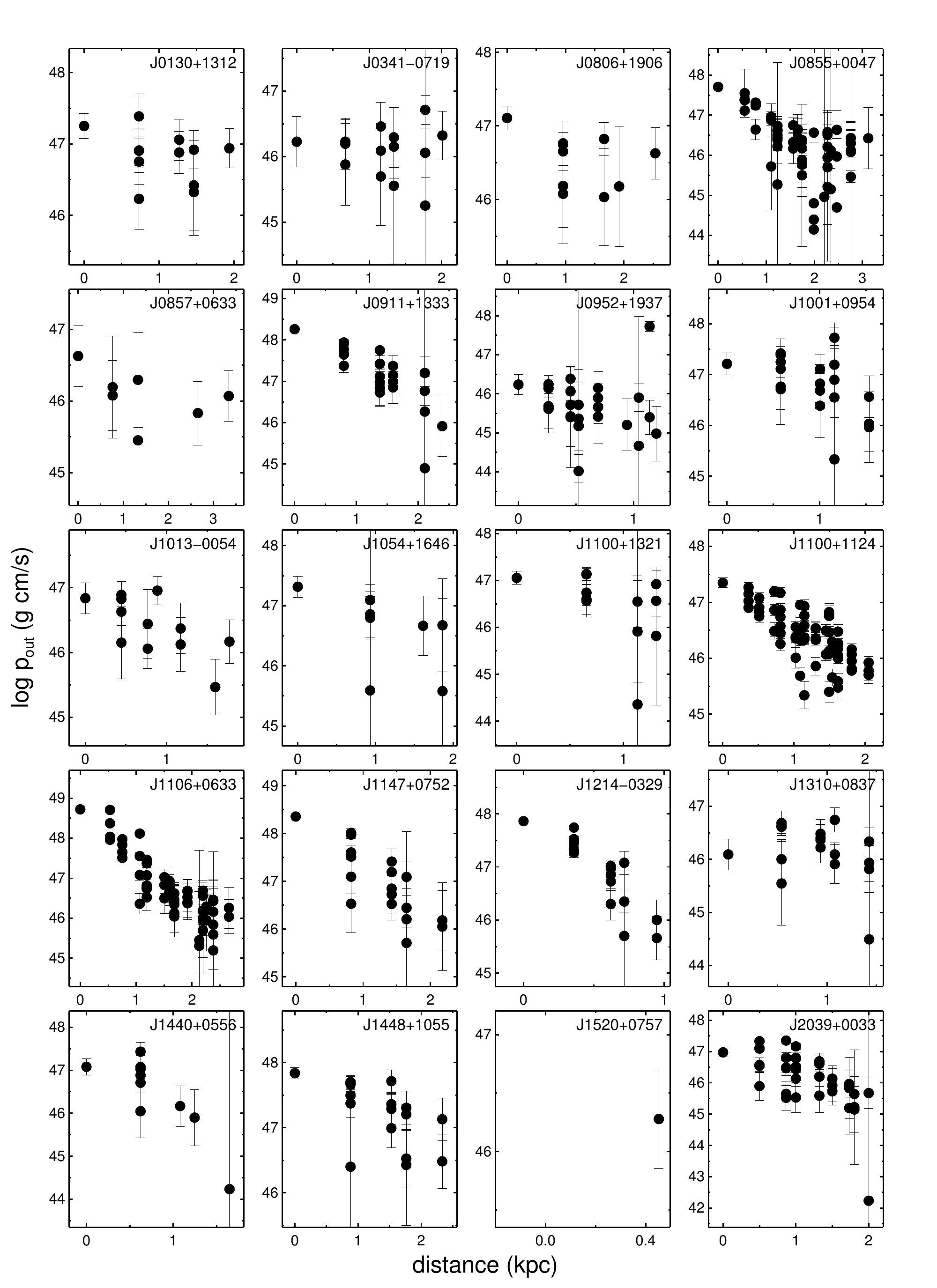}
\caption{The energy (left) and momentum (right) distributions as a function of distance. The error bars denote 1$\sigma$ uncertainties. The AGNs are listed in order of accending R.A from top-left to bottom-right.}
\label{fig:em_dist_total}
\end{figure*}

\begin{figure}
\centering
\includegraphics[width=0.48\textwidth]{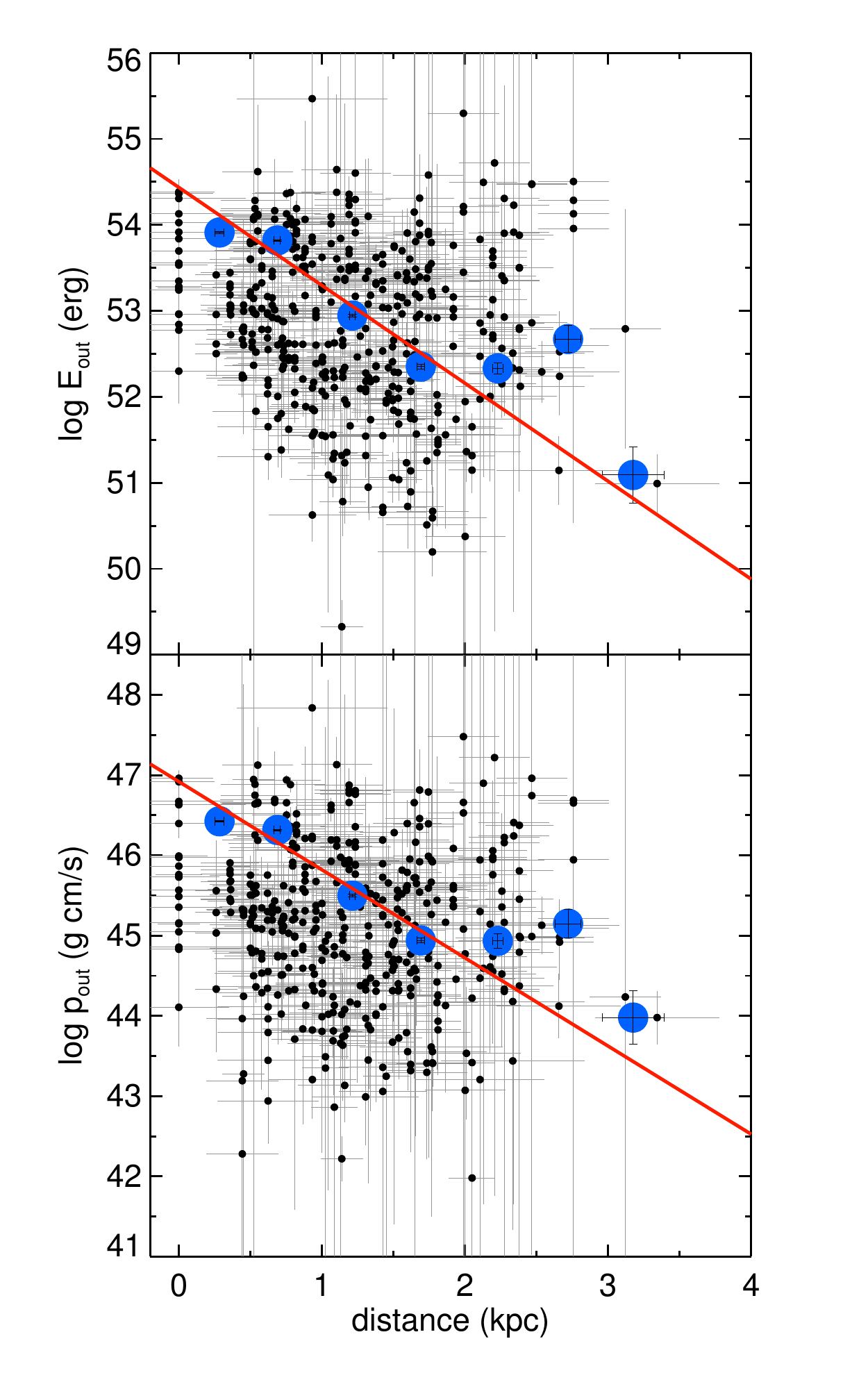}
\caption{The combined energy (top) and momentum (bottom) distributions as a function of distance. Blue dots are the error-weighted values at each bin of 0.5 kpc distance, and the red lines are the result of linear regression for the error-weighted values. The error bars denote 1$\sigma$ uncertainties.}
\label{fig:em_dist_comb}
\end{figure}

\begin{figure*}
\centering
\includegraphics[width=0.33\textwidth]{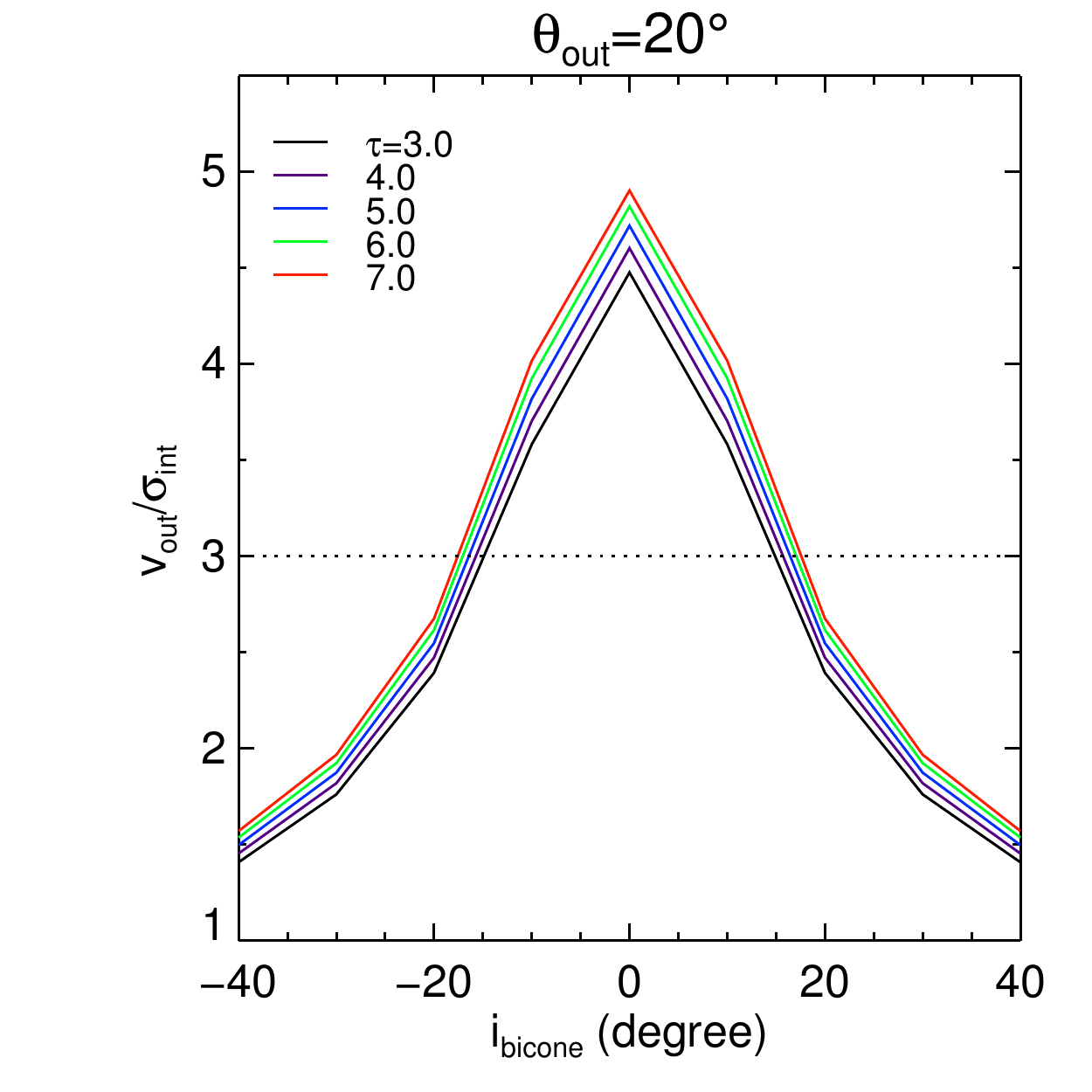}
\includegraphics[width=0.33\textwidth]{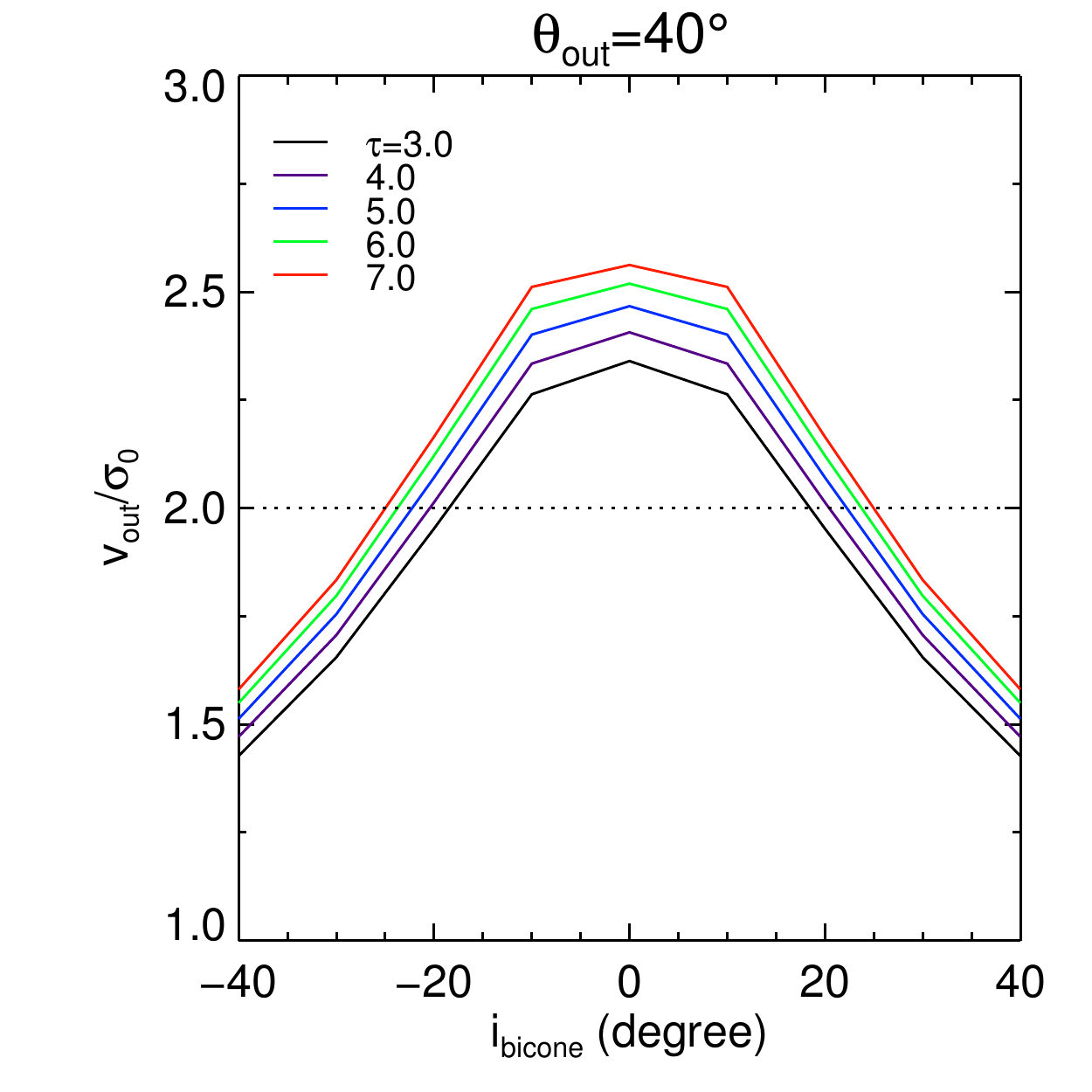}
\includegraphics[width=0.33\textwidth]{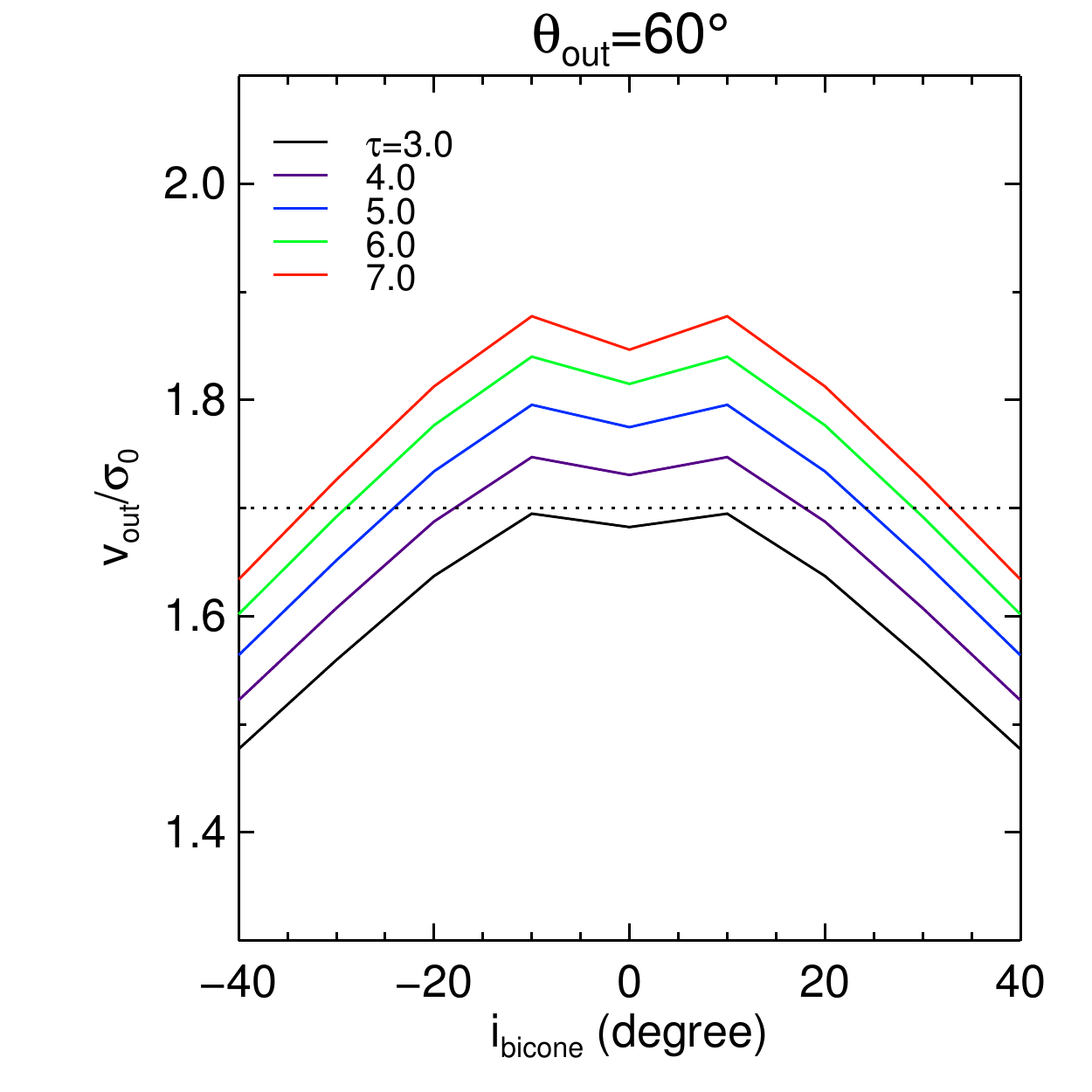}
\caption{Distributions of $v_{\text{out}}/\sigma_0$ as a function of bicone inclination with a linearly decreasing velocity profile for bicone outer- and inner half-opening angle = [40\degree, 20\degree] (left), [40\degree, 20\degree] (middle), and [60\degree, 30\degree] (right), respectively. Lines of different colors represent the different radial flux profiles of the bicone, i.e., $f(d) = f_{n}e^{-\tau d}$, where $f_{n}$ is the flux of the nucleus.}
\label{fig:f_test}
\end{figure*}

\section{Energetics of gas outflows}
\label{energetics}
Measuring the physical properties of gas outflows (mass, energy, and momentum) is highly uncertain due to the complex nature of the NLR. However, a proper measurement of those quantities is a crucial step towards understanding the co-evolution of galaxies and SMBHs. In the following, we describe the methods and results of estimates for the mass of ionized gas, energy, and momentum by using both spatially-resolved and integrated spectra. For simplicity, here we assume case B recombination and the biconical geometry for gas outflows \citep[e.g.,][]{2010ApJ...708..419C,2013ApJS..209....1F,2016ApJ...828...97B} 

\subsection{Spatially-resolved energetics}
\label{sr_energetics}
First, we calculate the energetics in each spaxel and examine the variation as a function of radial distance from the center. The ionized gas mass ($M_{\text{gas}}$) can be estimated as
\begin{eqnarray}
M_{\text{gas}} = (9.73 \times 10^8 M_{\odot} ) \times  L_{\text{H}\alpha,43}  \times n_{e,100} ^{-1},
\end{eqnarray}
where $L_{\text{H}\alpha,43}$ is the \Ha\ luminosity in units of $10^{43}$ \ergs, and $n_{e,100}$ is the electron density in units of 100 cm$^{-3}$ \citep{2006ApJ...650..693N}. We estimate $n_e$ for each spaxel by using the [\SII] line ratio \citep{2006agna.book.....O}. Assuming a gas temperature of $10^4$ K in the NLR, $n_e$ in the central spaxel ranges from 54 to 854 cm$^{-3}$, with the mean value of $n_e\sim$360$\pm$230 cm$^{-3}$, which is consistent with previous studies \citep[e.g.,][]{2006ApJ...650..693N,2016ApJ...833..171K}. We find that both $n_e$ and $L_{\text{H}\alpha}$ radially decrease with radial distance, thus $M_{\text{gas}}$ also decreases outwards. 

The kinetic energy of gas outflow ($E_{\text{out}}$) is the summation of bulk energy ($E_{\text{bulk}}$) and turbulence energy ($E_{\text{turb}}$) as
\begin{eqnarray}
E_{\text{out}} = E_{\text{bulk}} + E_{\text{turb}} = {1\over 2} M_{\text{gas}} ( v_{\text{gas}}^2 + \sigma_{\text{gas}} ^2 ), 
\end{eqnarray}
where $M_{\text{gas}}$ is the estimated ionized gas mass, and $v_{\text{gas}}$ and $\sigma_{\text{gas}}$ are the velocity offset and the velocity dispersion measured from the total [\OIII] profile, respectively. The momentum ($p$) can be estimated as $p_{\text{out}} = M_{\text{gas}} v_{\text{gas}}$. For the estimation, we only use the spaxels where the [\OIII] and [\SII]-doublet fluxes have S/N $>$ 3. Since the [\SII] line flux is in general weaker than the \Ha\ flux, the spatial extent of the estimation is mostly focused on the inner $\sim$2 kpc of the AGNs, which is close to the size of $R_{\text{out}}$ of the sample.

We find a general trend that the energy and momentum decrease $\sim$1 order of magnitude per $\sim$1 kpc increase in the radial distance in most of sample (Figure \ref{fig:em_dist_total}), Thus, we obtain distance-energy and distance-momentum relations by combining the energy and momentum per spaxel for the sample (Figure \ref{fig:em_dist_comb}). We obtain the error-weighted values of energy and momentum at each bin of 0.5 kpc distance. By assuming the uncertainty of distance as a half of spaxel ($\sim$0.3\arcsec), we also obtain the error-weighted values of distance at each bin. Then, we apply a forward regression method using the FITEXY code for the error-weighted values, obtaining:

\begin{eqnarray}
\log E_{\text{out}} = (54.44\pm 0.05) - (1.14\pm 0.03)D_{kpc}, 
\end{eqnarray}
\begin{eqnarray}
\log p_{\text{out}} = (46.92\pm 0.06) - (1.10\pm 0.03)D_{kpc}
\end{eqnarray}
where $D_{kpc}$ is the distance in units of kpc. The relationships show that both energy and momentum steeply decrease as a function of distance.    

\begin{table*}
\centering
\caption{Integrated physical properties of the type 2 AGNs within $R_{\text{NLR}}$ \label{tbl-4}}
\begin{tabular}{ccccccccccccccc}
\tableline
\tableline
SDSS name & log $L_{\text{[OIII]}}$ & log $L_{\text{bol}}$ & $\dot{M}_{\text{acc}}$ & $v_{\text{[OIII]}} $ & $\sigma_{\text{[OIII]}}$ & $\sigma_{0}$ & $v_{\text{out}}$ & log $L_{\text{H}\alpha} $ & $n_e$ & $M_{\text{gas}}$ & log $E_{\text{kin}}$ & $\dot{M}_{\text{out}}$ &  log $\dot{E}_{\text{out}}$ & log $\dot{p}_{\text{out}}$  \\
(1) & (2) & (3) & (4) & (5) & (6) & (7) & (8) & (9) & (10) & (11) & (12) & (13) & (14) & (15)\\
\tableline
J0130+1312    & 40.34 & 43.88 & 1.35 &  -175 & 193 & 260 & 520 & 40.54 & 370       & 9.2 & 54.40 & 1.5 & 41.12 & 33.70\\
J0341$-$0719 & 40.39 & 43.93 & 1.50 & +144 & 246 & 285 & 570 & 40.58 &122$^a$& 30.3 & 54.99 & 6.6 & 41.83 & 34.38\\
J0806+1906    & 40.69 & 44.23 & 3.03 & -86 & 187 & 205 & 411 & 40.72 & 344          & 14.8 & 54.39 & 1.1 & 40.78 & 33.47 \\
J0855+0047   & 40.26 & 43.80 & 1.13 & -230  & 285 & 366 & 732 & 40.41 & 171        & 14.6 & 54.89 & 2.3 & 41.59 & 34.02 \\
J0857+0633    & 40.11 & 43.66 & 0.80 & +122 & 177 & 215 & 430 & 40.24 & 129       & 13.1 & 54.38 & 1.2 & 40.86 & 33.53 \\
J0911+1333    & 40.77 & 44.31 & 3.61 & -132 & 429 & 449 & 898 & 41.19 & 583        & 26.2 & 55.32 & 7.3 & 42.27 & 34.62\\
J0952+1937    & 39.47 & 43.01 & 0.18 & -63 & 345 & 351 & 702 & 40.49 & 78            & 38.5 & 55.28 & 17.7 & 42.44 & 34.89\\
J1001+0954    & 40.33 & 43.87 & 1.32 & -25 & 548 & 548 & 1096 & 40.48 & 399          & 7.3 & 54.94 & 3.0 & 42.06 & 34.32 \\
J1013$-$0054 & 40.11 & 43.65 & 0.80 & +12 & 300 & 300 & 600 & 40.14 & 236         & 5.7 & 54.31 & 1.7 & 41.29 & 33.81\\
J1054+1646    & 40.72 & 44.27 & 3.27 & -282 & 377 & 471 & 942 & 40.64 & 156        & 27.6 & 55.39 & 9.5 & 42.42 & 34.75 \\
J1100+1321    & 40.56 & 44.10 & 2.24 & -137 & 470 & 490 & 980 & 40.32 & 237        & 8.6 & 54.91 & 3.4 & 42.01 & 34.32 \\
J1100+1124   & 39.53 & 43.07 & 0.21 & +158 & 254 &  299 & 598 & 39.95& 341$^b$& 2.1 & 53.88  & 0.3 & 40.53 & 33.06 \\
J1106+0633   & 40.46 & 44.00 & 1.78 & -176 & 234 & 294 & 587 & 40.82 & 520         & 12.3 & 54.63  & 2.5 & 41.43 & 33.96 \\
J1147+0752    & 40.65 & 44.19 & 2.73 & -358 & 437 & 564 & 1129 & 40.64 & 327        & 12.8 & 55.21 & 7.6 & 42.48 & 34.73 \\ 
J1214$-$0329 & 40.30 & 43.84 & 1.23 & -281 & 331 & 434 & 869 & 40.70 & 761        & 6.4 & 54.69 & 4.9 & 42.06 & 34.43 \\
J1310+0837    & 40.42 & 43.97 & 1.63 & +172 & 250 & 304 & 608 & 40.09 & 321       & 3.7 & 54.13 & 1.0 & 41.05 & 33.56 \\ 
J1440+0556    & 40.04 & 43.58 & 0.67 & +42 & 480 & 482 & 964 & 39.96 & 854         & 1.0 & 53.98 & 0.6 & 41.24 & 33.56 \\
J1448+1055    & 40.94 & 44.48 & 5.37 & -293 & 576 & 646 & 1292 & 40.74 & 620      & 8.7 & 55.16 & 3.7 & 42.29 & 34.48 \\
J1520+0757    & 39.16 & 42.70 & 0.09 & -486 & 430 & 649 & 1297 & 39.77 & 54        & 10.6 & 55.25 & 9.2 & 42.69 & 34.88 \\
J2039+0033    & 40.55 & 44.10 & 2.20 & -71 & 468 & 474 & 947 & 40.71 & 251          & 19.8 & 55.25 & 7.8  & 42.35 & 34.67\\
\tableline
\end{tabular}
\tablecomments{(1) SDSS name of AGN; (2) Extingtion-uncorrected [\OIII] luminosity in logaritm (\ergs); (3) AGN bolometric luminosity in logarithm as $L_{\text{bol}} = 3500 \times L_{[\text{O III}]}$ \citep[\ergs,][]{Heckman:2004js}; (4) Mass accretion rate as $\dot{M}_{\text{acc}} = L_{\text{bol}}/\mu c^2$ ($10^{-2}M_{\odot}$ yr$^{-1}$); (5) [\OIII] velocity offset with respect to the systemic velocity of the stellar component (\kms); (6) [\OIII] velocity dispersion (\kms); (7) dust-corrected [\OIII] velocity dispersion (\kms); (8) bulk velocity of the outflows (\kms) (9) \Ha\ luminosity in logarithm (\ergs); (10) electron density estimated from the [\SII] line ratio (cm$^{-3}$); (11) ionized gas mass in units of 10$^5$ $M_{\odot}$ (see Equation 5); (12) kinetic energy in logarithm (erg); (13) mass outflow rate ($M_{\odot}$ yr$^{-1}$); (14) energy injection rate (\ergs); (15) momentum flux (dyne). \\
$^a$The estimated electron density from the [\SII] ratio is saturated to the lowest density allowed in the calculation (1 cm$^{-3}$). Hence we adopt the electron density from SDSS.\\
$^b$The [\SII]-doublet is affected by a telluric absorption band, so we adopt the electron density from SDSS.}
\end{table*}

\begin{table*}
\centering
\caption{Integrated physical properties of the type 2 AGNs within $R_{\text{out}}$ \label{tbl-5}}
\begin{tabular}{ccccccccccccccc}
\tableline
\tableline
SDSS name & log $L_{\text{[OIII]}}$ & log $L_{\text{bol}}$ & $\dot{M}_{\text{acc}}$ & $v_{\text{[OIII]}} $ & $\sigma_{\text{[OIII]}}$ & $\sigma_{0}$ & $v_{\text{out}}$ & log $L_{\text{H}\alpha} $ & $n_e$ & $M_{\text{gas}}$ & log $E_{\text{kin}}$ & $\dot{M}_{\text{out}}$ &  log $\dot{E}_{\text{out}}$ & log $\dot{p}_{\text{out}}$  \\
(1) & (2) & (3) & (4) & (5) & (6) & (7) & (8) & (9) & (10) & (11) & (12) & (13) & (14) & (15)\\
\tableline
J0130+1312    & 38.82 & 42.36 & 0.04 &  -106 & 200 & 226 & 453 & 38.90 & 57       & 1.4   & 53.45 & 0.8 & 40.72 & 33.36\\
J0341$-$0719 & 40.53 & 44.07 & 2.09 &+126 & 239 & 270 & 541 & 40.82 &--         & --      & -- & -- & -- & --\\
J0806+1906    & 39.80 & 43.34 & 0.39 & -38 & 231 & 234 & 469 & 39.92 & 416          & 1.9 & 53.63 & 0.7 & 40.68 & 33.31 \\
J0855+0047   & 40.42 & 43.96 & 1.63 &-132  & 277 & 307 & 615 & 41.02 & 93        & 108.2 & 55.61 & 3.6 & 41.63 & 34.15 \\
J0857+0633    & 39.73 & 43.27 & 0.33 & +134 & 222 & 259 & 518 & 39.85 & --       & --       & -- & -- & -- & --\\
J0911+1333    & 40.99 & 44.54 & 6.10 & -98 & 355 & 369 & 737 & 41.42 & 516        & 49.3 & 55.43 & 3.7 & 41.80 & 34.23\\
J0952+1937    & 39.64 & 43.18 & 0.27 & -70 & 180 & 193 & 386 & 40.82 & --            & --      & -- & -- & -- & --\\
J1001+0954    & 40.63 & 44.17 & 2.62 & -57 & 489 & 492 & 984 & 40.88 & 536          & 13.7 & 55.12 & 1.9 & 41.77 & 34.08 \\
J1013$-$0054 & 40.14 & 43.69 & 0.86 & -5 & 293 & 293 & 587 & 40.19 & 187         & 8.1 & 54.44 & 2.2 & 41.38 & 33.92 \\
J1054+1646    & 41.00 & 44.54 & 6.18 & -260 & 357 & 442 & 884 & 40.95 & 151        & 57.2 & 55.65 & 5.7 & 42.15 & 34.50 \\
J1100+1321    & 40.82 & 44.37 & 4.09 & -126 & 432 & 450 & 901 & 40.70 & 122        & 39.7 & 55.51 & 5.7 & 42.17 & 34.51 \\
J1100+1124   & 39.64 & 43.19 & 0.27 & +140 & 223 &  264 & 528 & 40.12& --           & --      & --  & -- & -- & -- \\
J1106+0633   & 40.64 & 44.19 & 271 & -129 & 237 & 270 & 540 & 41.14 & 250         & 53.3 & 55.19  & 1.6 & 41.18 & 33.74 \\
J1147+0752    & 41.17 & 44.71 & 9.07 & -314 & 373 & 488 & 976 & 41.31 & 428        & 46.8 & 55.65 & 6.6 & 42.30 & 34.61 \\ 
J1214$-$0329 & 40.49 & 44.03 & 1.89 & -231 & 314 & 390 & 779 & 40.92 & 686        & 11.8 & 54.85 & 4.5 & 41.93 & 34.34 \\
J1310+0837    & 40.25 & 43.79 & 1.09 & +165 & 236 & 288 & 576 & 39.94 & 289       & 2.9 & 53.99 & 1.2 & 41.09 & 33.63 \\ 
J1440+0556    & 40.25 & 43.79 & 1.09 & +28 & 451 & 452 & 905 & 40.23 & 803         & 2.1 & 54.23 & 0.5 & 41.15 & 33.49 \\
J1448+1055    & 41.22 & 44.76 & 10.22 & -254 & 561 & 616 & 1232 & 41.10 & 729      & 16.7 & 55.40 & 2.2 & 42.02 & 34.23 \\
J1520+0757    & 39.49 & 43.03 & 0.19 & -505 & 466 & 687 & 1374 & 40.18 & --        & --    & -- & -- & -- & -- \\
J2039+0033    & 40.53 & 44.07 & 2.08 &-59 & 408 & 413 & 825 & 40.70 & 255          & 19.3 & 55.12 & 6.5  & 42.15 & 34.53\\
\tableline
\end{tabular}
\tablecomments{((1) SDSS name of AGN; (2) Extingtion-uncorrected [\OIII] luminosity in logaritm (\ergs); (3) AGN bolometric luminosity in logarithm as $L_{\text{bol}} = 3500 \times L_{[\text{O III}]}$ \citep[\ergs,][]{Heckman:2004js}; (4) Mass accretion rate as $\dot{M}_{\text{acc}} = L_{\text{bol}}/\mu c^2$ ($10^{-2}M_{\odot}$ yr$^{-1}$); (5) [\OIII] velocity offset with respect to the systemic velocity of the stellar component (\kms); (6) [\OIII] velocity dispersion (\kms); (7) dust-corrected [\OIII] velocity dispersion (\kms); (8) bulk velocity of the outflows (\kms) (9) \Ha\ luminosity in logarithm (\ergs); (10) electron density estimated from the [\SII] line ratio (cm$^{-3}$); (11) ionized gas mass in units of 10$^5$ $M_{\odot}$ (see Equation 5); (12) kinetic energy in logarithm (erg); (13) mass outflow rate ($M_{\odot}$ yr$^{-1}$); (14) energy injection rate (\ergs); (15) momentum flux (dyne). }
\end{table*}

\begin{figure*}
\centering
\includegraphics[width=0.33\textwidth]{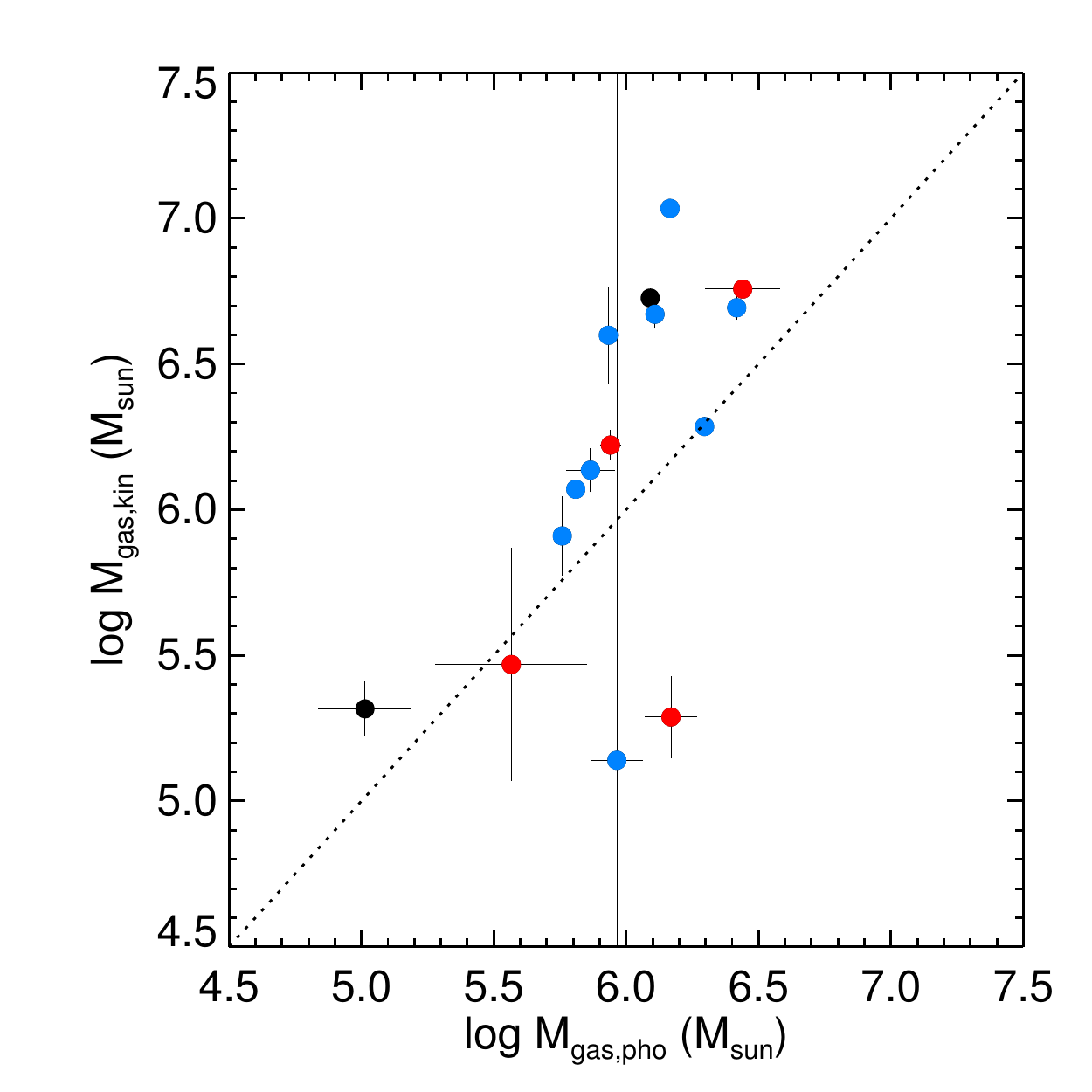}
\includegraphics[width=0.33\textwidth]{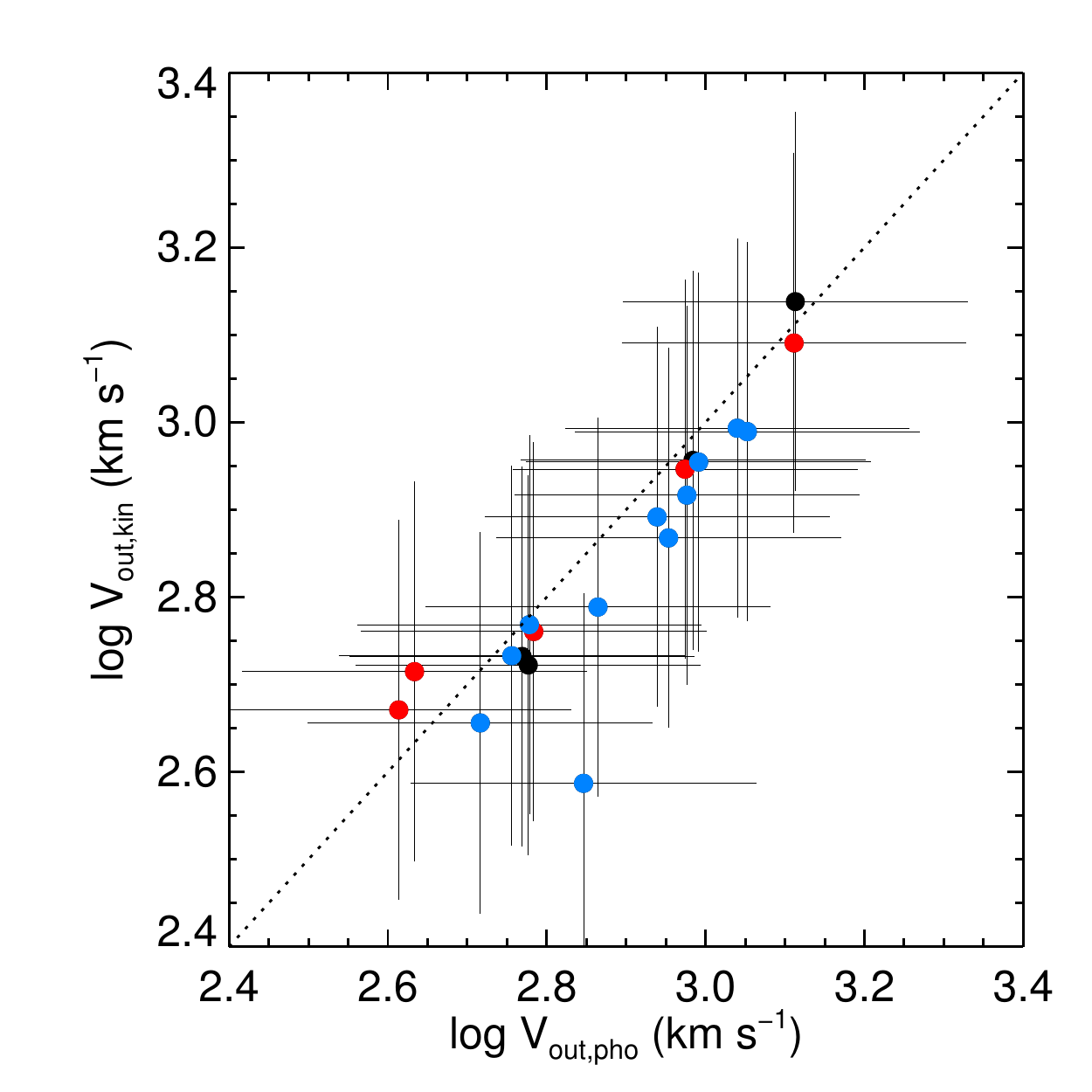}
\includegraphics[width=0.33\textwidth]{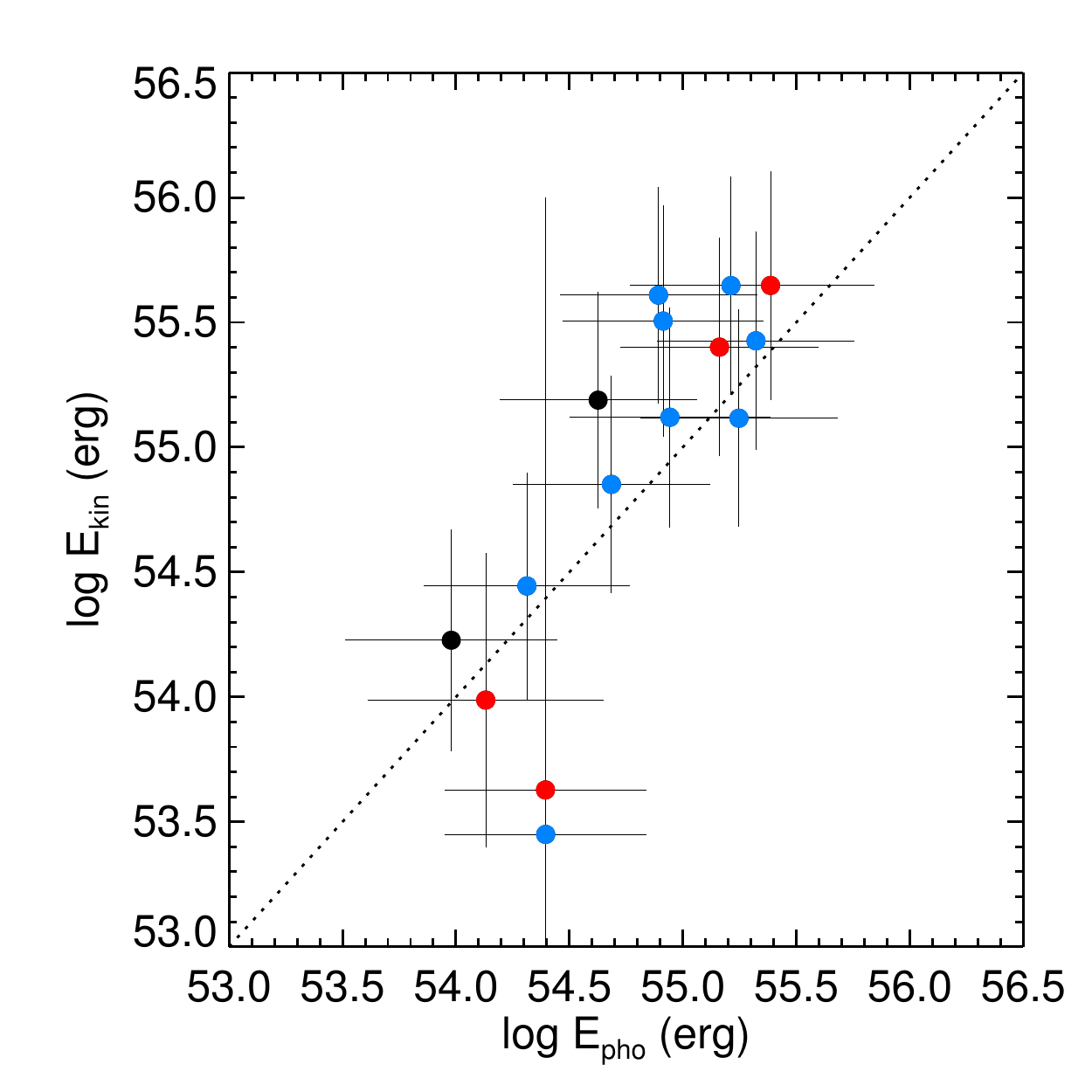}
\includegraphics[width=0.33\textwidth]{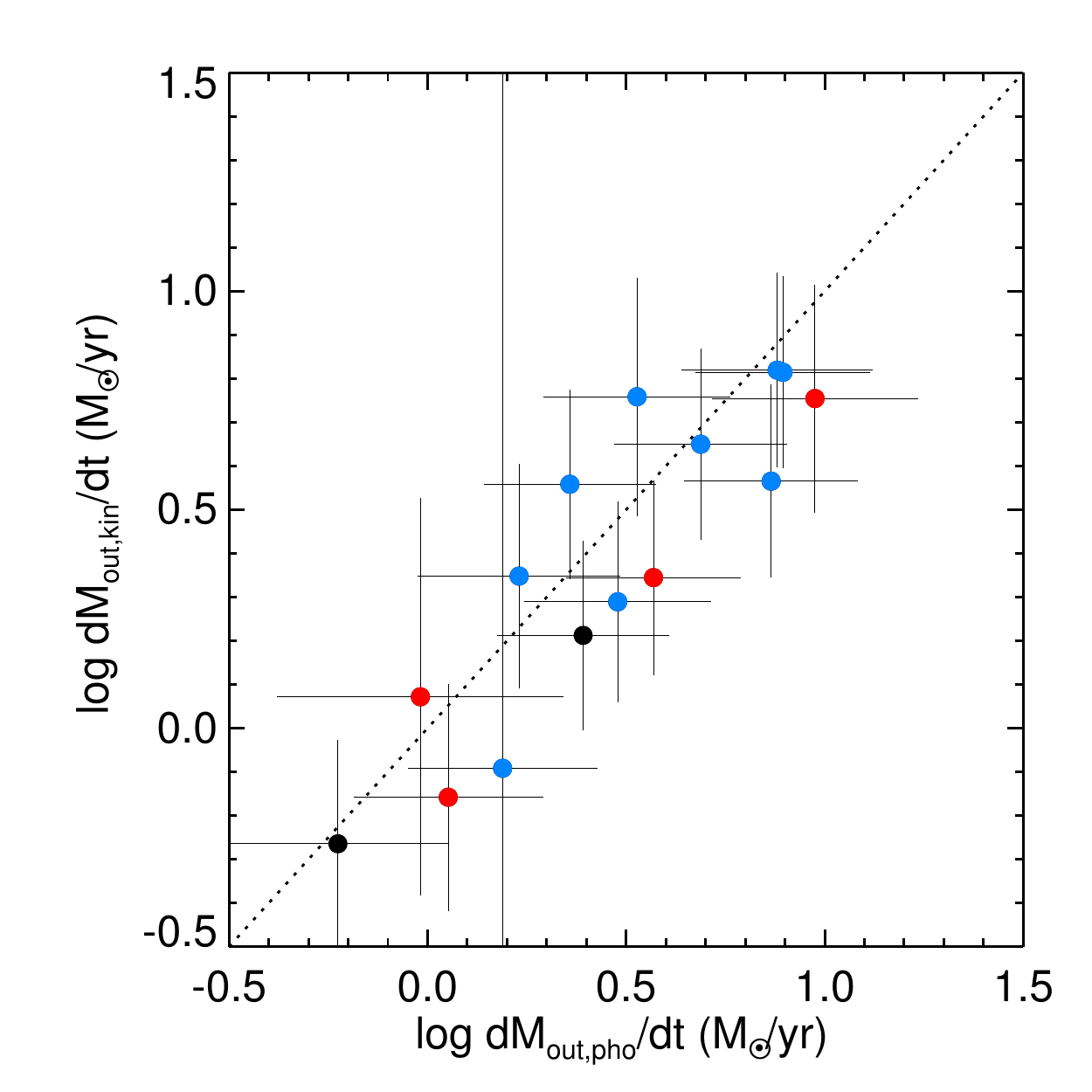}
\includegraphics[width=0.33\textwidth]{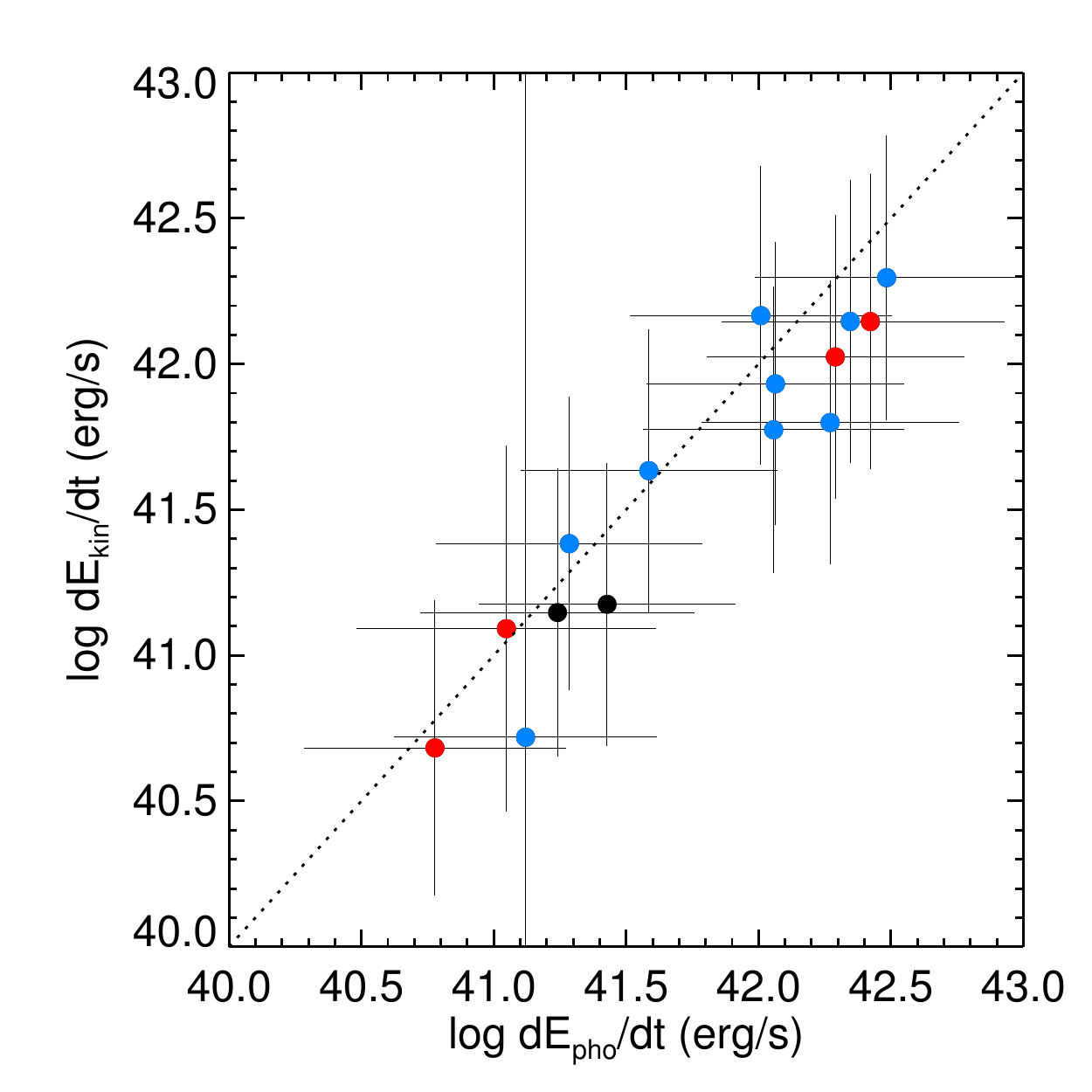}
\includegraphics[width=0.33\textwidth]{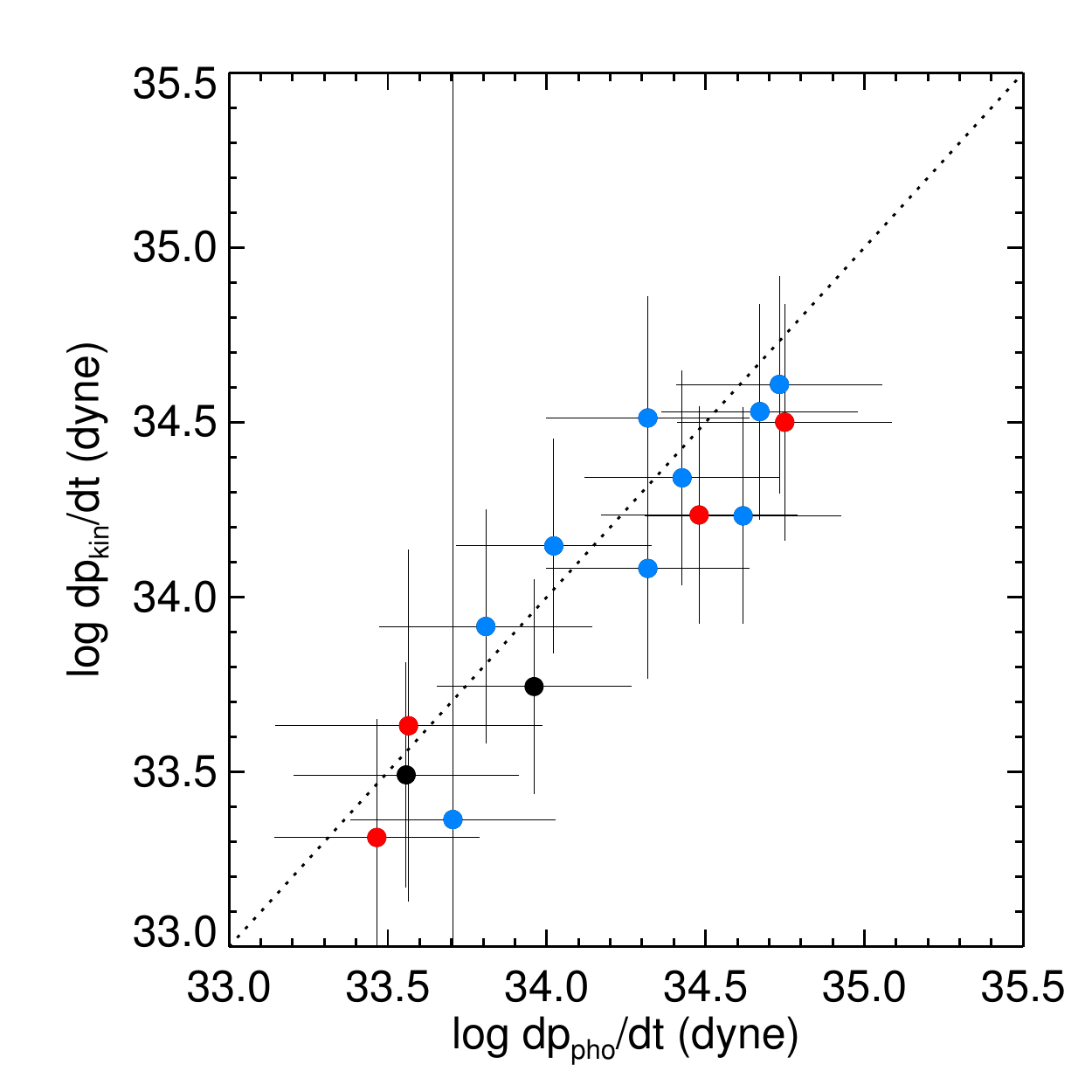}
\caption{Comparisons of the integrated energetics when using two different sizes, i.e., flux-weighted NLR size (x-axis) and kinematics-based outflow size (y-axis): ionized gas mass (upper-left), intrinsic outflow velocity (upper-middle), total energy (upper-right), mass outflow rate (lower-left), energy injection rate (lower-middle), and momentum flux (lower-right). Blue, red, and black dots represent the different types of AGNs with SF-type disk, AGN-type disk, and no/ambiguous rotation, respectively. Error bars represent the 1$\sigma$ uncertainties of the estimation. }
\label{fig:e_comparison}
\end{figure*}

\begin{figure}
\centering
\includegraphics[width=0.45\textwidth]{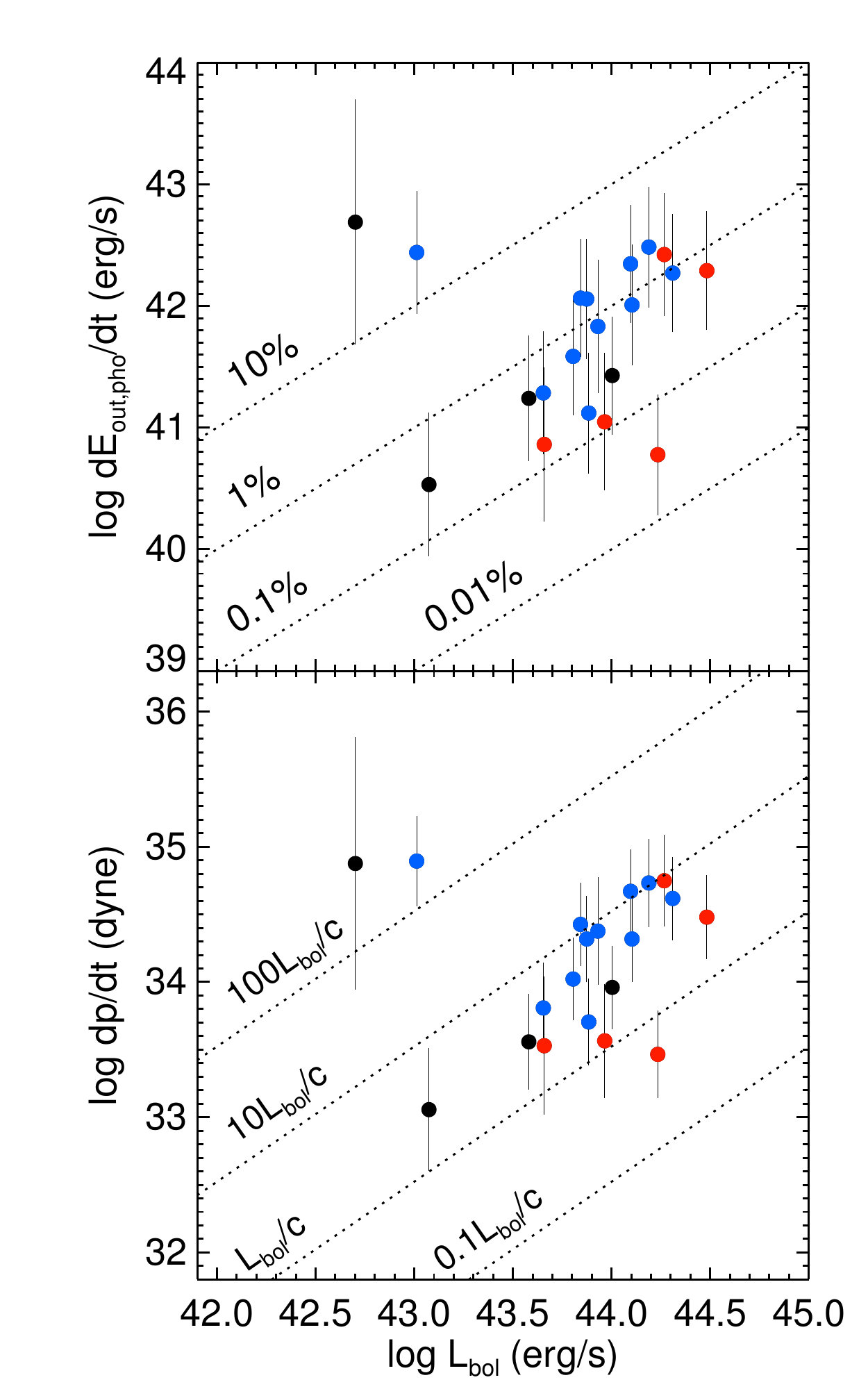}
\caption{The energy injection rate (upper panel) and the momentum flux (lower panel) as a function of AGN bolometric luminosity for 20 AGNs. Blue, red, and black dots represent the different types of AGNs with SF-type disk, AGN-type disk, and no/ambiguous rotation, respectively. Error bars represent the 1$\sigma$ uncertainties of the estimation.}
\label{fig:lbol_vs_em}
\end{figure}

\subsection{Integrated energetics: impact on the nuclear star-formation}
\label{int_energetics}
As we determined two different sizes based on the flux distribution ($R_{\text{NLR}}$) and kinematics of the NLR ($R_{\text{out}}$), here we obtain the total energetics of the outflows (i.e., mass outflow rate, energy injection rate, and momentum flux) for the two different sizes and compare the results. We also compare the energetics of the groups of AGNs with different disk properties (Section \ref{disk}) in order to investigate whether the nuclear star-formation is affected by AGN outflows. 

\subsubsection{Energetics for two different sizes}
To estimate the total energetics of AGN outflows for two different sizes, we construct an integrated spectrum for both $R_{\text{NLR}}$ and $R_{\text{out}}$ of each AGN, then we calculate $M_{\text{gas}}$ and $n_e$ in the same manner as for the spatially resolved case. We also calculate the [\OIII] velocity offset and velocity dispersion from the integrated spectra. As we estimate the integrated energetics using two different sizes, we can directly compare the values and examine the effect of different sizes estimates on the energetics calculation.

If we assume biconical outflows with radius $r$ and the ionized gas mass $M_{gas}$ within $r$, the mass outflow rate $\dot{M}_{\text{out}}$ can be calculated as 
\begin{eqnarray}
\dot{M}_{\text{out}} = {M_{\text{gas}} \over {\tau_{\text{dyn}}}} = { 3M_{\text{gas}} v_{\text{out}}   \over  r},
\end{eqnarray}
where $\tau_{\text{dyn}}$ is the dynamical time for the ionized gas with $v_{\text{out}}$ to reach a distance $r$, and $v_{\text{out}}$ is the flux-weighted intrinsic outflow velocity, or bulk velocity of the outflows \citep{2012MNRAS.425L..66M}. Since we use the \Ha\ luminosity from the total profile of \Ha\ for our $M_{\text{gas}}$ estimation (Equation 5), the results can be regarded as an upper limit of $M_{\text{gas}}$. 

Then the energy injection rate $\dot{E}_{\text{out}}$ and the momentum flux $\dot{p}_{\text{out}}$ can be calculated as follows, respectively,
\begin{eqnarray}
\dot{E}_{\text{out}} = {1\over 2} \dot{M}_{\text{out}} v_{\text{out}} ^2,
\end{eqnarray}
\begin{eqnarray}
\dot{p}_{\text{out}} = \dot{M}_{\text{gas}}  v_{\text{out}}.
\end{eqnarray}

\subsubsection{Estimation of the intrinsic outflow velocity}
To calculate the energy injection rate and the momentum flux (Equations 9--11), we need to properly estimate the intrinsic outflow velocity $v_{\text{out}}$ from the observations. We measured $\sigma_\text{[O III]}$ and $v_\text{[O III]}$, which are closely related to AGN outflows. However, these values are far from the intrinsic $v_{\text{out}}$, because the measured values were affected by dust extinction and projection effects, which can be more severe in type 2s than type 1s \citep[e.g.,][]{2016ApJ...828...97B}. Hence, we need to assume proper outflow models to estimate $v_{\text{out}}$, which is difficult to estimate from observations.

For example, \citet{2013MNRAS.436.2576L} assumed spherical symmetric or wide-angle biconical outflow models, resulting in the relationship of $v_{\text{out}}=\sim$1.9$\times \sigma_{\text{[O III]}}$. This number may not be directly applicable to this study, however, since our sample consists less luminous than theirs, which may indicate a different outflow geometry \citep{2016ApJ...828...97B}. Our sample are also type 2 AGNs, which can be more affected by projection effects. To obtain the relationship of $v_{\text{out}}$ and $\sigma_{\text{[O III]}}$ for our sample, hence, we exploit 3D biconical models suggested by \citet{2016ApJ...828...97B}, which successfully reproduced the distributions of the [\OIII] kinematics of type 2 AGNs in SDSS \citep{2016ApJ...817..108W}. 

In our models, we assume the exponentially decreasing radial flux profile $f(d) = A f_{n}e^{-\tau d}$, where $f_{n}$ is the flux of the nucleus, and $A$ is the amount of dust extinction (0 < A < 1) for each position in 3D. We adopt the range of $\tau$=3--7 and the size of the bicone as unity. We also assume different velocity profiles $v(d)$ (e.g., linear decrease or increase). Then, the intrinsic, flux-weighted $v_{\text{out}}$ can be calculated as 
\begin{eqnarray}
v_{\text{out}} = \frac{\int f(d)v(d) dp}{\int f(d)dp},
\end{eqnarray}
where $p$ represents each position in 3D.
 The bicone has outer- and inner half-opening angles, and we use three different outer half-opening angles as 20\degree, 40\degree, and 60\degree. The inner half-opening angles are fixed as an half of the outer half-opening angle \citep{2016ApJ...828...97B}. In calculations, we use the value of $\sigma_{0}$=$(v_{\text{[O III]}}^2+\sigma_{\text{[O III]}}^2)^{0.5}$ rather than $\sigma_{\text{[O III]}}$, because $\sigma_{0}$ can be a good proxy of the line width without dust extinction \citep{2016ApJ...828...97B}. 
 
By integrating the 3D bicone models, we calculate the intrinsic $v_{\text{out}}$ and $\sigma_0$ as a function of bicone inclination (Figure \ref{fig:f_test}). While $v_{\text{out}}$ is a fixed value, $\sigma_0$ is minimized when the bicone axis is parallel to the plane of the sky ($i_{\text{bicone}}$=0\degree) due to the projection effects. On the other hand, if the bicone is inclined to $\pm$40\degree, which is about the maximum inclination for type 2 AGNs \citep{2014MNRAS.441..551M}, $\sigma_0$ become larger due to lower projection effects. As a result, $v_{\text{out}}/\sigma_0$ has a large range from $\sim$1.5 to $\sim$2.5 if the outer half-opening angle of the bicone is 40\degree (middle panel). 

For the IFU sample, we adopt a bicone half-opening angle of 40\degree, which is consistent with the mean value of half-opening angle estimated for 17 Seyfert galaxies based on HST/STIS data \citep{2014ApJ...785...25F}. Hence, we adopt the relationship of $v_{\text{out}}$=(2.0$\pm$0.5)$\times$$\sigma_{0}$ in the energetics calculation for our sample. Alternatively, if the outer half-opening angle of the bicone is 20\degree (left panel) or 60\degree (right panel), $v_{\text{out}}/\sigma_0$ ranges from $\sim$1.5 to $\sim$4.5 and from $\sim$1.5 to $\sim$1.9, respectively. Also, if we use different velocity profiles, the range of $v_{\text{out}}/\sigma_0$ becomes smaller than in the case of linear decrease.

\subsubsection{Estimated energetics}
First, we compare the ionized gas mass and outflow energetics obtained by using two different sizes (Figure \ref{fig:e_comparison}). The estimated physical quantities within $R_{\text{NLR}}$ and $R_{\text{out}}$ are listed in Tables \ref{tbl-4} and \ref{tbl-5}, respectively. Note that we could not measure the electron density for two AGNs (J0341$-$0719 and J1100+1124) when using $R_{\text{NLR}}$, and five AGNs (J0341$-$0719, J0857+0047, J0952+1937, J1100+1124, and J1520+0757) when using $R_{\text{out}}$. It is because the estimated electron densities based on [\SII]-line ratios are saturated to the lowest density (1 cm$^{-3}$) (J0341$-$0719, J0857+0047, J0952+1937, and J1520+0757), or the [\SII]-doublet is affected by the telluric absorption band (J1100+1124).

We find that the ionized gas mass using $R_{\text{out}}$ is $\sim$2.3$\pm$1.9 times larger than the mass using $R_{\text{NLR}}$ (top-left panel), because $R_{\text{out}}$ is a factor of $\sim$2 larger than $R_{\text{NLR}}$ (Section \ref{size}). However, the estimated energetics, i.e., outflow velocity (top-middle), total energy (top-right), mass outflow rate (bottom-left), energy injection rate (bottom-middle), and momentum flux (bottom-right), are overall consistent with each other within uncertainties. The reason for this consistency is mainly due to relatively large uncertainties for the estimated quantities, e.g., outflow velocity and electron density. Also, the estimated outflow velocity does not change much when we increase the size since we estimate the velocity using the flux-weighted kinematics of [\OIII], which is exponentially decreasing from the nucleus. Hence, we will use $R_{NLR}$-based physical properties in the following analysis of the energetics of outflows. Note that we use the electron density from SDSS for two AGNs lacking this information from our data (J0341$-$0719 and J1100+1124).

The estimated $M_{\text{gas}}$ is (1.0--38.5)$\times 10^5 M_{\odot}$ and the mass outflow rate $\dot{M}_{\text{out}}$ is 0.3--17.7 $M_\odot$ yr$^{-1}$. The mass accretion rate $\dot{M}_{\text{acc}}$ can be calculated as $\dot{M}_{\text{acc}}$ = $L_{\text{bol}} / \eta c^2$, where the $\eta$ is the accretion efficiency typically assumed to be 0.1, and $L_{\text{bol}}$ is the AGNs bolometric luminosity estimated as $L_{\text{bol}}$ = 3500$\times$$L_{[\text{O III}]}$, where $L_{[\text{O III}]}$ is the extinction-uncorrected [\OIII] luminosity \citep{Heckman:2004js}. The mean $\dot{M}_{\text{out}}$ $\sim$4.6$\pm$4.3 $M_\odot$ yr$^{-1}$ is consistent with the values of nearby AGNs \citep[0.1--10 $M_\odot$,][]{2005ARA&A..43..769V}. The estimated mean mass outflow rate is about a factor of $\sim$260 higher than the estimated mean $\dot{M}_{\text{acc}} \sim$0.02$\pm$0.01 $M_\odot$ yr$^{-1}$, indicating powerful mass loading of the AGN outflows by the ISM \citep[e.g.,][]{2005ARA&A..43..769V,2009MNRAS.396....2B}.  

We also compare the energy injection rate and the momentum flux as a function of AGN bolometric luminosity (Figure \ref{fig:lbol_vs_em}). The estimated $\dot{E}_{\text{out}}$ is 10$^{40.5-42.7}$ \ergs\ and the estimated $\dot{p}_{\text{out}}$ is 10$^{33.1-34.9}$ dyne. The majority of AGNs (18 out of 20) have relatively low energy injection rate, about 0.8$\pm$0.6\% of $L_{\text{bol}}$, and also have relatively low momentum flux, about (5.4$\pm$3.6)$\times$$L_{\text{bol}}/c$. Both the energy injection rate and the momentum flux for the AGNs are in general increasing as a function of $L_{\text{bol}}$. Interestingly, we find two AGNs (J0952+1937 and J1520+0757) with much higher energetics than the majority of AGNs in our sample. Although these two AGNs have relatively low bolometric luminosities ($L_{\text{bol}}\sim 10^{42.7-43.0}$\ergs), their energy injection rate is 27--97\% of $L_{\text{bol}}$ and the momentum flux is 228--449 $L_{\text{bol}}/c$.

\section{Discussion}
\label{discussion}

\subsection{a mixture of the NLR kinematcis: gravitational vs. non-gravitational}
\label{mixture}
In Section \ref{oiii} and \ref{ha}, we investigate various properties, e.g., flux, velocity, and velocity dispersion, of [\OIII] and \Ha, respectively, in 1D and 2D, finding that the NLR is a mixture of non-gravitational, i.e., AGN outflows, and gravitational kinematic, i.e., rotation or random motion \citep[see also ][]{2016ApJ...819..148K}. 
The non-gravitational kinematics are commonly detected in the broad component of both [\OIII] and \Ha, which are revealed by their large velocity dispersion compared to the stellar velocity dispersion and/or their large velocity offset with respect to the systemic velocity of the host galaxy. While the broad components of [\OIII] and \Ha\ show qualitatively similar non-gravitational kinematics, the absolute value of velocity and velocity dispersion of \Ha\ are relatively smaller than those of [\OIII].

We find a clear sign of gravitational kinematics (i.e., Keplerian rotation) in the narrow component of \Ha\ for most objects (16 out of 20 AGN). The other four AGNs, which do not show any rotation in the narrow component of \Ha, may also have gravitational kinematics such as a random motion. In comparison, only 8 out of 20 AGNs show the gravitational kinematics (i.e., rotation or random motion) in the narrow component of [\OIII], implying that the gravitational kinematics are more significantly presented in \Ha\ than [\OIII]. 
Consider a simple picture that biconical outflows and a star-forming disk co-exist in the nucleus. Then, the star-forming disk is responsible for the gravitational kinematics, while the biconical outflows are responsible for the outflow kinematics in the NLR. In general, the star-forming region emits the stronger Balmer emission than the [\OIII] emission, so the rotational feature is more distinguishable in \Ha\ than in [\OIII]. For the same reason, the outflowing region, which is ionized by AGN, emits stronger [\OIII] than Balmer lines. Thus, the outflow kinematics are more noticeable in [\OIII] than in \Ha.

In Section \ref{disk}, we further focus on 
the origin of rotational features 
the relation of the narrow component of \Ha\ with star formation. We find that 11 AGNs with SF-type disks show lower $D_{n}(4000)$ and higher $H_{\delta}$ indices than the AGNs with AGN-type disks, showing that the AGNs with SF-type disks have on-going star formation, while five AGNs with AGN-type disks do not. The results further support that, even in the AGNs with strong outflows, the star-forming disk and AGN outflows co-exist in the nucleus ($\sim$central kpc), as several studies pointed out \citep[e.g.,][Woo et al. 2017]{Davies:2016ig}

\begin{figure}
\centering
\includegraphics[width=0.5\textwidth]{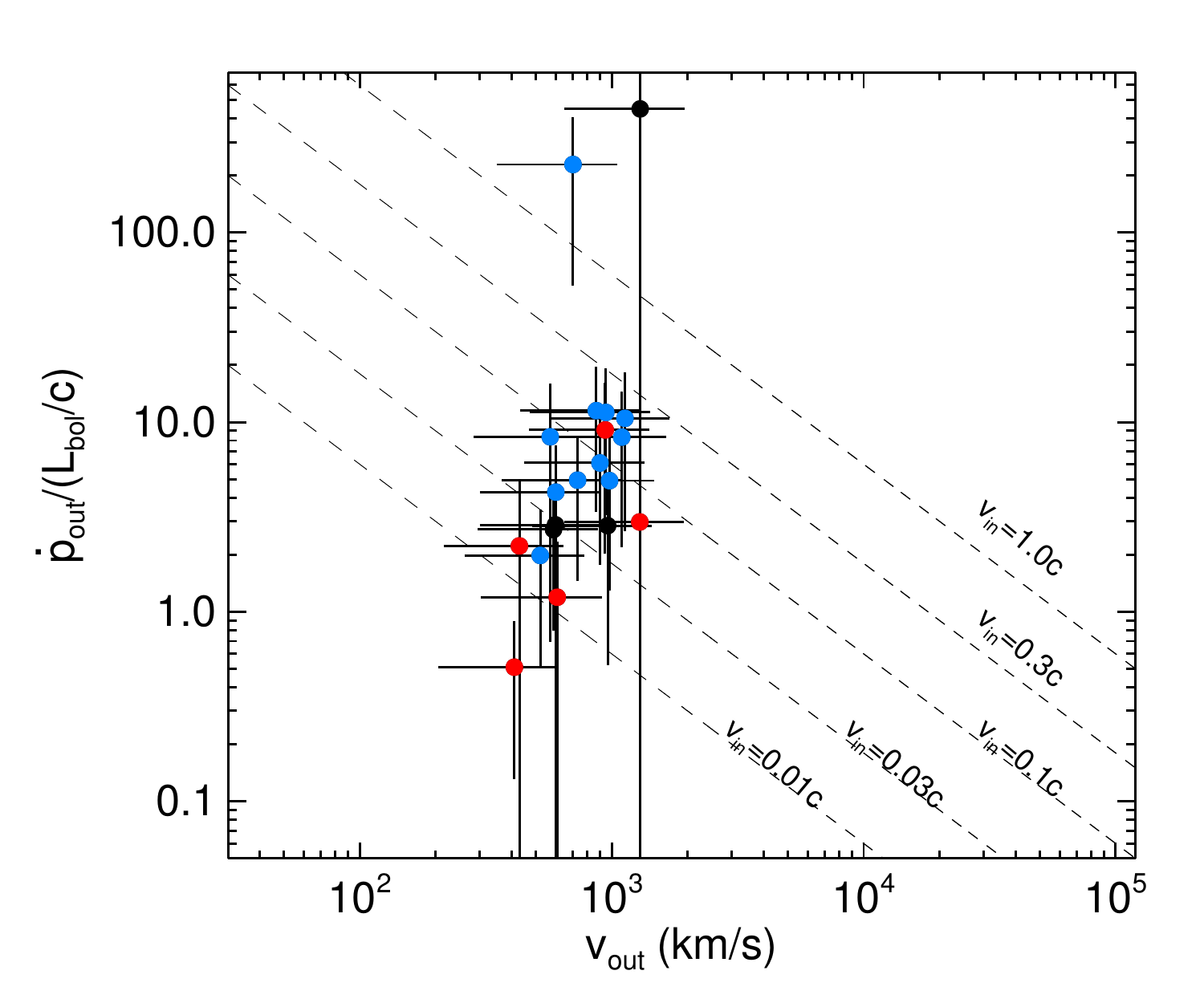}
\caption{The momentum flux of gas outflows $\dot{p}_{\text{gas}}$ normalized to $L_{\text{bol}}/c$ as a function of the gas outflow velocity $v_{\text{gas}}$. Blue, red, and black dots represent the different types of AGNs with SF-type disk, AGN-type disk, and no/ambiguous rotation, respectively. Error bars represent the 1$\sigma$ uncertainties of the estimation. Dashed lines indicate the initial velocity $v_{\text{in}}$ of small-scale wind as $v_{\text{in}}$ = (0.01, 0.03, 0.1, 0.3, 1.0)$\times$c respectively from bottom-left to top-right, where c is the speed of light.}
\label{fig:momentum}
\end{figure}

\subsection{Driving mechanism of the outflows}
\label{origin}
While it is clear that the ionized gas outflows of our sample are mainly due to AGN activity (see Section \ref{int_energetics}), we further investigate the driving mechanism of the outflows. Theoretical studies expect that the outflows is energy conserving if the outflowing gas expand as a hot bubble without efficient cooling, while the outflow is momentum conserving if the outflows rapidly cool down and lose their energy \citep[see][for a review]{2015ARA&A..53..115K}. AGN outflows would generate a different impact on the host galaxy depending whether the outflows are energy- or momentum conserving. 

Since the driving mechanism depends on the properties of the accretion wind and the ISM of the host galaxy, it is still under debate which phase is dominant when and at what physical scales \citep[e.g.,][]{1998A&A...331L...1S,2012MNRAS.425..605F,2014MNRAS.440.2625Z}. For example, \citet{2015ARA&A..53..115K} imagine a scenario where the outflows are initially momentum conserving due to rapid cooling of the wind. In this case, the outflows lose their energy and can not remove the ISM, so the SMBH grows until
it reaches the $M_{\text{BH}}$--$\sigma_{\star}$ relation. When the SMBH has grown enough, the energy-conserving phase starts. This phase is more energetic than the momentum-conserving phase. As the energy-conserving outflows boost the momentum at larger scales, the outflows sweep the ISM and may suppress the growth of the SMBH. \citet{2012MNRAS.425..605F} argue that the energy-conserving outflows are plausible if the wind is fast ($>\sim$10,000 \kms), or even if the wind is slow ($>\sim$1,000 \kms) with more restricted conditions. 

Since the momentum flux compared to $L_{\text{Bol}}/c$ of our sample are much larger than unity in most cases, we assume that the wind-driven outflows at the nucleus expand to large-scale ($\sim$kpc) outflows via the energy-conserving phase as 
\begin{eqnarray}
\dot{M}_{\text{gas}} v_{\text{gas}} ^2 \approx f \dot{M}_{\text{in}} v_{\text{in}} ^2,
\end{eqnarray}
where $\dot{M}_{\text{in}}$ and $v_{\text{in}}$ are the initial mass outflow rate and initial velocity in the nucleus, respectively. $f$ is an energy transfer efficiency from small-scale wind to large-scale outflow. We adopt an efficiency of 0.2 based on recent observational results \citep{2015Natur.519..436T}. 

The momentum boost, i.e., $\dot{p}_{\text{gas}}/\dot{p}_{\text{in}}$, where $\dot{p}_{\text{gas}} =\dot{M}_{\text{gas}} v_{\text{gas}}$ and $\dot{p}_{\text{in}} =\dot{M}_{\text{in}} v_{\text{in}}$ can be calculated by combining with Equation 13 as follows
\begin{eqnarray}
{{\dot{p}_{\text{gas}}} \over {\dot{p}_{\text{in}}} } \approx 0.2 { {{v_{\text{in}}}} \over {v_{\text{gas}}} }.
\end{eqnarray}
By using this equation, we estimate the values of $v_{\text{gas}}$ and $\dot{p}_{\text{gas}}$ with the expectations from the energy-conserving outflow by assuming $\dot{p}_{\text{in}} = L_{\text{bol}}/c$ (Figure \ref{fig:momentum}). We find that most AGNs lie within $v_{\text{in}} = (0.01-0.3)c$, which is broadly consistent with ultra fast outflows (UFOs, $v_{\text{in}} \approx$ (0.1--0.4)c) detected in X-ray observations \citep[e.g.,][]{2015Natur.519..436T}. We do not find, however, any clear evidence for instantaneous quenching of the star formation due to outflows, since 11 out of 20 AGNs still show a sign of on-going star formation (see Sections \ref{disk} and \ref{mixture}). 

On the contrary, if it were the momentum-conserving phase, the momentum boost would be expected to be 0.2 regardless of $v_{\text{gas}}$, which is not the case for our sample. These results imply that energy-conserving outflows due to accretion-disk wind might be the driving mechanism of the ionized gas outflows observed in the sample.  

Note that we have two extreme AGNs with a high momentum boost (J0952+1937 and J1520+0757), which lead to un-physically large $v_{\text{in}} > c$, although the uncertainties are large. One possibility is that the $f$ factor is larger for these AGNs due to largely different ISM composition. For J1520+0757, however, it is difficult to obtain $v_{\text{in}} < c$ even with $f=1$. Another possibility is that the two AGNs may have much lower $L_{\text{bol}}$ at present compared to the time when the outflow was launched, resulting in an overestimated momentum boost compared to the true value. We will discuss this scenario in the following section.

\subsection{Time-delayed AGN outflows}
In Sections \ref{int_energetics} and \ref{origin}, we find two AGNs with extreme energetics, while their bolometric luminosities are relatively low (J0952+1937 and J1520+0757). These two AGNs have the lowest electron density among the sample ($n_e$=78 and 54 cm$^{-3}$, respectively), implying that the ionized gas has possibly been swept out, which is consistent with their high mass outflow rate ($\dot{M}_{\text{out}}$=17.7 and 9.2 $M_{\odot}$ yr$^{-1}$, respectively). Then, what can cause the discrepancy between their energetics, e.g., energy injection rate and momentum flux, and the bolometric luminosity? Since we measured the bolometric luminosity of AGNs based on the [\OIII] luminosity in the NLR, the difference between the photoionization timescale ($\sim$$10^4$ years) and the dynamical timescale of the outflows ($\sim$$10^6$ years) to reach the NLR may cause the discrepancy. We note that this is the same explanation for the weak relationship between $R_{\text{out}}$ and the [\OIII] luminosity (see Figure \ref{fig:size_lum_kin}). 

Let's consider the energy-conserving outflow expanding from high velocity, small-scale winds close to the nucleus (see Section \ref{origin}). As the winds expand, the hot bubble would sweep out the surrounding ISM and the outflow velocity would become smaller. If we simply consider the constant outflow velocity of 1000 \kms, it will take $10^6$ years, which is the dynamical timescale of outflows, to reach the ISM up to $R_{\text{out}}$ of 1 kpc. Also, we find that the mean mass outflow rate is $\sim$4.6 $M_\odot$ yr$^{-1}$ for the sample, so the ionized gas within $R_{\text{NLR}}$ will be removed after $\sim$10$^5$ years. Since the outflow would sweep out the ISM in the vicinity of the SMBH within a much shorter time scale $\ll$10$^5$ years, it is possible that the AGN becomes inactive while the outflows propagate in the NLR ($\sim$kpc). Hence, the AGN can be inactive or less luminous when the powerful outflow reaches the NLR.

\subsection{Do AGN outflows affect star formation? }
\label{feedback}
From the integrated spectra within $R_{\text{NLR}}$, we estimate the physical properties related to the energetics of gas outflows, such as mass outflow rate, energy injection rate, and momentum flux (Section \ref{int_energetics}). We find the mean outflow velocity $v_{\text{out}}=\sim$800 \kms\, and the mean $R_{\text{NLR}}=\sim$900 pc. The dynamical time $t_d$ is given by $t_d = R_{\text{NLR}} / v_{\text{out}} \sim $10$^6$ years, which is an order of magnitude smaller than the theoretical expectation of 10$^7$ years for quasar lifetimes \citep[e.g.,][]{2006ApJS..163....1H}. The AGNs have a mass outflow rate $\dot{M}_{\text{out}}$ = 0.3--17.7 $M_\odot$ yr$^{-1}$ with a mean value of $\sim$4.6 $M_\odot$ yr$^{-1}$, which is a factor of $\sim$300 larger than the mean mass accretion rate $\dot{M}_{\text{acc}} \sim$0.02 $M_\odot$ yr$^{-1}$. For this mean gas mass for AGNs ($\sim$1.4$\times$10$^6$ $M_\odot$) and mean mass outflow rate ($\sim$4.6 $M_\odot$ yr$^{-1}$), it would take $\sim$(4.2$\pm$3.1)$\times$10$^5$ years to remove the ionized gas from the $R_{NLR}\sim$1 kpc, which is less than the dynamical timescale of AGNs $\sim$10$^6$ years. Such gas removal timescale ($\sim$4.2$\times$10$^5$ years) is comparable to the AGN flickering timescale $\sim$10$^5$ years \citep{2015MNRAS.451.2517S}. 

From the quantitative estimations, we show that the AGN outflows can remove the ionized gas within $R_{\text{NLR}}$ in a reasonably short timescale, implying that star formation within $\sim$1 kpc of AGN can be affected by AGN outflows. In Section \ref{disk}, for example, we find that $\sim$30\% of AGNs with a disk are AGN-type disks, which is possibly affected by AGN photoionization. The higher $D_{n}(4000)$ and smaller H$\delta_{A}$ of the AGNs with AGN-type disks indicate a relatively low star-formation rate within the central kpc of the AGNs. We do not find, however, any differences between the AGNs with AGN-type and SF-type disks in terms of outflow energetics (Figures \ref{fig:e_comparison}, \ref{fig:lbol_vs_em}, and \ref{fig:momentum}). 

Nonetheless, the most kinematically energetic AGNs of our sample, i.e., J0952+1937 and J1520+0757, may provide an insight into AGN feedback. For example, J1520+0757 has no detectable rotational feature in \Ha, residing in the ``green valley'' with log SSFR=$-$10.9 yr$^{-1}$. On the other hand, J0952+1937 has a SF-type disk and harbors on-going star formation with log SSFR=$-$9.5 yr$^{-1}$ (see Figure \ref{fig:ssfr}). If the outflows in these AGNs are powerful enough to affect the star formation in the host galaxy, their different SSFRs are possibly due to different geometry of outflows with respect to the star-forming disk. This speculation can be tested using high spatial resolution IFU observations combined with a proper modeling of a mixed kinematics of outflows and disk rotation. Alternatively, the outflows may not be powerful enough to immediately quench the star formation in the host galaxy, although the ionized gas shows evidence of powerful outflows. Recent numerical simulations show the inefficiency of AGN outflows in quenching star formation, because the outflows are mostly affecting ionized gas, rather than dense molecular gas \citep[e.g.,][]{2014MNRAS.441.1615G}. This scenario can be tested whether the object shows outflow features in molecular gas as well as ionized gas via, e.g., ALMA observations.

Do AGN outflows also affect star formation on galactic scales ($\sim$10 kpc)? From our results, it would be difficult for our intermediate-luminosity AGNs, because of the steeply decreasing energy and momentum as a function of distance (see Section \ref{sr_energetics}). Also, \citet{2016ApJ...833..171K} showed that $\sim$90\% of outflow energy and momentum are within a $\sim$1--2 kpc region ($\sim$$R_{\text{out}}$) in type 2 AGNs, implying that energy injected by AGN outflows would become smaller and eventually negligible at larger distance, e.g., $\sim$10 kpc, for the AGN luminosities of our sample. 

In Section \ref{disk}, we find that the AGNs with SF-type disks are located on the main sequence of star-forming galaxies, while the AGNs with AGN-type disks are below the main sequence. The global SSFR of host galaxies of the AGNs with AGN-type disks is on average $\sim$1 order of magnitude smaller than that in the AGNs with SF-type disks. Does the low SSFR in the sample of AGN-type disks support the AGN feedback in galactic scales? By considering that the AGNs with AGN-type disks are on average a factor of $\sim$3.5 more massive than the AGNs with SF-type disks (see Section \ref{disk}), it is difficult to conclude that the low SSFRs support the galactic scale AGN feedback, although it is a plausible scenario. It is because such low SSFRs in massive galaxies can also be accounted for alternative quenching mechanisms, e.g., environmental quenching \citep[e.g.,][]{Smith:2012bn,Peng:2015bq}. More IFU-observed samples within similar mass scales may provide a hint for the primary quenching mechanism.

\section{Summary \& Conclusion}
\label{summary}
Using the Magellan/IMACS-IFU and the VLT/VIMOS-IFU, we obtained the spatially resolved kinematics of ionized gas in the NLR for a luminosity-limited sample of 20 local type 2 AGNs, which are selected from a large sample of $\sim$39,000 type 2 AGNs from SDSS DR7 \citep{2016ApJ...817..108W}, 
based strong outflow signatures, i.e., velocity dispersion $>$ 300 \kms\ and/or large velocity offset $>$ 100 \kms\ in [\OIII], with [\OIII] luminosity log $L_{\text{[O III]}} > 41.5$ \ergs. These AGNs are arguably best suited for studying AGN outflows. Here we summarize the main results.

\smallskip

$\bullet$ By performing a decomposition on the emission-line profile, we successfully obtained the flux and kinematic maps of the narrow- and the broad components of the [\OIII] and \Ha\ lines. The broad components in both [\OIII] and \Ha\ represent the non-gravitational kinematics, i.e., gas outflows, while the narrow components, especially in \Ha, represent the gravitational kinematics.

\smallskip

$\bullet$ We measured the photometric size of the NLR ($R_{\text{NLR}}$) based on the [\OIII] flux distribution. By combining our sample and 29 luminous quasars from the literature, we obtained the photometric size-luminosity relation as $R_{\text{NLR}} \propto L_{\text{[O III]}} ^{0.41}$, which is consistent with the results from the literature \citep[e.g.,][]{2003ApJ...597..768S, 2013MNRAS.430.2327L}. 

\smallskip

$\bullet$ We obtained the outflow size ($R_{\text{out}}$) based on the spatially resolved [\OIII] kinematics as \citet{2016ApJ...819..148K}. 
We found no clear outflow size--luminosity relation, presumably due to the dynamical timescale of the outflows ($\sim$$10^6$ years).  

\smallskip

$\bullet$ By using the integrated spectra within $R_{\text{NLR}}$, we estimated the physical quantities of the outflow energetics. The estimated ionized gas mass is (1.0--38.5)$\times 10^5 M_{\odot}$ while the mean mass outflow rate $\dot{M}_{\text{out}}$ is 4.6$\pm$4.3 $M_\odot$ yr$^{-1}$, which is factor of $\sim$260 higher than the mean $\dot{M}_{\text{acc}}\sim$0.02$\pm$0.01 $M_\odot$ yr$^{-1}$. The result implies powerful mass loading of the AGN outflows by the ISM \citep[e.g.,][]{2005ARA&A..43..769V,2009MNRAS.396....2B}.    

\smallskip

$\bullet$ The majority (18 out of 20) of AGNs have relatively low energy injection rate, which is about 0.8$\pm$0.6\% of $L_{\text{bol}}$, and also have relatively low momentum flux, which is about $\sim$5.4$\pm$3.6$\times$$L_{\text{bol}}/c$. Both the energy injection rate and the momentum flux correlate, in general, with $L_{\text{bol}}$. The estimated outflow parameters are consistent with the expectations from the energy-conserving outflow scenario with outflow velocities of $\sim$0.01-0.3c near the accretion disk \citep[e.g.,][]{2012MNRAS.425..605F}. However, we find no supporting evidence for instantaneous quenching of the star formation due to the outflows. 

\acknowledgments
We thank the anonymous referee for his/her valuable comments and suggestions. The work of HJB was supported by NRF (National Research Foundation of Korea) Grant funded by the Korean Government (NRF-2010-Fostering Core Leaders of the Future Basic Science Program). JHW acknowledges the support by the NRF grant funded by the Korea government (No. 2016R1A2B3011457 and No. 2010-0027910). This work is based on observations made with ESO Telescopes at the La Silla Paranal Observatory under program ID 091.B-0343(A) (PI: Flohic), and data gathered with the 6.5 meter Magellan Telescopes located at Las Campanas Observatory, Chile. This work is also supported by K-GMT Science Program ID GN-2015A-Q-204 (PI:Woo) of Korea Astronomy and Space Science Institute (KASI). Based on observations obtained at the Gemini Observatory processed using the Gemini IRAF package, which is operated by the Association of Universities for Research in Astronomy, Inc., under a cooperative agreement with the NSF on behalf of the Gemini partnership: the National Science Foundation (United States), the National Research Council (Canada), CONICYT (Chile), Ministerio de Ciencia, Tecnolog\'{i}a e Innovaci\'{o}n Productiva (Argentina), and Minist\'{e}rio da Ci\^{e}ncia, Tecnologia e Inova\c{c}\~{a}o (Brazil).

\appendix
\renewcommand\thefigure{\thesection.\arabic{figure}}
\section{Additional Figures}

\setcounter{figure}{0}

\begin{figure*}[!hb]
\centering
\includegraphics[width=0.8\textwidth]{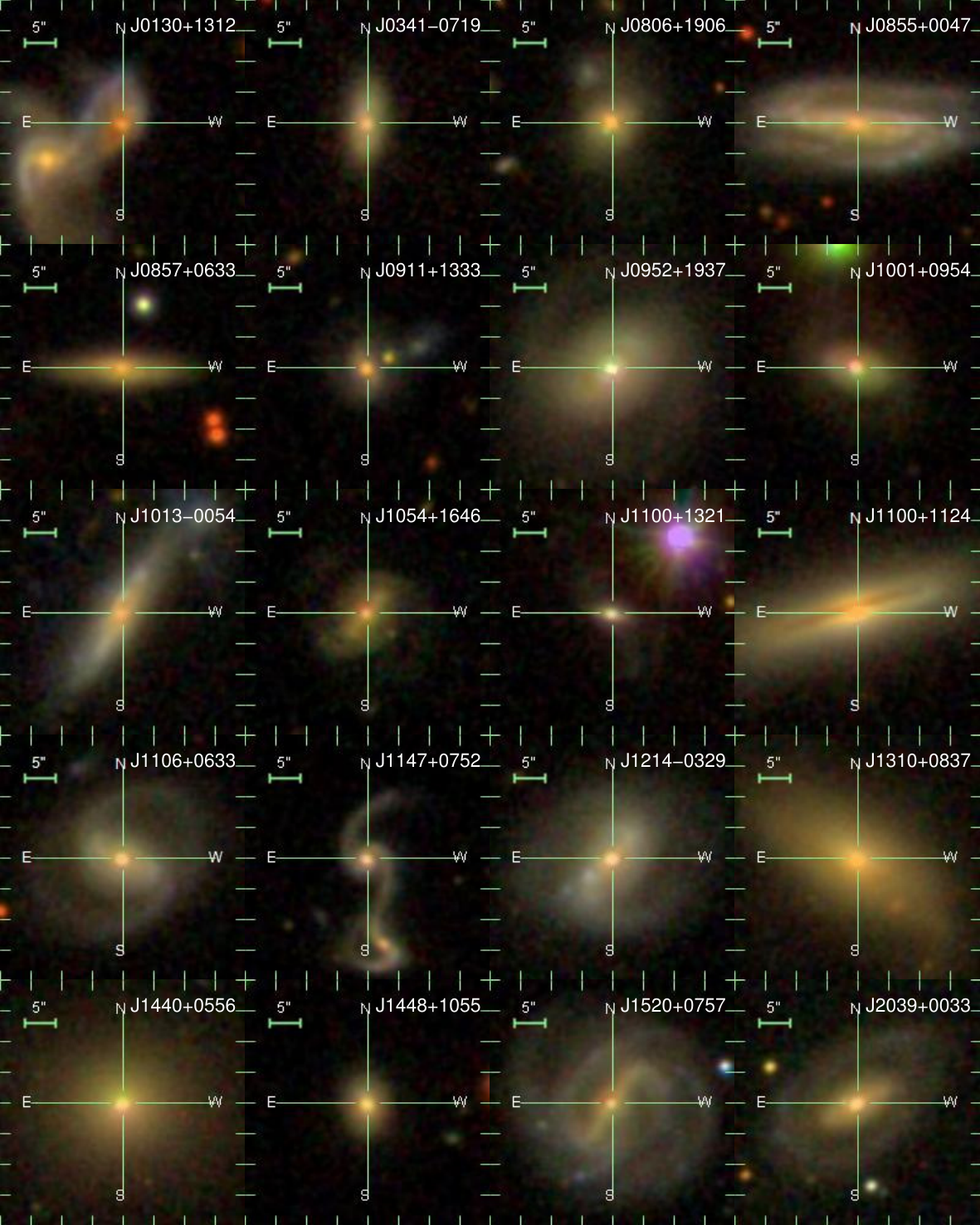}
\caption{SDSS $gri$-composite images (40\arcsec$\times$40\arcsec) of the 20 AGNs listed in accending R.A. from top-left to bottom-right. The name of the AGN is shown at the top-right corner of each image.}
\label{fig:sdss_image}
\end{figure*}

\begin{figure*}
\centering
\includegraphics[width=0.48\textwidth]{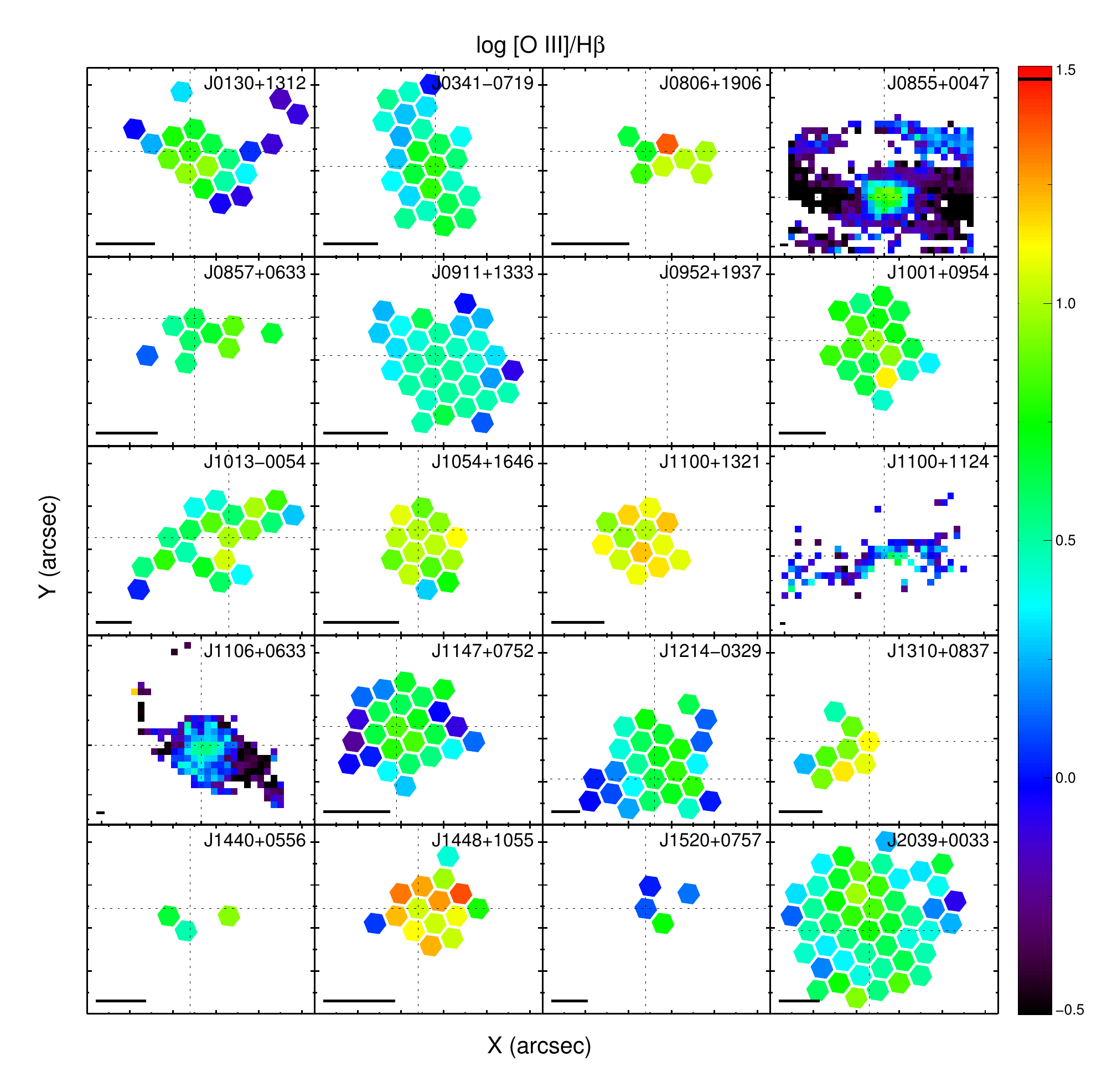}
\includegraphics[width=0.48\textwidth]{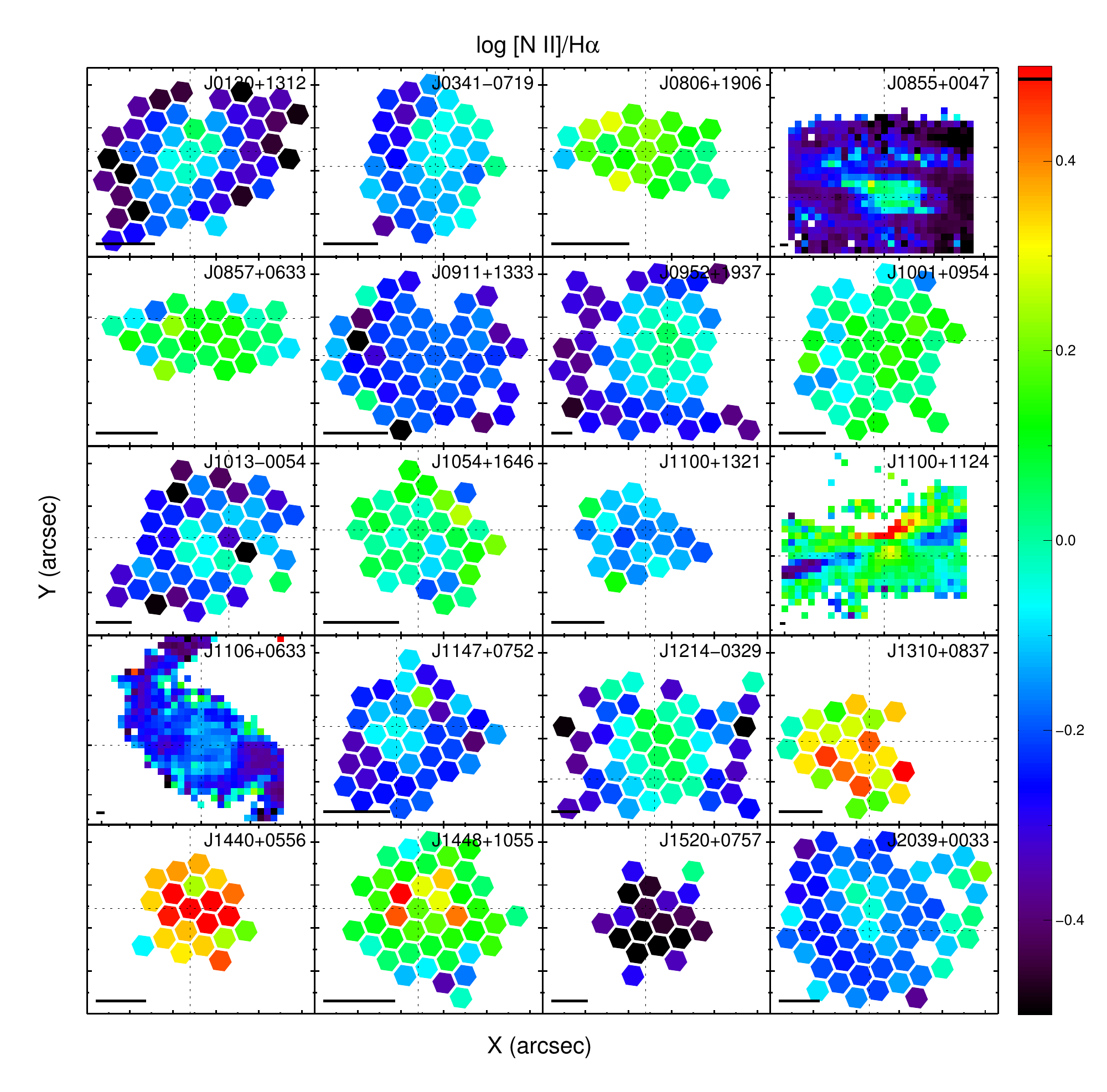}
\includegraphics[width=0.48\textwidth]{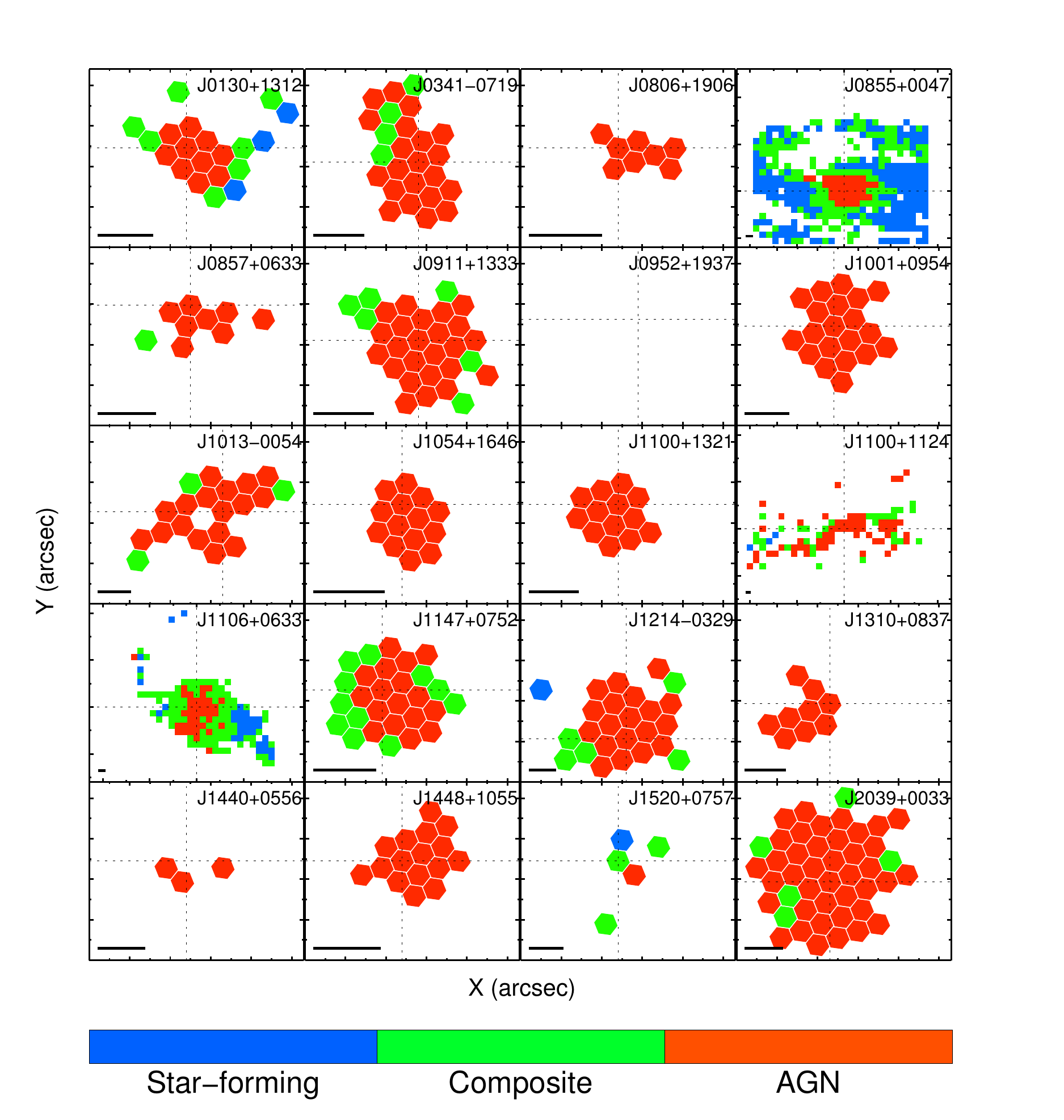}
\caption{The maps for the emission-line ratios of [\OIII]/H$\beta$ (top-left) and [\NII]/H$\alpha$ (top-right), and the emission-line diagnostics (bottom). The major ticks in both x- and y-axes represent 1\arcsec\ for the Magellan targets, while the major ticks denote 5\arcsec\ for the VLT targets, i.e., J0855+0047, J1100+1124, and J1106+0633. The AGNs are listed in order of accending R.A from top-left to bottom-right as in Figure \ref{fig:sdss_image}. We note that J0952+1937 has no maps either [\OIII]/H$\beta$ nor in emission-line diagnostics since the H$\beta$ line is in the gap between the CCDs.}
\label{fig:line_ratio}
\end{figure*}

\end{document}